\documentclass[12pt,oneside,final]{amsart}

\usepackage[utf8]{inputenc}
\usepackage[T1]{fontenc}
\usepackage[BCOR=0mm]{typearea}
\usepackage{amsmath,amsfonts,amssymb,amsthm}
\usepackage{mathtools}
\usepackage{stackrel}
\usepackage{showkeys}
\usepackage[foot]{amsaddr}
\usepackage[color=green!50]{todonotes}

\usepackage{tensor}
\usepackage{cite}

\usepackage[a4paper, textwidth=16cm,textheight=21cm]{geometry}

\DeclareMathAlphabet{\mathdutchcal}{U}{dutchcal}{m}{n}
\SetMathAlphabet{\mathdutchcal}{bold}{U}{dutchcal}{b}{n}
\DeclareMathAlphabet{\mathdutchbcal}{U}{dutchcal}{b}{n}
\newcommand{\ST}{\mathcal{S}}
\newcommand{\brstb}{\mathdutchcal{b}}
\newcommand{\brsts}{\mathdutchcal{s}}

\def\be{\begin{equation}}
\def\ee{\end{equation}}
\def\dm{\int d^4x\sqrt{-g}}
\def\uJ{\underline{J}}

\allowdisplaybreaks
\DeclareRobustCommand{\SkipTocEntry}[4]{}

\numberwithin{equation}{section}

\begin{document}

\title[Einstein-Hilbert Gravity Embedded in a Higher Derivative Model]{On the Perturbative Quantization of Einstein-Hilbert Gravity Embedded in a Higher Derivative Model}

\author{Steffen Pottel}
\address{K\"uhne Logistics University, Hamburg, Germany. \emph{Email address: }\rm \texttt{steffen.pottel(at)the-klu.org}.}

\author{Klaus Sibold}
\address{Institute for Theoretical Physics, Leipzig University, Germany. \emph{Email address: }\rm \texttt{sibold(at)physik.uni-leipzig.de}.}

\date{\today}

\begin{abstract}
In a perturbative approach Einstein-Hilbert gravity is quantized about a flat background.
In order to render the model power counting renormalizable, higher order curvature terms are added to the action. 
They serve as Pauli-Villars type regulators and require an expansion in the number of fields in addition to the standard expansion in the number of loops.
Renormalization is then performed within the BPHZL scheme, which provides the action principle to construct the Slavnov-Taylor identity and invariant differential operators.
The final physical state space of the Einstein-Hilbert theory is realized via the quartet mechanism of Kugo and Ojima. 
Renormalization group and Callan-Symanzik equation are derived for the Green functions and, formally, also for the $S$-matrix.
\end{abstract}

\maketitle
{
\tableofcontents
}

\setlength{\parindent}{0em}
\setlength{\parskip}{2ex}

\section{Introduction\label{se:introduction}}

In the perturbative construction of Einstein-Hilbert (EH) gravity on four dimensional spacetime one splits the metric $g^{\mu\nu}$ into a background $\hat{g}^{\mu\nu}$ and oscillations $h^{\mu\nu}$ around it which are quantized.
Back in the 1970's quite a few attempts were undertaken to formulate such models of quantized gravity. 
Most influential were the pioneering papers of 'tHooft and Veltman \cite{tHooft:1974toh}, in which explicit calculations showed that in higher than one-loop order the theory becomes intractable due to power counting non-renormalizability. 
Many more papers dealt with the problem without surmounting these difficulties (see e.g.\ \cite{Goroff:1985th}). 
Out of these early papers we concentrate on two in which important progress had been achieved and which were very helpful for our own understanding.\\
Kugo and Ojima \cite{KuOj} provided a quantized model of EH general relativity.
In order to deal with the indefinite metric problem which results after having replaced diffeomorphism invariance by an appropriate Becchi-Rouet-Stora-Tyutin (BRST) invariance they use their quartet mechanism. 
Hence they realize unitarity. 
They base their reasoning, remarkably enough, on a general solution of the Slavnov-Taylor identity (ST)
associated with the BRST transformation without restriction by power counting.
This is, of course, motivated by the fact that the model is power counting non-renormalizable, hence quite reasonable. The renormalization problem is left
open. \\
Stelle \cite{Stelle} presented a complementary approach to quantize
classical relativity: he added the square of the Ricci tensor and the
square of the curvature scalar to the EH action. This model is 
power counting renormalizable, but it is not unitary. Looking at the propagator 
which has a fall-off like $1/(p^2)^2$ for large $p$ it is obvious that the lack 
of unitarity has nothing to do with the gauge dependence of the model, but 
originates from the {\sl invariants} which contain four derivatives of the metric. \\
Calculations to be presented below show that the gauge fixing of \cite{KuOj}
can also be used in the context, where the square of the Ricci tensor and the square of the scalar curvature are present in the action. 
Hence one has the quartet mechanism at one's disposal. 
Since the higher derivative terms render the model power counting renormalizable, we could be led to interpret the regularizing effect as Pauli-Villars type, which can be removed after renormalization with a suitable scheme \cite{Zimmermann:1975gk}.
This turns out to be wrong. 
We rather arrive at the conclusion that the higher derivatives
are tied fundamentally to the EH theory. Their seemingly disastrous effect
of causing negative metric in state space can be overcome by a
suitable LSZ-projection. The dependence of the resulting theory from the 
additional 
two coupling parameters however remains. Since this enlarged model is power 
counting renormalizable, but depends crucially on a field of canonical 
dimension zero, it contains infinitely parameters, which are associated with
the redefinition of this field as a function of itself. These generalized field
amplitudes are, fortunately, of gauge parameter type, hence do not contribute
to physical quantities.\\
Before going into details of the realization of the model we would like to present
the argument why we are convinced that the higher derivatives are necessary
ingredients for the definition of EH in quantum field theory.\\
Suppose we would like to gauge the translations in a matter model, say a massless
scalar field of canonical dimension one, with the usual Noether procedure, then
one lets the parameter $a_\mu$ of the translations depend on $x$ and couples
the respective conserved current, the energy-momentum-tensor $T_{\mu\nu}$ (EMT), 
to an external tensor field field $h^{\mu\nu}$. This entails the field $h$ with
transformations dictated by the local translations. These turn out to be
just the general coordinate transformations known from general relativity (GR).
If then the field $h$ becomes a dynamical field with its own invariant kinetic
term, this kinetic term has to involve four derivatives, if one wants to keep
power counting renormalizability: after all the EMT has canonical dimension four.
I.e. the field $h$ must have dynamical dimension zero. The metric $g^{\mu\nu}$
which arises also in the course of the Noether procedure is given by
$g^{\mu\nu}=\eta^{\mu\nu}+h^{\mu\nu}$ -- without any parameter carrying 
dimension. Quite reasonable in QFT. In classical GR the metric may depend on 
parameters which have
mass dimension, but that is the engineering dimension and not the dynamical one
which it has to be in quantum field theory, where the dimensions are dictated
by the kinetic terms. (The details of this derivation can be found in 
\cite{EKKSI, EKKSII, EKKSIII}. However many other authors have considered gauging translations,
concluding that the resulting gauge theory is a gravitational theory with
higher derivatives, e.g. \cite{Hehl:2020hhp} and citations therein.) \\
We therefor continue with quantization, renormalization and analysis of the
implications.\\

We choose the Bogoliubov-Parasiuk-Hepp-Zimmermann-Lowenstein (BPHZL) renormalization scheme \cite{Zimmermann:1969jj,Lowenstein:1975ps} for our purposes. 
The auxiliary mass which is required in this scheme, is put in by hand, but it serves very well to construct finite Green functions since
the higher derivatives rendered the model power counting renormalizable.
The main gain of this version to deal with the UV-infinities is that one has an 
action principle \cite{Lowenstein:1971jk} 
at one's disposal which one would not have in the power counting
non-renormalizable EH model. The hurdle that this scheme is not
BRST invariant can be overcome by cohomology results existing in the literature
since the 1980's (see \cite{Baulieu:1983tg}).
They become now powerful tools because -- supplemented by power counting -- they exist also analytically.\\ 
Even in this rather modest approach of quantizing gravity, namely perturbation theory and flat background, one encounters quite 
a few difficulties: the interaction is non-polynomial and the main field to start 
with has canonical dimension zero, hence in a perturbative approach one has
an expansion in the number of loops and in the number of fields -- a situation
familiar from supersymmetric gauge theories \cite{Piguet:1984mv}. The presence of a field with
vanishing canonical dimension, which goes hand in hand with propagators 
falling off as $1/(p^{2})^2$ for large $p$, points to possible infrared problems
already off-shell. Those will be controlled by infrared power counting
which is a built-in instrument of the scheme.

The paper is structured according to the use of the fundamental field $h^{\mu\nu}$. In  
sections \ref{se:treeapproximation}$\to$\ref{se:removingregulators} we take $h$ at face value and formulate in terms of it the standard invariants of general relativity related to: $R,R^2,R^{\mu\nu}R_{\mu\nu}$ -- expanded in terms of $h$. We call this the ``special solution'' (of diffeomorphism invariance). In the tree approximation we set up the model, construct propagators, the ST identity, prove unitarity of the $S$-matrix, make explicit the parameters of the model and look at gauge parameter independence. In Sect.\ \ref{se:renormalization} we start the renormalization by introducing an auxiliary mass required in the BPHZL scheme which we use. Central is then power counting: in the ultraviolet (UV) and infrared (IR) region of
momentum space integrations, and convergence. It guarantees the existence of normal product insertions and thus of Green functions: one-particle-irreducible (1PI) or vertex functions, connected and general one's. We then establish the ST to all orders of perturbation theory. 
Thereby formal unitarity of the $S$-matrix is established. Sections \ref{se:invdiffop}$\to$\ref{se:invparadiffeq} are devoted to the derivation and use of symmetric differential operators which yield parametric differential equations: the Lowenstein-Zimmermann (LZ) equation which shows that the Green functions are ultimately independent of the auxiliary mass; the renormalization group (RG) equation which governs the change of the normalization parameter; the Callan-Symanzik equation (CS) which yields the scaling properties of Green functions. 
In Sect.\ \ref{se:removingregulators} we project down to the EH-theory.
In Sect.\ \ref{se:generalsolutionSTI} we study the ``general'' solution, i.e.\ we replace the
original field $h$ by an arbitrary function of itself 
$h^{\mu\nu}\to \mathcal{F}^{\mu\nu}(h)$. This is possible due to the 
vanishing canonical dimension of $h$ and this space of functions $\mathcal{F}$ is swept out in the course of renormalization, hence the study is necessary.
Sect.\ \ref{se:DisCon} is devoted to discussions and conclusions.

\section{Tree approximation\label{se:treeapproximation}}
For a decent perturbative treatment it is mandatory to set up the first orders
carefully. In the present context this refers to the zero-loop order and the
first and second order in the number of fields.

\subsection{The model and its invariances\label{se:modelandinvariances}}
As explained in the introduction we base our study of EH in the more general context of permitting invariants under diffeomorphisms up to fourth order in the derivatives. Restricting ourselves to spacetimes which are topologically equivalent to flat one's we may use the Gau\ss-Bonnet theorem and express the square of the Riemann tensor
in terms of the Ricci tensor and the curvature scalar
\begin{equation}\label{GaBo}
\int\sqrt{-g}R^{\mu\nu\rho\sigma}R_{\mu\nu\rho\sigma}
=\int\sqrt{-g}(4R^{\mu\nu}R_{\mu\nu}-R^2).
\end{equation}
Together with the cosmological constant a basis of invariants is then provided by the terms in the following action 
\begin{equation}\label{ivc}
	\Gamma^{\rm class}_{\rm inv}=\dm(c_0\kappa^{-4}
	+c_3\kappa^{-2}R+c_2R^2+c_1R^{\mu\nu}R_{\mu\nu}) \, .
\end{equation}
Here $\kappa$ denotes the gravitational constant.
The invariance under general coordinate transformations is to be translated into
Becchi-Rouet-Stora-Tyutin invariance (BRST) with respective gauge fixing.
The field $h^{\mu\nu}$ is defined via
\begin{equation}\label{dfh}
h^{\mu\nu}=g^{\mu\nu}-\eta^{\mu\nu} .
\end{equation}
The propagators of $h$ (s.b.) will tell us that 
$h$ has canonical dimension $0$, hence $\kappa$ must not show up in its definition.

The classical action 
\begin{eqnarray}\label{clssct}
\Gamma^{\rm class}&=& \Gamma^{\rm class}_{\rm inv} + \Gamma_{\rm gf} 
				  + \Gamma_{\phi\pi}+\Gamma_{\rm e.f.}\\
\Gamma_{\rm gf}&=&-\frac{1}{2\kappa}\int g^{\mu\nu}
                                (\partial_\mu b_\nu+\partial_\nu b_\mu) 
	                     -\frac{1}{2}\alpha_0\int \eta^{\mu\nu}b_\mu b_\nu \label{gf1} \\
	\Gamma_{\phi\pi}&=&-\frac{1}{2}\int(D^{\mu\nu}_\rho c^\rho)
      (\partial_\mu \bar{c}_\nu +\partial_\nu\bar{c}_\mu)\label{gf2}\\
D^{\mu\nu}_\rho&\equiv&-g^{\mu\lambda}\delta^\nu_\rho\partial_\lambda
		       -g^{\nu\lambda}\delta^\mu_\rho\partial_\lambda
		       +\partial_\rho g^{\mu\nu}\\
	\Gamma_{\rm e.f.}&=&\int (K_{\mu\nu}\brsts h^{\mu\nu}+L_\rho \brsts c^\rho)	   
\end{eqnarray}
is invariant under the BRST-transformation
\begin{eqnarray}\label{brst}
	\brsts g^{\mu\nu}&=&\kappa D^{\mu\nu}_\rho c^\rho
	\qquad \brsts c^\rho=-\kappa c^\lambda \partial_\lambda c^\rho	 \\
	\brsts \bar{c}_\rho &=& b_\rho
	\qquad \brsts b_\rho=0\\
	\brsts_0 h^{\mu\nu}&=&-\kappa(\partial^\mu c^\nu+\partial^\nu c^\mu)\\
	\brsts_1 h^{\mu\nu}&=&
	      -\kappa(\partial_\lambda c^\mu h^{\lambda\nu}
			     +\partial_\lambda c^\nu h^{\lambda\mu}
			     -c^\lambda\partial_\lambda h^{\mu\nu}) .
\end{eqnarray}
In accordance with the expansion in the number of fields we
have introduced the transformations $\brsts_0,\brsts_1$ which maintain the number, resp.\ raise it by one.
$K_{\mu\nu}, L_{\rho}$ are external fields to be used for generating insertions 
of non-linear field transformations. The Lagrange multiplier 
$b_\mu$ 
couples to $\partial_\lambda h^{\mu\lambda}$ and thus
fixes eventually these derivatives (deDonder like gauge fixing). 
Since the terms
$R^2, R^{\mu\nu}R_{\mu\nu}$ contain however four derivatives one might be
tempted to fix also the higher derivatives in a corresponding manner, or
only those. It turns out that this is superfluous or even contradictory when using a Lagrange multiplier field $b$,
so we stick to (\ref{gf1}),(\ref{gf2}) which is the gauge fixing chosen in \cite{KuOj}.

\subsection{Propagators\label{se:propagators}}
The definition of the propagators as inverse of vertex functions requires the
knowledge of first and second orders in the number of fields of (\ref{clssct}). 
Since the cosmological term contributes at first order in the field $h$ we suppress it here in the tree approximation by
putting $c_0=0$ and in higher orders by a normalization 
condition. (A classical argument for this demand is that flat space should be a solution to the h-field equations.)
In Fourier space one arrives at 
\begin{eqnarray}\label{bln}
	\Gamma_{h_{\mu\nu}h_{\rho\sigma}}&=& 
	\frac{1}{4}\sum_{KLr}\gamma^{(r)}_{KL}(P_{KL}^{(r)})_{\mu\nu\rho\sigma} \\
	\Gamma_{b_\rho h_{\mu\nu}}&=&
	-\frac{i}{\kappa} \Big(\frac{1}{2}(\theta_{\rho\mu} p_\nu+\theta_{\rho\nu}p_\mu) 
                                                 +\omega_{\mu\nu}p_\rho \Big) \\
	\Gamma_{b_\rho b_\sigma}&=& -\alpha_0\eta_{\rho\sigma} \\
	\Gamma_{c_\rho\bar{c}_\sigma}&=& -ip^2 \big( \theta_{\rho\sigma}\xi(p^2)+
	\omega_{\rho\sigma}\frac{1}{2}\eta(p^2) \big) .
\end{eqnarray}
For the $h$-bilinear terms we introduced projection operators $P$ (see App.\ A)
and general coefficient functions $\gamma$. It will turn out that the
propagators can be uniquely determined for general scalar functions 
$\gamma(p^2)$ with the projectors taking care of the spin structure
inherent in the terms of $(\ref{ivc})$. In tree approximation the values for
$\gamma$ are given by 
\begin{eqnarray}\label{coffs}
\gamma^{(2)}_{TT} &=&-p^2(c_1p^2-c_3\kappa^{-2})\\
\label{eq:coffs0}
\gamma^{(0)}_{TT} &=&p^2 \big((3c_2+c_1)p^2+\frac{1}{2}c_3\kappa^{-2} \big)\\ 
	\gamma^{(1)}_{SS}&=&\gamma^{(0)}_{WW}=\gamma^{(0)}_{TW}=\gamma^{(0)}_{WT}=0 .
\end{eqnarray}
The coefficients of $\Gamma_{bh}$ and
$\Gamma_{bb}$ will turn out to be fixed, whereas those of $\Gamma_{c\bar{c}}$
again can be very general with tree values $\xi=\eta=1$.

The inversion equations to obtain the propagators read for the bosonic fields
\begin{eqnarray}\label{bosinver}
\Gamma_{h_{\mu\nu}h_{\alpha\beta}}G^{h^{\alpha\beta}h^{\rho\sigma}}
+\Gamma_{h_{\mu\nu}b_\lambda}G^{b^\lambda h^{\rho\sigma}}&=&
      \frac{i}{2}(\tensor{\eta}{_\mu^\rho} \tensor{\eta}{_\nu^\sigma} + \tensor{\eta}{_\mu^\sigma} \tensor{\eta}{_\nu^\rho})\\
	      \Gamma_{hh}G^{hb}+\Gamma_{hb}G^{bb}&=&0\\
	      \Gamma_{bh}G^{hh}+\Gamma_{bb}G^{bh}&=&0\\
      \Gamma_{b_\rho h^{\alpha\beta}}G^{h^{\alpha\beta} b^\sigma}
      +\Gamma_{b_\rho b_\lambda}G^{b^\lambda b^\sigma}&=&-i\tensor{\eta}{_\rho^\sigma}.
\end{eqnarray}
\noindent
For the ghosts they have the form
\be\label{ghostinver}
\Gamma_{c^\rho\bar{c}^\lambda}G^{c^{\lambda}\bar{c}^\sigma}=i \tensor{\eta}{_\rho^\sigma}.
\ee
For the $\langle hh \rangle$-propagators we introduce like for the 2-point-vertex functions
an expansion in terms of projection operators
\be\label{hhpropproj}
	G^{hh}_{\mu\nu\rho\sigma}= 
		4\sum_{KLr} \langle hh \rangle^{(r)}_{KL}(P_{KL}^{(r)})_{\mu\nu\rho\sigma}.
\ee
In order to solve the inversion equations we introduce 
\begin{eqnarray}\label{bhprp}
	G^{bh}_{\rho\mu\nu}&=&\frac{\kappa}{p^2} \big((p_\mu \theta_{\nu\rho}
				    +p_\nu \theta_{\mu\rho})b_1
                +p_\rho \omega_{\mu\nu}b_2
		+p_\rho \theta_{\mu\nu}b_3 \big)\\
	G^{hb}&=&G^{bh} .
\end{eqnarray}
Here $b_1,b_2$, and $b_3$ are arbitrary scalar functions such that this is the most
general expression compatible with Lorentz invariance and naive dimensions.\\
The gauge parameter independent solutions $\langle hh \rangle^{(r)}_{KL}$ turn 
out to be 
\be\label{bosprop}
	\langle hh \rangle^{(2)}_{TT}=\frac{i}{\gamma^{(2)}_{TT}}			       
	\qquad\quad\langle hh \rangle^{(0)}_{TT}=\frac{i}{\gamma^{(0)}_{TT}},		       
\ee
whereas the ``gauge parameter multiplet'' is given by
\begin{eqnarray}\label{gaugeprop}
	\langle hh \rangle^{(1)}_{SS}&=&\frac{4i\alpha_0\kappa^2}{p^2} \qquad
	\langle hh \rangle^{(0)}_{WW}=\frac{4i\alpha_0\kappa^2}{p^2}\\		       
	\qquad \langle hh \rangle^{(0)}_{TW}&=&\langle hh \rangle^{(0)}_{WT}=0.       
\end{eqnarray}
It is important to observe that
the gauge parameter independent part is determined by the coefficient
functions $\gamma$, which depend on the model, i.e.\ by the invariants
and -- as will be seen later -- by higher orders, whereas the gauge 
multiplet is essentially fixed and only determined by the specific gauge fixing.
The remaining bosonic propagators read
\be\label{bprop}
\langle b_\rho h_{\mu\nu} \rangle =\frac{\kappa}{p^2} \big( (p_\mu \theta_{\nu\rho}+p_\nu 
      \theta_{\mu\rho})b_1+p_\rho \omega_{\mu\nu}b_2+p_\rho \theta_{\mu\nu}b_3 \big)
\ee
and
\be
\langle b_\rho b_\sigma\rangle =0   .   
\ee
In the tree approximation $b_1=b_2=1$ and $b_3=0$.
The antighost/ghost propagator has the general form 
\be\label{ggpropKO}
\langle \bar{c}_\rho c_\sigma \rangle=\frac{-1}{p^2}
    \Big( \frac{\theta_{\rho\sigma}}{\xi(p^2)}+ \frac{1}{2} \frac{\omega_{\rho\sigma}}{\eta(p^2)} \Big).
\ee
The tree approximation values are $\xi=\eta=1$, s.t.\
\be\label{ggpropEK}
\langle \bar{c}_\rho c_\sigma \rangle =-i \big(\theta_{\rho\sigma}
                        +\frac{1}{2}\omega_{\rho\sigma} \big) \frac{1}{p^2}.
\ee
We note that $\langle bb \rangle =0$, in accordance with the field $b_\rho$ to be
a Lagrange multiplyer.

Another general remark is in order. In the Landau gauge $\alpha_0 = 0$
the two-point functions $\langle hh \rangle$ fall off for large $|p|$ like $|p|^{-4}$,
hence one has to associate to the field $h$ the canonical dimension zero.
This implies that field monomials $\partial^{\boldsymbol{\mu}}h\cdots h$ always have
canonical dimension $|\boldsymbol{\mu}|= \hbox{\rm degree}$ of the multiderivative 
$\partial^{\boldsymbol{\mu}}$, independent of the number of fields $h$ in the monomial. 

\subsection{The Slavnov-Taylor identity in tree approximation\label{se:STidentitytree}}
Since the $\brsts$-variations of $h,c$ are non-linear in the fields, they are best
implemented in higher orders via coupling to external fields 
(cf. \eqref{clssct}), hence the ST identity then reads
\be\label{fbrst}
\mathcal{S}(\Gamma)\equiv
\int(\frac{\delta\Gamma}{\delta{K}}\frac{\delta\Gamma}{\delta h}
+\frac{\delta\Gamma}{\delta L}\frac{\delta\Gamma}{\delta c}
+b\frac{\delta\Gamma}{\delta\bar{c} })=0 .
\ee
Since the $b$-equation of motion
\be\label{beq}
\frac{\delta \Gamma}{\delta b^\rho}=
                     \kappa^{-1}\partial^\mu h_{\mu\rho}-\alpha_0b_\rho
\ee
is linear in the quantized field $b$, it can be integrated trivially to the original
gauge fixing term. Thus it turns out to be useful to introduce a functional
$\bar{\Gamma}$ which does no longer depend on the $b$-field:
\be\label{Gmmbr}
\Gamma=\Gamma_{\mathrm{gf}}+\bar{\Gamma} .
\ee
One finds
\be\label{rstc}
\kappa^{-1}\partial_\lambda\frac{\delta\bar{\Gamma}}{\delta K_{\mu\lambda}}
+\frac{\delta\bar{\Gamma}}{\delta\bar{c}_\mu} =0
\ee
as restriction. Hence $\bar{\Gamma}$ depends on $\bar{c}$ only via
\be\label{sceH}
H_{\mu\nu}=K_{\mu\nu} - \frac{1}{2\kappa}(\partial_\mu\bar{c}_\nu+\partial_\nu\bar{c}_\mu)
\ee
and the ST identity takes the form
\begin{eqnarray}\label{brGm}
\mathcal{S}(\Gamma)&=&\frac{1}{2}\mathcal{B}_{\bar{\Gamma}}\bar{\Gamma}=0\\
	\mathcal{B}_{\bar{\Gamma}}&\equiv& 
	\int(
  \frac{\delta\bar{\Gamma}}{\delta H}\frac{\delta}{\delta h}
+ \frac{\delta\bar{\Gamma}}{\delta h}\frac{\delta}{\delta H} 
+ \frac{\delta\bar{\Gamma}}{\delta L}\frac{\delta}{\delta c}
+ \frac{\delta\bar{\Gamma}}{\delta c}\frac{\delta}{\delta L}
	) .
\end{eqnarray}
This form shows that $\mathcal{B}_{\bar{\Gamma}}$ can be interpreted as a variation und thus
(\ref{brGm}) expresses an invariance for $\bar{\Gamma}$.

\subsection{Unitarity in the tree aproximation\label{se:unitaritytree}}
The $S$-operator can be defined \cite{Itzykson:1980rh} via
\begin{eqnarray}\label{sma}
	S&=&:\Sigma: Z(\uJ)|_{\uJ=0}, \\
	\Sigma &\equiv& \exp\left\{ {\int dx\,dy\, \Phi_{\rm in}(x)K(x-y)z^{-1}
	\frac{\delta}{\delta \uJ(y)}}\right\}, 
\end{eqnarray}
where $\uJ$ denotes the sources 
$J_{\mu\nu},j_{\bar{c}}^\rho,j_c^\rho,j^\rho_b$ for the fields 
$h^{\mu\nu},\bar{c}^\rho,c^\rho,b_\rho$, respectively, and
their in-field versions are collected in $\Phi_{\rm in}.$ 
$K(x-y)z^{-1}$ refers to all in-fields and stands for the higher derivative wave operator,
hence removes the complete (tree approximation) propagator matrix.
$\Sigma$ would then map onto the respective large Fock space of the higher derivative model. As mentioned already the dynamical degrees of freedom which originate from the
higher derivatives are definitely unphysical,
therefore they have to be removed before we consider the S-matrix for 
the Einstein-Hilbert theory. Here in the tree approximation this is trivial
because all Green functions are well-defined. So we put simply $c_1=c_2=0$.
With this the massive poles are absent, the wave operator is the one of Einstein-Hilbert
and we study just those unphysical degrees of freedom which go along with that 
model. These differ slightly from those studied by \cite{KuOj} because we employ a different field $h$, but the general structure is the same (cf.\ (\ref{Gldbrgv})).
Here we follow \cite{Becchi:1985bd}
and would like to show, that the $S$-matrix
commutes with the BRST-charge $Q$ by establishing the equations
\be\label{brscomm}
[\mathcal{S},:\Sigma:]Z_{|\uJ=0}=-[Q,:\Sigma:]Z_{|\uJ=0}=[Q,S]=0,
\ee
where
\begin{equation}
    \mathcal{S}\equiv \int \Big(J_{\mu\nu}\frac{\delta}K_{\mu\nu}
                           -j^\rho_c\frac{\delta}{\delta L^\rho}
 	  -j^\rho_{\bar{c}}\frac{\delta}{\delta j^\rho_b} \Big)
 	  \quad \mbox{with} \quad
 	  \mathcal{S}Z=0 \, .
\end{equation}
The lhs of \eqref{brscomm} is a commutator in the space of functionals, i.e.\ of $\mathcal{S}$, the ST-operator, with the $S$-matrix defined on the functional level via $Z$, the generating functional for general Green functions. Now
\be\label{2brscomm}
[\mathcal{S},:\Sigma:]Z_{|\uJ=0}=0 
\ee
since the first term of the commutator vanishes because $\mathcal{S}=0$ for
vanishing sources, the second term of the commutator vanishes due to the
validity of the ST-identity.\\
The rhs of (\ref{brscomm}) is an equation in terms of (pre-)Hilbert 
space operators: $S$-operator and BRST-charge, both defined on the indefinite
metric Fock space of creation and annihilation operators. The claim is 
that we can find an operator $Q$ such that the rhs holds true.\\
We then know that a subspace defined by $Q|\mathrm{phys}\rangle=0$ is stable under $S$, hence
physical states are mapped into physical states.\\
To show that (\ref{2brscomm}) indeed holds, we observe first that the commutator
$[\mathcal{S},:\Sigma:]$ is of the form $[\mathcal{S},e^Y]$. If 
$[\mathcal{S},Y]$ commutes with $Y$, one can reorder
the series into $[\mathcal{S},e^Y]=[\mathcal{S},Y]e^Y$. This has to be 
evaluated.
Since in the tree approximation $z=1$, hence $K(x-y)_{\Phi\Phi'}= \Gamma_{\Phi\Phi'}$
we define for the explicit calculation 
\be\label{auxy}
Y\equiv \int\Big(
h^{\mu\nu}\Gamma^{hh}_{\mu\nu\rho\sigma}\frac{\delta}{\delta J_{\rho\sigma}}
+h^{\mu\nu}\Gamma^{hb}_{\mu\nu\rho}\frac{\delta}{\delta j_\rho^b}
+ b^{\rho}\Gamma^{bh}_{\rho\alpha\beta}\frac{\delta}{\delta J_{\alpha\beta}}
+ b^{\rho}\Gamma^{bb}_{\rho\sigma}\frac{\delta}{\delta j_\sigma^b}
+ c^{\rho}\Gamma^{c\bar{c}}_{\rho\sigma}\frac{\delta}{\delta j_\sigma^{\bar{c}}}
+ \bar{c}^{\rho}\Gamma^{\bar{c}c}_{\rho\sigma}\frac{\delta}{\delta j_\sigma^c}\Big) .
\ee
For the desired commutator one finds
\be\label{XYcomm}
[\mathcal{S},Y]=-\int\Big(
h^{\mu\nu}\Gamma^{hh}_{\mu\nu\rho\sigma}\frac{\delta}{\delta K_{\rho\sigma}}
-c^\rho\Gamma^{c\bar{c}}_{\rho\sigma}\frac{\delta}{\delta j^b_\sigma}
-\bar{c}^\rho\Gamma^{\bar{c}c}_{\rho\sigma}\frac{\delta}{\delta L_\sigma}\Big),
\ee
so it clearly commutes with $Y$.\\
In the next step we have to consider $:[\mathcal{S},Y]e^Y:Z$, i.e.\ terms of 
the type
\begin{align}\label{XYcomm1}
-\int:\Big(h^{\mu\nu}\Gamma^{hh}_{\mu\nu\rho\sigma}
	\frac{\delta}{\delta K_{\rho\sigma}}
-c^\rho\Gamma^{c\bar{c}}_{\rho\sigma}\frac{\delta}{\delta j^b_\sigma}
-\bar{c}^\rho\Gamma^{\bar{c}c}_{\rho\sigma}\frac{\delta}{\delta L_\sigma} \Big) 
	:Y(1)\cdots Y(n)\cdot Z(\uJ)_{|\uJ=0}\phantom{asdfg} 
\end{align}
i.e.
\begin{align}
-\int:\Big(h^{\mu\nu}\Gamma^{hh}_{\mu\nu\rho\sigma}\kappa D^{\rho\sigma}_\lambda 
	                                           c^\lambda
-c^\rho\Gamma^{c\bar{c}}_{\rho\sigma}b^\sigma
-\bar{c}^\rho\Gamma^{\bar{c}c}_{\rho\sigma}c^\lambda\partial_\lambda c^\sigma\Big) 
	:Y(1)\cdots Y(n)\cdot Z(\uJ)_{|\uJ=0} . \phantom{asdfg}\nonumber 
\end{align}	
These terms constitute insertions into the functional $Z$. A closer look 
in terms of Feynman diagrams reveals that due to momentum conservation
from $D^{\rho\sigma}_\lambda c^\lambda$ only terms linear in the fields
survive and also the
last term bilinear in $c$ cannot contribute -- when going on mass shell they
cannot develop particle poles. We arrive thus at
\be\label{transeq}
	:[\mathcal{S},Y]:Z=
     :\Sigma\Big[ \int(-h^{\mu\nu}\Gamma^{hh}_{\mu\nu\alpha\beta}
                 \kappa (\partial^\alpha c^\beta+\partial^\beta c^\alpha)
	 +c^\rho\Gamma^{c\bar{c}}_{\rho\sigma}b^\sigma)
       \Big]:\cdot Z(\underline{J})_{|\underline{J}=0}.
\ee
The second factors in the insertion are just the linearized BRST-variations
of $h^{\alpha\beta}$, resp.\ $\bar{c}^\sigma$. This suggests to introduce
 a corresponding BRST operator $Q$ which generates these
transformations
\begin{eqnarray}\label{linBRS}
	Q\Gamma&\equiv&\int\Big\lbrack
	\kappa(\partial^\mu c^\nu+\partial^\nu c^\mu)
	\frac{\delta}{\delta h^{\mu\nu}}
		   + b^\rho\frac{\delta}{\delta \bar{c}^\rho}\Big\rbrack \Gamma\\
	QZ_c&\equiv&-i\int\Big\lbrack
	\kappa(\partial^\mu\frac{\delta Z_c}{\delta j_\nu^c} +	
	\partial^\nu\frac{\delta Z_c}{\delta j_\mu^c})J_{\mu\nu}+
	\frac{\delta Z_c}{\delta j^b_\rho}j_\rho^{\bar{c}}
	\Big\rbrack\\
	QZ&\equiv&-\int\Big\lbrack
	J_{\mu\nu}\kappa(\partial^\mu\frac{\delta}{\delta j^c_\nu} +	
		   \partial^\nu\frac{\delta}{\delta j_\mu^c})+
	j^{\bar{c}}_\rho\frac{\delta}{\delta j^b_\rho}\Big\rbrack Z ,
\end{eqnarray}
and to calculate the commutator $[Q,:\Sigma:]Z_{|\uJ=0}$. And, indeed it
coincides with the rhs of (\ref{transeq}). Following in detail the 
aforementioned diagrammatic analysis we have a simple interpretation: 
in the Green functions $G(y;z_1,...,z_n)$ a field
entry has been replaced by the linearized BRST-transformation of it.
Having established (\ref{brscomm}) one can continue along the lines of 
\cite{KuOj}, form within the linear subspace of physical states equivalence
classes by modding out states with vanishing norm with the well-known result that
these factor states have non-vanishing norm and the $S$-matrix is unitary.

\subsection{Parametrization and gauge parameter independence.\label{se:paraandgpi}}
It is a necessary preparation for higher orders to clarify, which parameters the model contains and how they are fixed.
Also a glance at the free propagators, (\ref{bosprop}) versus
(\ref{gaugeprop}), shows that they differ in their fall-off properties depending from the value of the gauge parameter 
$\alpha_0$. Since Landau gauge $\alpha_0=0$ simplifies calculations enormously we would like to show that it is stable against perturbations. Since these two issues are closely linked
we treat them here together.
Obvious parameters are the couplings $c_0,c_1,c_2$, and $c_3$. In the next subsection we give a prescription, how to fix them by
appropriate normalization conditions. Also obvious is the gauge parameter $\alpha_0$. It will be fixed by the equation of motion for the $b$-field. Since this equation is linear in the $b$-field it also determines its amplitude.
Less obvious is the normalization of the fields
$h^{\mu\nu},c^\rho$ and of the external fields $K,L$. In order to find
their amplitudes it is convenient to inquire under which linear redefinitions of them the ST (\ref{fbrst}) stays invariant.
We define
\begin{align}\label{frdfs}  
\hat{h}^{\mu\nu}&=z_1(\alpha_0)h^{\mu\nu} &\hat{c}^\rho&=y(\alpha_0)c^\rho\\
\hat{K}_{\mu\nu}&=\frac{1}{z_1(\alpha_0)}K_{\mu\nu}  &\hat{L}_\rho&=\frac{1}{y(\alpha_0)}L_\rho, 
\end{align}
where we admitted a dependence on the gauge parameter because we
would like to vary it and detect in this way $\alpha_0$-dependence algebraically. Clearly, the values for $z_1$ and $y$ have to be prescribed. It is also clear that with 
$\alpha_0$-independent values for $z_1$ and $y$ the ST-identity is maintained.
In order to make changes of $\alpha_0$ visible we differentiate (\ref{clssct}) with respect to it, i.e.\
\be\label{varalph}
\frac{\partial}{\partial\alpha_0}\Gamma=\frac{\partial}{\partial{\alpha_0}}
        \Gamma_{\rm gf}=\int(-\frac{1}{2})b_\mu b_\nu\eta^{\mu\nu}=
	\brsts\int(-\frac{1}{4})(\bar{c}_\mu b_\nu+\bar{c}_\nu b_\mu)\eta^{\mu\nu} .
\ee
We observe that this is an $\brsts$-variation and thus, if we introduce a fermionic partner $\chi= \brsts\alpha_0$ and perform the change
\be\label{chG}
\Gamma_{\rm gf} +\Gamma_{\phi\pi}\to\Gamma_{\rm gf}+\Gamma_{\phi\pi}
    +\int(-\frac{1}{4})\chi(\bar{c}_\mu b_\nu+\bar{c}_\nu b_\mu)\eta^{\mu\nu} .
\ee
we have
\be\label{chidouble}
\ST(\Gamma)+\chi\partial_{\alpha_0}\Gamma=0 .
\ee
We carry over this extended BRST-transformation to $Z$
\be\label{chiZ}
\hat{\mathcal{S}}Z\equiv\mathcal{S}Z+\chi\partial_{\alpha_0}Z=0,
\ee
with the implication
\be
\partial_\chi(\hat{\mathcal{S}}Z)=0 \quad\Rightarrow\quad 
    \partial_{\alpha_o} Z=-\mathcal{S}\partial_\chi Z 
\ee
showing that $\alpha_0$-dependence is a BRST-variation, hence unphysical. 
This last equation
can be easily checked on the free propagators (for propagators connected and 
general Green functions coincide).

Using for $Z(\uJ)$ the form
\be\label{znt1}
Z(\uJ)=\exp \Big\{i\int \mathcal{L}_{\rm int}\big(\frac{\delta}{i\delta\uJ}\big)\Big\}Z_0 \qquad Z_0=\exp\Big\{\int i\uJ \langle \Phi\Phi \rangle i\uJ \Big\}
\ee
one obtains
\be\label{znt2}
\partial_{\alpha_0}Z(\uJ)=\partial_{\alpha_0}Z_0\cdot Z(\uJ)
=\Big(\partial_{\alpha_0}\int i\uJ \langle \Phi\Phi \rangle i\uJ \Big)\cdot Z .
\ee
(Here $\uJ$ stands for the sources of all propagating fields 
$\Phi$.)
Hence $\alpha_0$-dependence remains purely at external lines, if one does not add
$\alpha_0$-dependent counterterms, and then vanishes on the $S$-matrix where these
lines are amputated. It also means that the power counting for the gauge multiplet
is irrelevant because this multiplet shows up only as external 
lines.

We now step back and analyze $\alpha_0$-dependence more systematically. Equations (\ref{chidouble}), (\ref{chiZ}) and the analogous one for connected Green functions
\be\label{chiZc}
\mathcal{S}Z_c+\chi\partial_{\alpha_0}Z_c=0,
\ee
where $\alpha_0$ undergoes the change
\be\label{vrgp}
\brsts\alpha_0=\chi \qquad \brsts\chi=0
\ee
have to be solved. 
The rhs of (\ref{chG}) is solution of the extended gauge condition
\be\label{egc}
\frac{\delta \Gamma}{\delta b^\rho}=
\kappa^{-1}\partial^\mu h_{\mu\rho} -\alpha_0 b^\nu-\frac{1}{2}\chi\bar{c}_\rho.
\ee 
Acting with $\delta/\delta b^\rho$ on the ST (\ref{chidouble}) we find
that the ghost equation of motion has changed accordingly
\be\label{gee}
G^\rho\Gamma\equiv \Big(\kappa^{-1}\partial^\mu\frac{\delta}{\delta K_{\mu\rho}}
  +\frac{\delta}{\delta \bar{c}_\rho} \Big) \Gamma=\frac{1}{2}\chi b^\rho
\ee
As in (\ref{Gmmbr}) and (\ref{sceH}) we introduce 
$H_{\mu\nu}=K_{\mu\nu} - \frac{1}{2\kappa}(\partial_\mu\bar{c}_\nu
     +\partial_\nu\bar{c}_\mu)$ and $\bar{\Gamma}$ by
\be\label{pbrGm}
\Gamma=\bar{\Gamma}
      +\int \Big(-\frac{1}{2}\alpha_0 b_\mu b_\nu \eta^{\mu\nu}
  -\frac{1}{2\kappa}h^{\mu\nu}(\partial_\mu b_\nu+\partial_\nu b_\nu)
-\frac{1}{4}\chi(\bar{c}_\mu b_\nu+\bar{c}_\nu b_\mu)\eta^{\mu\nu} \Big) .
\ee
The extended ST reads in terms of $\bar{\Gamma}$
\be\label{ebrG}
\ST (\Gamma)=\mathcal{B}(\bar{\Gamma})=0
\ee
with
\be\label{pbrGm2}
\mathcal{B}(\bar{\Gamma})\equiv
\int \Big(\frac{\delta\bar{\Gamma}}{\delta K}\frac{\delta\bar{\Gamma}}
                                                      {\delta h}
      +\frac{\delta\bar{\Gamma}}{\delta L}
       \frac{\delta\bar{\Gamma}}{\delta c}
+\chi\frac{\partial\bar{\Gamma}}{\partial \alpha_0} \Big).
\ee
$\bar{\Gamma}$ satisfies the homogeneous ghost equation of motion
\be\label{pgem}
G\bar{\Gamma}=0.
\ee
We now have to find the most general solution of ghost equation
(\ref{gee}) and the new ST (\ref{ebrG}). Due to dimension and
$\phi\pi$-charge neutrality $\bar{\Gamma}$ can be decomposed
as
\be\label{gsa}
\bar{\Gamma}=\bar{\bar{\Gamma}}(h,c,K,L,\alpha_0)
                 +\chi\int(f_K(\alpha_0)Kh+f_L(\alpha_0)Lc)
\ee            
With the choice of linear dependence from $h$, however, we
certainly do not cover the most general case:     
due to the vanishing dimension of $h^{\mu\nu}$ one could 
replace the linear factor $h^{\mu\nu}$ by an arbitrary function 
$\mathcal{F}^{\mu\nu}(h)$ in $K_{\mu\nu}h^{\mu\nu}$.
For simplicity we discuss here the linear case, which continues (\ref{frdfs}), whereas the non-linear one will be treated below (see Sect.\ \ref{se:generalsolutionSTI}).

From (\ref{ebrG}) and (\ref{pbrGm2}) we deduce that
\be\label{ntrm}
0=\mathcal{B}(\bar{\Gamma})=\mathcal{B}(\bar{\bar{\Gamma}})|_{\chi=0}+\chi\int(
-f_Hh^{\mu\nu}\frac{\delta\bar{\bar{\Gamma}}}{\delta h^{\mu\nu}}
+f_HH^{\mu\nu}\frac{\delta\bar{\bar{\Gamma}}}{\delta H^{\mu\nu}}
+f_Lc\frac{\delta\bar{\bar{\Gamma}}}{\delta c}
-f_LL\frac{\delta\bar{\bar{\Gamma}}}{\delta L})
+\chi\frac{\partial\bar{\bar{\Gamma}}}{\partial \alpha_0} .
\ee
At $\chi=0$ follows first
\be\label{ntrm2}
\mathcal{B}(\bar{\bar{\Gamma}})|_{\chi=0}=0,{}
\ee
and then
\be\label{ntrm3}
\int \Big(
-f_Hh^{\mu\nu}\frac{\delta\bar{\bar{\Gamma}}}{\delta h^{\mu\nu}}
+f_HH^{\mu\nu}\frac{\delta\bar{\bar{\Gamma}}}{\delta H^{\mu\nu}}
+f_Lc\frac{\delta\bar{\bar{\Gamma}}}{\delta c}
-f_LL\frac{\delta\bar{\bar{\Gamma}}}{\delta L} \Big)
+\frac{\partial\bar{\bar{\Gamma}}}{\partial \alpha_0}=0.
\ee
(\ref{ntrm2}) corresponds to (\ref{fbrst}), hence we know that the general solution (of the linear case) is given by
\begin{eqnarray}\label{frt}
\bar{\bar{\Gamma}}&=&
    \hat{c}_3\kappa^{-2}\int\sqrt{-g}R(z_1(\alpha_0)h)
   +\hat{c}_1\int\sqrt{-g}R^{\mu\nu}R_{\mu\nu}(z_1(\alpha_0)h)
   +\hat{c}_2\int\sqrt{-g}R^2(z_1(\alpha_0)h)\nonumber\\
 &&+\hat{c}\int(\kappa H_{\mu\nu}(\frac{y(\alpha_0)}{z_1(\alpha_0)}
                  (-\partial^\mu c^\nu-\partial^\nu c^\mu)
                  -y(\alpha_0)(\partial_\lambda c^\mu h^{\lambda\nu}
                  -c^\lambda\partial_\lambda h^{\mu\nu}
                  +c^\lambda\partial_\lambda  h^{\mu\nu})) \\
  &&-\kappa y(\alpha_0) L_\rho\partial^\lambda c^\rho)     . \nonumber
\end{eqnarray}
(\ref{frt}) inserted into (\ref{ntrm3}) implies after some calculations that all $\hat{c}$ 
are independent of $\alpha_0$, whereas the functions $f_{H,L}$
satisfy the relations
\be\label{rsfrt}
     \partial_{\alpha_{0}}z_1=f_H z_1 \qquad
     \partial_{\alpha_{0}}y=-f_L y 
\ee
All parameters $\hat{c}$ can therefore be fixed by normalization conditions
independent of $\alpha_0$. Since we shall work in Landau gauge,
$\alpha_0=0$, the functions $f_H,f_L$ will be independent of
$\alpha_0$, as well as $z_1$ and $y$, hence numbers. 

\subsection{Normalization conditions I\label{se:nc1}}
In the tree approximation as studied in this section the free parameters of the model
can be prescribed by the following conditions
\begin{eqnarray}
	\frac{\partial}{\partial p^2}\,\gamma^{(2)}_{\rm TT}|_{p^2=0}&
	=&c_3\kappa^{-2}\qquad ({\rm coupling\,\, constant})\label{trnorm3}\\
\frac{\partial}{\partial p^2}\frac{\partial}{\partial p^2}\,
\gamma^{(2)}_{\rm TT}&
	=&-2c_1\qquad({\rm coupling\,\, constant})\label{trnorm1}\\
\frac{\partial}{\partial p^2}\frac{\partial}{\partial p^2}\,\gamma^{(0)}_{\rm TT}
	&=&2(3c_2+c_1)\qquad (\rm{coupling\,\,constant})\label{trnorm2}\\
	\Gamma_{h^{\mu\nu}}&
  =&-\eta_{\mu\nu}c_0\doteq0\qquad(\rm{coupling\,\,constant})\label{trnorm0}\\
\frac{\partial}{\partial p_\sigma}\Gamma_{K^{\mu\nu}c_\rho}&=&
                                -i\kappa(\eta^{\mu\sigma}\delta^\nu_\rho
	                          +\eta^{\nu\sigma}\delta^\mu_\rho
	-\eta^{\mu\nu}\delta^\sigma_\rho)\qquad ({\rm amplitude\,\,of\,\,h\,\,and\,\, K})\label{trnorm4}\\
	\frac{\partial}{\partial p^\lambda}\Gamma_{L_\rho c^\sigma c^\tau}&=&
				-i\kappa(\delta^\rho_\sigma\eta_{\lambda\tau}
-\delta^\rho_\tau\eta_{\lambda\sigma})\qquad({\rm amplitude\,\,of\,\,c\,\,and\,\,L})\label{trnorm5}
\end{eqnarray}
Imposing the $b$-equation of motion (\ref{beq}) fixes $\alpha_0$ and the
$b$-amplitude.
It is worth mentioning that the $c_1$-contribution to $\gamma^{(0)}$ in \eqref{eq:coffs0} is an implication of the invariance under $\brsts_1 h$, hence must not be postulated via some normalization condition.

\section{Renormalization\label{se:renormalization}}
At first we have to specify the perturbative expansion in which we would
like to treat the model. Due to the vanishing canonical dimension of the
field $h^{\mu\nu}$ we have to expand in the number of this field. Second
we expand as usual in the number of loops. Next we have to choose a
renormalization scheme in order to cope with the divergences of the loop
diagrams. We shall use the Bogoliubov-Parasiuk-Hepp-Zimmermann-Lowenstein
(BPHZL) scheme \cite{Lowenstein:1975ps} which is based on momentum subtractions and an auxiliary
mass in order to avoid spurious infrared divergences which otherwise
would be introduced by the momentum subtractions when dealing with massless 
propagators.\\
The key ingredients of this scheme are the subtraction operator
acting on one-particle-irreducible diagrams (1PI) and the forest formula 
which organizes the subtractions.  The subtraction operator reads
\be\label{sbtr}
(1-\tau_\gamma)=(1-t^{\rho(\gamma)-1}_{p^\gamma(s^\gamma-1)})
                (1-t^{\delta(\gamma)}_{p^\gamma s^\gamma}).
\ee
Here $t^d_{x_1...x_n}$ denotes the Taylor series about $x_i=0$ to order
$d$ if $d\ge 0$ or $0$ if $d<0$. $\gamma$ denotes a 1PI diagram, $p^\gamma$
refers to its external momenta, and $s^\gamma$ to an auxiliary subtraction
variable to be introduced. 
$\rho(\gamma)$ and  $\delta(\gamma)$ are the infrared and ultraviolet subtraction degrees of $\gamma$, respectively.
Those will be specified below. As far as the forest formula is concerned we 
refer to the literature (cf. \cite{Lowenstein:1975ug, Lowenstein:1975ps}).
For later use we note that
\be\label{rmss}
(1-\tau_\gamma)=(1-t^{\delta(\gamma)}_{p^\gamma})
			\qquad {\rm for}\quad \rho(\gamma)=\delta(\gamma)+1 .
\ee

\subsection{Auxiliary mass\label{se:auxmass}}
In the BPHZ subtraction scheme one removes UV divergences by suitable subtractions at vanishing external momenta. In the massless case those would introduce artificial (off-shell) IR divergences. Hence in an extension, the BPHZL scheme, one introduces an auxiliary mass term of type $M^2(s-1)^2$ for every massless propagator. Subtractions with respect to $p,s$ performed at $p=0,s=0$ take care of the UV divergences. Subtractions with respect to $p,s-1$ thereafter establish correct normalizations for guaranteeing
 poles at $p=0$ and vanishing of three-point functions (of massless fields) at  $p=0$ .\\
When trying to introduce such an auxiliary mass term for the massless pole in the double pole propagators one encounters difficulties. Neither with a naive $hh$-term nor with a Fierz-Pauli type mass term can one invert $\Gamma_{hh}$ to propagators 
$G_{hh}$ such that the Lagrange multiplier field $b_\rho$ remains non-propagating. But its propagation would prevent its use in the quartet formalism of \cite{KuOj}. A glance at the propagators (\ref{bosprop}) and the coefficients 
$\gamma^{({\rm r})}_{\rm KL}$, (\ref{coffs}) suggests to replace
the overall factor $p^2$ in the $\gamma$’s by 
\be\label{axms}
p^2 - m^2 \equiv p^2-M^2(s-1)^2 .
\ee
Here $m^2$ denotes the auxiliary mass contribution.
This \emph{Push} in $p^2$ still maintains restricted invariance, i.e.\ under $\brsts_0 h$, (see Sect.\ \ref{se:push}
and App.\ \ref{se:brst0inv}), and is 
fairly easy to carry along as we shall see.\\
Accepting this change of vertices and propagators one has to analyze in some
detail what it implies. For the propagators it is clear that the pole at $p^2=0$
is shifted, as desired to a pole at $p^2=m^2$. It affects not only the invariant
parts, but also the gauge fixing dependent propagators $\langle bh \rangle$ and $\langle \bar{c}c \rangle$.
This can be seen when performing Push in $\Gamma$ and having a look at the 
inversion equations. The $\gamma$'s (\ref{coffs}) then read
\begin{eqnarray}\label{mcoffs}
	\gamma^{(2)}_{TT} &=&-(p^2-m^2)(c_1p^2-c_3\kappa^{-2})\\
	&\Rightarrow& 
	m^2\hat{\gamma}^{(2)}_{TT}(m^2)=m^2(c_1p^2-c_3\kappa^{-2})\label{gmps1}  \\
\gamma^{(0)}_{TT} &=&(p^2-m^2)((3c_2+c_1)p^2+\frac{1}{2}c_3\kappa^{-2})\\
    &\Rightarrow&
m^2\hat{\gamma}^{(0)}_{TT}(m^2)=-m^2((3c_2+c_1)p^2+\frac{1}{2}c_3\kappa^{-2})\label{gmps2}\\ 
 \gamma^{(1)}_{SS}&=& \gamma^{(0)}_{WW}=\gamma^{(0)}_{TW}=\gamma^{(0)}_{WT}
	             =0 .
\end{eqnarray}
In the inversion equations one has products of $\gamma^{(r)}_{KL}$ with
its direct counterpart $\langle hh \rangle^{(r)}_{KL}$, such that this change is not a change there.\\
For gauge fixing terms we find the effect of Push as follows
\begin{eqnarray}\label{mfg}
\Gamma^{hb}_{\mu\nu\rho}G^{bh}&=&\frac{i}{2\kappa}(\eta_{\rho\mu}p_\nu+\eta_{\rho\nu} p_\mu)\frac{\kappa}{p^2}(p^\mu\theta^{\nu\rho}+p^\nu\theta^{\mu\rho}+p^\rho\omega^{\mu\nu})                              \qquad{\rm (local)}\\
&=&\frac{i}{2\kappa}(\eta_{\rho\mu}p_\nu+\eta_{\rho\nu} p_\mu)
       \frac{p^2}{p^2}\frac{\kappa}{p^2}(p^\mu\theta^{\nu\rho}+p^\nu
       \theta^{\mu\rho}+p^\rho\omega^{\mu\nu})
       \qquad{\rm (local)}                                          \\
		&\stackrel{\rm Push}{\rightarrow}&\frac{i}{2\kappa}(\eta_{\rho\mu}p_\nu
+\eta_{\rho\nu}p_\mu)\frac{p^2-m^2}{p^2}\frac{\kappa}{p^2-m^2}
(\theta^{\rho\mu}p^\nu+\theta^{\rho\nu} p^\mu
                                      +p^\rho\omega^{\mu\nu})\\
\Rightarrow\Gamma(m^2)^{hb}_{\mu\nu\rho}&=&\frac{-im^2}{2\kappa p^2}(\eta_{\rho\mu}p_\nu+\eta_{\rho\nu} p_\mu)   \qquad({\rm non-local}),\label{gmgf}\\
\Rightarrow G^{bh}_{\rho\mu\nu}&=&\frac{\kappa}{p^2-m^2}(p_\mu\theta_{\rho\nu}
+ p_\nu\theta_{\rho\mu}+p_\rho\omega_{\mu\nu})\qquad({\rm massive\,\, propagator})
\end{eqnarray}
i.e.\ there appears an additional term in $\Gamma^{hb}$ and the $\langle bh \rangle$-propagator
becomes massive (with the auxiliary mass). 
In $x$-space the complete gauge fixing term reads
\begin{eqnarray}\label{cmgf}
	\Gamma_{\rm{gf}}&=&-\frac{1}{2\kappa}\int dxdy\, h^{\mu\nu}(x)
(\partial_\mu b_\nu+\partial_\nu b_\mu)(y)\Big\lbrace\delta(x-y)
+\frac{m^2}{(x-y)^2}\Big\rbrace -\frac{\alpha_0}{2}\int\eta^{\mu\nu} b_\mu b_\nu\nonumber\\
	            &=&-\frac{1}{2\kappa}\int dxdy\, h^{\mu\nu}(x)
	(\partial_\mu b_\nu+\partial_\nu b_\mu)(y)\Big\lbrace\big( \frac{\Box}{4\pi^2} + m^2 \big)
\frac{1}{(x-y)^2}\Big\rbrace -\frac{\alpha_0}{2}\int\eta^{\mu\nu} b_\mu b_\nu.
\end{eqnarray}
A suitable Faddeev-Popov (FP) term is then
\begin{eqnarray}\label{FaPo}
	\Gamma_{\phi\pi}&=&-\frac{1}{2}\int dxdy\, D^{\mu\nu}_\rho c^\rho(x)
(\partial_\mu \bar{c}_\nu+\partial_\nu \bar{c}_\mu)(y)\lbrace\delta(x-y)
	               +\frac{m^2}{(x-y)^2}\rbrace\nonumber\\
	&=&-\frac{1}{2}\int dxdy\, D^{\mu\nu}_\rho c^\rho(x)
(\partial_\mu \bar{c}_\nu
	+\partial_\nu \bar{c}_\mu)(y)\lbrace\big( \frac{\Box}{4\pi^2} + m^2 \big)\frac{1}{(x-y)^2}\rbrace ,
\end{eqnarray}
because it maintains the BRST-doublet structure within the gauge fixing procedure.\\
A comment to the ``non-local'' terms is in order. Our writing is
symbolic shorthand in order to have a simple handling of these terms. Using the explicit form of $\brsts_0 h$ and integration by parts one may observe that the actual non-local part is of projector type in terms of differential operators -- quite in line with its first appearance in $p$-space. 
There the projectors lead formally to direction dependent integrals. However Zimmermann's $\varepsilon$, introduced as
\begin{align}
	p^2 \to p^2+i\epsilon({\mathbf p}^2) \, , \label{psl}
\end{align}
guarantees absolute convergence, hence no serious problem will arise once we have reliable power counting and appropriate correct subtractions. Of course, at the physical value $s=1$ it disappears anyway.\\ 
We therefore discuss in the next subsection power counting and convergence with positive
outcome, and return thereafter to a discussion of the $m^2$-dependent terms.
Before starting with the presentation of power counting we have to have a look at the basis of naively symmetric insertions
once we have introduced an auxiliary mass term. 
Obviously we can introduce the following \emph{Shift}
\be\label{ivce}
\int\sqrt{-g}c_3\kappa^{-2}R \to \int\sqrt{-g}(c_{30}\kappa^{-2}
+c_{31}\kappa^{-1}m+c_{32}\frac{1}{2}m^2)R.
\ee
In the tree approximation these terms are invariant (and for 
$s=1$ reduce to the original term), but in higher orders they represent new and independent elements in the basis of symmetric normal products
with $\delta=\rho=4$ (cf. \cite{Zimmermann:1972tv}). 
So, we have to carry them along as vertices when studying power counting.

\subsection{Power counting and convergence\label{se:powercounting}}
In the Landau gauge, $\alpha_0=0$, the only non-vanishing propagators are
the following one's:
\begin{eqnarray}\label{sprops}
	\langle hh \rangle^{(2)}_{TT}&=&\frac{i}{(p^2-m^2)c_1( p^2-\frac{c_3\kappa^{-2}}
	{c_1})}\\
	\langle hh \rangle^{(0)}_{TT}&=&\frac{i}{(p^2-m^2)(3c_2+c_1)(p^2+\frac{c_3\kappa^{-2}}{2(3c_2+c_1)})}\\
	\langle b_\rho h_{\mu\nu} \rangle&=&\frac{1}{p^2-m^2}(p_\mu\theta_{\nu\rho}
	+p_\nu\theta_{\mu\rho}+p_\rho\omega_{\mu\nu})\\
	\langle \bar{c}_\rho c_\sigma  \rangle&=&
	-i\big(\theta_{\rho\sigma}+\frac{1}{2}\omega_{\rho\sigma} \big)\frac{1}{p^2-m^2}
\end{eqnarray}
In addition to $m=M(s-1)$ one needs also Zimmermann's $\varepsilon$-prescription (\ref{psl}). This will guarantee absolute convergence of diagrams, once power counting
is established and subtractions are correctly performed.\\
Important note: in all formulas to follow in this section
the replacement of $c_3$ by the sum given in (\ref{ivce}) is to be understood. Relevant for power counting arguments is never
a coefficient in front of a vertex, but the number of lines and derivatives at the vertex and its associated subtraction degree.
The $\langle bh \rangle$ propagator will be of no relevance for reasons spelled out after (\ref{znt2}). 

Power counting is based on ultraviolet (UV) and infrared (IR) degrees
of propagators and vertices. The upper degree $\overline{\rm deg}_{p,s}$ gives the asymptotic power for $p$ and $s$ tending to infinity; the lower degree 
$\underline{\rm deg}_{p,(s-1)}$ gives the asymptotic power for $p$ and $s-1$ tending to zero.
For propagators they read
\begin{eqnarray}\label{dgpr}
	{\overline{\rm deg}}_{p,s}(\langle hh \rangle^{(2)}_{TT})&=&-4 \qquad{}
	{\underline{\rm deg}}_{p,s-1}(\langle hh \rangle^{(2)}_{TT})=-2 \\
	\label{dgpr2}
	{\overline{\rm deg}}_{p,s}(\langle hh \rangle^{(0)}_{TT})&=&-4 \qquad
	{\underline{\rm deg}}_{p,s-1}(\langle hh \rangle^{(0)}_{TT})=-2 \\
        {\overline{\rm deg}}_{p,s}(\langle \bar{c}c \rangle)&=&
	           {\underline{\rm deg}}_{p,s-1}(\langle \bar{c}c \rangle)=-2 .
\end{eqnarray}
As shorthand we write also 
$\overline{\rm deg}\equiv \overline{D}_L$ and $\underline{\rm deg}\equiv \underline{D}_L$.
The degrees of the vertices thus have the values
\begin{eqnarray}\label{dgve}
	\overline{D}_{V^{(c_1)}}&=&\overline{D}_{V^{(c_2)}}=4, 
	\quad \overline{D}_{V^{(c_3)}}=2
	\quad \overline{D}_{V^{(\phi\pi)}}=2\\
\underline{D}_{V^{(c_1)}}&=&\underline{D}_{V^{(c_2)}}=4, 
	      \quad \underline{D}_{V^{(c_3)}}=2
	      \quad \underline{D}_{V^{(\phi\pi)}}=2 .
\end{eqnarray}

Let us now consider a one-particle-irreducible (1PI) diagram $\gamma$ with $m$ 
loops, $I_{ab}$ internal lines, $a,b = h,c,\bar{c}$, and $V$
vertices of type $V^{(c_1,c_2,c_3,\phi\pi)}$ or insertions $Q_i$ as well as $N$ amputated 
external lines. In the subsequent considerations a more detailed notation is useful: $N_a$ are of type $\Phi_a$,
$n_{ai}$ are of type $a$ and are attached to the $i^{\rm th}$ vertex. Then 
with $Q_i$
\be\label{ins} 
Q_i(x)=(\frac{\partial}{\partial x})^{|\mu_i|}\prod_a(\Phi_a^{c_{ai}}(x)) ,
\ee
we first find for the UV- and IR-degrees of $\gamma$ 
\begin{eqnarray}\label{gmmadeg}
d(\gamma)&=&4m(\gamma)+\sum_{V\in\gamma}\overline{D}_V+\sum_{L\in\gamma}\overline{D}_L\\
         &=&4m(\gamma) +4V^{(c_1,c_2)}+2V^{(c_3)}+2V^{(\phi\pi)}
	               -4I_{hh}-2I_{c\bar{c}},\\
r(\gamma)&=&4m(\gamma)+\sum_{V\in\gamma}\underline{D}_V
	                             +\sum_{L\in\gamma}\underline{D}_L\\
	 &=&4m(\gamma)+4V^{(c_1,c_2)}+2V^{(c_3)}+2V^{(\phi\pi)}
	 -2I_{hh}-2I_{\bar{c}c} .
\end{eqnarray}
The topological relations
\begin{eqnarray}\label{topform}
	m&=&I-V+1\\
	N_a&=&\sum_i n_{ai} 
	\qquad 2I_{aa}=\sum_i(c_{a i}-n_{ai})=\sum_ic_{ai}-N_a
\end{eqnarray}
permit to rewrite these degrees as
\begin{eqnarray}\label{uvirdeg}
	d(\gamma)&=&4+\sum_{V\in\gamma}(\overline{D}_V-4)
	                      +\sum_{L\in\gamma}(\overline{D}_L+4)\\
	d(\gamma)&=&4-N_{\tilde{c}}-2V^{(c_3)}\\ 
	r(\gamma)&=&4+\sum_{V\in\gamma}(\underline{D}_V-4)
			     +\sum_{L\in\gamma}(\underline{D}_L+4)\\
	r(\gamma)&=&4-2V^{(c_3)}-2V^{(\phi\pi)}+2I_{hh}+2I_{c\bar{c}} .
\end{eqnarray}
(Here $\tilde{c}$ stands for both, $c$ and $\bar{c}$.)  
The aim is now to associate subtraction degrees to them which are independent
of the detailed structure of the respective diagrams. An obvious choice is
\be\label{subtrdeg}
\delta(\gamma) = 4 \qquad \rho(\gamma) = 4
\ee
\noindent
Before proceeding,
a comment to $\delta(\gamma)=4$ is in order. Obviously there are infinitely many
divergent diagrams possible, even for every number $N$ of external $h$-lines. 
This requires infinitely many parameters as normalizations. Those are 
provided by the infinitely many arbitrary parameters which arise from the
redefinition of $h$ as a function of itself. They are gauge type parameters
and constitute only wave function
renormalizations hence are unphysical. This will be discussed in detail later (see Sect.\ \ref{se:generalsolutionSTI}).

We would like to prove convergence along the lines of theorems established in
\cite{Lowenstein:1975ps}. In order to do so we formulate a few conditions
which will later turn out to be sufficient for proving convergence. The first
one reads
\be\label{c1} 
	\delta(\gamma)= d(\gamma)+b(\gamma) 
			\quad\mbox{and}\quad \rho(\gamma)=  r(\gamma)-c(\gamma) \tag{C1}
\ee
with $b(\gamma)$ and $c(\gamma)$ being non-negative integers. 
$b(\gamma)\ge 0$ is obviously satisfied, but for $c(\gamma)$ 
we have to convince ourselves that the bracket terms in 
(\ref{uvirdeg}) are greater or equal to zero. Hence we need the more detailed 
information given by the line balances
\begin{eqnarray}\label{linetop}
	2I_{hh}&=& \sum_{i\in \gamma}(c_{h,i}-n_{h,i}) 
	      =  \sum_{i\in \gamma}(c_{h,i})-N_h 
		  \quad i\in\{V^{(c_1)},V^{(c_2}), V^{(c_3)},V^{(\phi\pi)}\}\\
2I_{c\bar{c}}&=&\sum_{i\in\phi\pi}(c_{c,i}-n_{c,i})
              = \sum_{i\in\phi\pi}c_{c,i}-N_{c} .
\end{eqnarray}
We find
\be\label{cg}
c(\gamma)=\sum_{i\in c_1,c_2}(c_{h,i}-n_{h,i})
           +\sum_{i\in c_3}(c_{h,i}-n_{h,i}-2)
           +\sum_{i\in \phi\pi}(c_{\tilde{c},i}-n_{\tilde{c},i}-2)
           +\sum_{i\in\phi\pi}(1-n_{h,\phi\pi})
\ee
If the vertex $i$ in question is not present in $\gamma$, the respective brackets
just vanish. If this vertex is present in $\gamma$, then $(c_{h,i}-n_{h,i})\ge 2$ and $(c_{h,i}-n_{h,i}-2)\ge 0$ -- both for 1PI $\gamma$.
Since $c_{\tilde{c},\phi\pi}=2$  the third bracket combines with the
fourth such that their sum is $\ge 0$ -- again for 1PI $\gamma$ -- we find two cases:
either $n_{h,i_0}=1$ at vertex $i_0$ s.t. $n_{\tilde{c},i_0}=0$ (otherwise 
$\gamma$ is not 1PI)
or $n_{h,i_0}=0$ at vertex $i_0$ s.t. $+1$ from here and from 
$n_{\tilde{c},i_0}$ at most 1, i.e. $-1$ in the sum (otherwise $\gamma$ is not 1PI), which
together is $0$, i.e. non-negative. 
Hence equations (\ref{c1}) are valid.

The next requirements refer to reduced diagrams 
$\bar{\Lambda}=\Lambda/\lambda_1,...\lambda_n$, which are obtained from
$\Lambda$
by contracting mutually disjoint, non-trivial 1PI subdiagrams $\lambda_i$
to points (reduced vertices) $V(\lambda_i)$ assigning (for the sake of power 
counting) the unit polynomial of momenta to each $V(\lambda_i)$. For 1PI 
$\gamma$ one has the relations
\begin{eqnarray}\label{subdeg}
	d(\gamma)&=&d(\gamma/\lambda_1...\lambda_n)+\sum_{i=1}^{n}d(\lambda_i)\\
	r(\gamma)&=&r(\gamma/\lambda_1...\lambda_n)+\sum_{i=1}^{n}r(\lambda_i).
\end{eqnarray}
Their analogues are also valid for connected diagrams.
Now one can formulate further conditions for convergence, i.e.\ 
\begin{align}\label{c2}
	\delta(\gamma)&\ge d(\gamma/\lambda_1...\lambda_n)
	          +\sum_{i=1}^{n}\delta(\lambda_i) \tag{C2}\\
	\rho(\gamma)&\le r(\gamma/\lambda_1...\lambda_n)
		  +\sum_{i=1}^{n}\rho(\lambda_i) \tag{C3} \label{c3}\\
	\rho(\gamma)&\le \delta(\gamma)+1 \tag{C4} \label{c4}
\end{align}
for arbitrary reduced 1PI subdiagrams $\gamma/\{\lambda_i\}$ of $\Gamma$.
In order to verify \eqref{c2} one just inserts the values for the respective 
degrees.
\begin{eqnarray}\label{cC2}
   \delta(\gamma)&=&4 \\
 \delta(\gamma_i)&=&4\\
d(\gamma)&=&4-2V^{(c_3)}(\gamma)-2V^{(\phi\pi)}(\gamma)+2I_{c\bar{c}}(\gamma)\\
d(\gamma_i)&=&4-2V^{(c_3)}(\gamma_i)-2V^{(\phi\pi)}(\gamma_i)
	                                    +2I_{c\bar{c}}(\gamma_i)\\
d(\bar{\gamma})&=&4-2V^{(c_3)}(\bar{\gamma})-2V^{(\phi\pi)}(\bar{\gamma})
	                  +I_{c\bar{c}}(\bar{\gamma})-4n\\
d(\bar{\gamma})+\sum_i\delta(\gamma_i)&=&4-2V^{(c_3)}(\bar{\gamma})
	+\sum_{i\in\phi\pi}(-2+c_{c,\phi\pi}-n_{c,\phi\pi})(\bar{\gamma})\\
\delta(\gamma)=4&\ge& 4-N_{\tilde{c}}(\bar{\gamma})-2V^{(c_3)}(\bar{\gamma}) .
\end{eqnarray}
(We have used that $c_{c,\phi\pi}=2$.)
The last inequality was to be proved.

For the proof of \eqref{c3} one treats first the 
case $\rho(\gamma)=\rho(\gamma_i)=4$ 
and uses the fact that the line balances used
for proving \eqref{c1} also hold for reduced diagrams. 
For the case 
$\rho(\gamma)=\rho(\gamma_i)=5=\delta(\gamma)+1=\delta(\gamma_i)+1$, 
which is 
the upper bound admitted for the IR-degrees, one finds also that the 
desired inequality holds. 
\eqref{c4} is satisfied by definition.

We can now refer to \cite[Theorem 4]{Lowenstein:1975ps} in which 
it is shown that these conditions being satisfied, Green functions exist as
tempered distributions, whereas for non-exceptional momenta (Euclidean sense)
vertex functions exist as functions. Due to a theorem of Lowenstein and Speer
\cite{Lowenstein:1975ku}
in the limit $\varepsilon \rightarrow 0$ Lorentz covariance is also satisfied.
An important improvement concerning Lorentz covariance has been provided by \cite{Clark:1976ym}. If one introduces Zimmermann's $\varepsilon$
via a change of metric 
$\eta_{\mu\nu} \to {\rm diag}(1,-(1-i\varepsilon),-(1-i\varepsilon), -(1-i\varepsilon) )$ in addition to multiplying each mass-square by 
$(1-i\varepsilon)$ then Lorentz covariance already holds for the rhs of ZI's before establishing the $\varepsilon\to 0$ limit. This is quite helpful for actual work with
ZI's.

The above proof of convergence refers to diagrams constructed out of vertices
with vanishing Faddeev-Popov (FP) charge. For installing the ST-identity in higher
orders one needs however diagrams which once contain the vertex $V^{(-)}$ of types
\begin{eqnarray}\label{fpm1}
\overline{D}(V^{(-)})=\left\{
	\begin{array}{ll}
3& {\rm for}\quad V^{(-)}\simeq \int c\,\partial\partial\partial\, h\cdots h\\  
5& {\rm for}\quad V^{(-)}\simeq \int c\,\partial\partial\partial\partial\partial\, 
	                                                   h\cdots h 
	\end{array}	\right. &
	\underline{D}(V^{(-)})=\overline{D}(V^{(-)}),
\end{eqnarray}
i.e.\ of FP-charge $-1$. The UV- and IR-degrees become resp.\
\begin{eqnarray}\label{gmmadeg2}
	d(\gamma)&=&4m(\gamma)+\sum_{V\in\gamma}\overline{D}_V
		       +\sum_{L\in\gamma}\overline{D}_L+\overline{D}_{V^{(-)}}  \\
r(\gamma)&=&4m(\gamma)+\sum_{V\in\gamma}\underline{D}_V
		       +\sum_{L\in\gamma}\underline{D}_L+\underline{D}_{V^{(-)}}  .
\end{eqnarray}
With (\ref{topform}) this results into $(V^{(-)}\in \gamma)$
\begin{eqnarray}\label{vld}
	d(\gamma)&=&4+\sum_{V\in \gamma}(\overline{D}_V-4)
		   +\sum_{L\in\gamma}(\overline{D}_L+4)\\
	 &=&4-N_{\tilde{c}}-2V^{(c_3)} +(\overline{D}_{V^{(-)}}-4)\\ 
	r(\gamma)&=&4+\sum_{V\in \gamma}(\underline{D}_V-4)
		   +\sum_{L\in\gamma}(\underline{D}_L+4)\\
 &=&4-2V^{(c_3)}-2V^{(\phi\pi)}+(\underline{D}_{V^{(-)}}-4) +2I_{hh}+2I_{c\bar{c}} .
\end{eqnarray}
As subtractions degrees we define
\begin{eqnarray}\label{sdr2}
	\delta(\gamma)&=d(\gamma)+b(\gamma)
	=\left\{\begin{array}{ll}    
		4& {\rm if}\quad V^{(-)}\notin \gamma\\  
		5& {\rm if}\quad V^{(-)}\in \gamma
		\end{array}	\right. \\
	\rho(\gamma)&=r(\gamma)-c(\gamma)
	=\left\{\begin{array}{ll}    
		4& {\rm if}\quad V^{(-)}\notin \gamma\\  
		5& {\rm if}\quad V^{(-)}\in \gamma .
		\end{array}	\right. 
\end{eqnarray}
	The line balances read now
\begin{eqnarray}\label{lbcst}
	2I_{hh}&=& \sum_{i\in \gamma}(c_{h,i}-n_{h,i}) 
	      =  \sum_{i\in \gamma}(c_{h,i})-N_h 
	  \quad i\in\{V^{(c_1)},V^{(c_2}), V^{(c_3)},V^{(\phi\pi)},V^{(-)}\}\\
2I_{c\bar{c}}&=&\sum_{i\in\gamma}(c_{c,i}-n_{c,i})
              = \sum_{i\in\gamma}c_{c,i}-N_{c}
	  \quad i\in\{V^{(\phi\pi)},V^{(-)}\} .
\end{eqnarray}
In order to verify \eqref{c1} we have to show that 
$b(\gamma)=\delta(\gamma)-d(\gamma)\ge 0$.
\begin{eqnarray}\label{c1dst}
	b(\gamma)&=&5-d(\gamma)\\
     &=&5-4+2V^{(c_3)}+2V^{(V_{\phi\pi})}-(\overline{D}^{V^{(-)}}-4)-2I_{c\bar{c}}\\
     &=&1+2V^{(c_3)}-1+\sum_{i\in\phi\pi}n_{\tilde{c},\phi\pi}-(1-n_{c,V^{(-)}})\\
     &=&2V^{(c_3)}+\sum_{i\in\phi\pi}n_{\tilde{c},\phi\pi}-(1-n_{c,V^{(-)}}) .
\end{eqnarray}
Here we have used the line balance for $I_{c\bar{c}}$ (\ref{linetop})and chosen 
the more dangerous case $\overline{D}_{V^{(-)}}=5$.
If $n_{c,V^{(-)}}=0$, there must a $+1$ coming from the $\phi\pi$-sum, because the
FP-charge is conserved. Hence the inequality holds.

The control of 
\begin{eqnarray}\label{2c1r}
c(\gamma)&=&r(\gamma)-\rho(\gamma)\\
	&=&4-2V^{(c_3)} -2V^{(\phi\pi)}+2I_{hh}+2I_{c\bar{c}}
               +(\underline{D}(V^{(-)})-4)-5\\
	&=&-2V^{(c_3)} -2V^{(\phi\pi)}+2I_{hh}+2I_{c\bar{c}}
               +(\underline{D}(V^{(-)})-4)-1\\
	&=&-2V^{(c_3)})-2V^{(\phi\pi)})+2I_{hh}+2I_{c\bar{c}}+ 
                \left\{\begin{array}{l} 
			-1 \,\,{\rm for}\,\,\underline{D}_{V^{(-)}}=3\\
			+1 \,\,{\rm for}\,\,\underline{D}_{V^{(-)}}=5 
	              \end{array} \right.\ge 0 .
\end{eqnarray}
is similar: 
On the vertices we have the information
\be\label{vrbl}
 \sum_{i\in c_1,c_2,c_3}(c_{h,i}-n_{h,i})
+\sum_{i\in\phi\pi}(c_{h,\phi\pi}-n_{h,\phi\pi})
+ (c_{h,V^{(-)}}-n_{h,V^{(-)}})+c_{c,V^{(-)}}\ge0,
\ee
where $c_{c,V^{(-)}}=1$: there is one $c$-field in $V^{(-)}$.
Inserting this into the more dangerous case $\underline{D}_{V^{(-)}}=3$
and taking into account the terms $-2V^{(c_3)}-2V^{(\phi\pi)}-2$ we get 
\begin{eqnarray}\label{cgaes}
	c(\gamma)&=&\sum_{i\in c_1,c_2}(c_{h,i}-n_{h,i})
	 +\sum_{i\in c_3}(c_{h,i}-n_{h,i}-2)\nonumber \\
	&&+\sum_{i\in\phi\pi}((1-n_{h,i}-2)+(2-n_{\tilde{c},i}))\\
	&&+(c_{h,V^{(-)}}-n_{h,V^{(-)}})+1-n_{c,V^{(-)}}-2\ge 0 .
\end{eqnarray}
The two sums in the first line are non-negative for $\gamma$ 1PI.
The same is true as before for the sum in the second line. 
In the third line we look at $1+c_{h,V^{(-)}}-n_{h,V^{(-)}}-n_{c,V^{(-)}}$.
\begin{eqnarray}
	n_{c,V^{(-)}}&=&1 \Rightarrow\, c_{h,V^{(-)}}-n_{h,V^{(-)}}\ge 2\\
	n_{c,V^{(-)}}&=&0 \Rightarrow\, c_{h,V^{(-)}}-n_{h,V^{(-)}}\ge 1,
\end{eqnarray}	
Hence in both cases is $1+c_{h,V^{(-)}}-n_{h,V^{(-)}}-n_{c,V^{(-)}}\ge 2$
and thus $c(\gamma)\ge 0$.

In order to check \eqref{c2} 
we start with the case $V^{(-)}\notin \gamma_i$, i.e.
\begin{eqnarray}\label{stC21}
d(\gamma)&=&4-2V^{(c_3)}(\gamma)-2V^{(\phi\pi)}(\gamma)+2I_{c\bar{c}}(\gamma)
       +\left\{\begin{array}{l}
	       -1\quad {\rm for}\quad \overline{D}_{V^{(-)}}=3 \\
	       +1\quad {\rm for}\quad \overline{D}_{V^{(-)}}=5 
       \end{array}\right.\\
d(\gamma_i)&=&4-2V^{(c_3)}(\gamma_i)-2V^{(\phi\pi)}(\gamma_i)
	                                    +2I_{c\bar{c}}(\gamma_i)\\
 \delta(\gamma)&=&5 \quad \mbox{and} \quad \delta(\gamma_i)=4\\
 d(\bar{\gamma})&=&4\mp1-2V^{(c_3)}(\gamma)-2V^{(\phi\pi)}(\gamma)
	                 +2I_{c\bar{c}}(\gamma)\\
                &&-\sum_i (4-2V^{(c_3)}(\gamma_i)
                     -2V^{(\phi\pi)}(\gamma_i)+2I_{c\bar{c}}(\gamma_i))\\
d(\bar{\gamma})+\sum_i\delta(\gamma_i)&=&
	4\mp 1-2V^{(c_3)}(\bar{\gamma})-2V^{(\phi\pi)}(\bar{\gamma})
	+2I_{c\bar{c}}(\bar{\gamma}) \stackrel{?}{\le} 5\,.\nonumber
\end{eqnarray}
The estimates for $b(\gamma)$ are also valid for $b(\bar{\gamma})$, hence this 
inequality is satisfied.

For the case $V^{(-)}\in \gamma_{i_0}$ the following equations are
relevant 
\begin{eqnarray}\label{stC22}
\delta(\gamma)&=&5 \quad \delta(\gamma_i)=4 \quad i\not=i_0 
		      \quad \delta(\gamma_{i_o})=5\\
         d(\gamma)&=&4-1(+1)-2V^{(c_3)}(\gamma)-2V^{(\phi\pi)}(\gamma)
	                         +2I_{c\bar{c}}(\gamma)\\
	d(\gamma_i)&=&4-2V^{(c_3)}(\gamma_i)-2V^{(\phi\pi)}(\gamma_i)
	                         +2I_{c\bar{c}}(\gamma_i) \quad i\not=i_0\\
 d(\gamma_{i_0})&=&4\mp1-2V^{(c_3)}(\gamma_{i_0})-2V^{(\phi\pi)}(\gamma_{i_0})
                        	+2I_{c\bar{c}}(\gamma_{i_0})\\
	d(\bar{\gamma})&=&4\mp1-2V^{(c_3)}(\gamma)-2V^{(\phi\pi)}(\gamma)
	                   +2I_{c\bar{c}}(\gamma)\\
&&- (4\mp1-2V^{(c_3)}(\gamma_{i_0})
	    -2V^{(\phi\pi)}(\gamma_{i_0})+2I_{c\bar{c}}(\gamma_{i_0}))\nonumber\\
	&&-\sum_{i\not={i_0}} (4-2V^{(c_3)}(\gamma_i)-2V^{(\phi\pi)}(\gamma_i)
			   +2I_{c\bar{c}}(\gamma_i))\nonumber\\
d(\bar{\gamma})+\sum_i\delta(\gamma_i)&=&
	5-2V^{(c_3)}(\bar{\gamma})-2V^{(\phi\pi)}(\bar{\gamma})
	             +2I_{c\bar{c}}(\bar{\gamma}) \stackrel{?}{\le} 5\,.		
\end{eqnarray}
Again: Since the estimate for $b(\gamma)$ is also valid for $b(\bar{\gamma})$
the inequality holds in this case, hence \eqref{c2} is verified.

We now have to verify \eqref{c3} 
For the case $V^{(-)}\notin \gamma_i$ we find

\begin{eqnarray}\label{stC31}
	r(\gamma)&=&4-2V^{(c_3)}(\gamma)-2V^{(\phi\pi)}(\gamma)\\
	&&+2I_{hh}(\gamma)+2I_{c\bar{c}}(\gamma)
         +\left\{\begin{array}{l}
	       -1\quad {\rm for}\quad \underline{D}_{V^{(-)}}=3 \\
	       +1\quad {\rm for}\quad \underline{D}_{V^{(-)}}=5 
       \end{array}\right.\\
r(\gamma_i)&=&4-2V^{(c_3)}(\gamma_i)-2V^{(\phi\pi)}(\gamma_i)
			  +2I_{hh}(\gamma_i)+2I_{c\bar{c}}(\gamma_i)\\
 \rho(\gamma)&=&5 \quad \mbox{and} \quad \rho(\gamma_i)=4\\
 r(\bar{\gamma})&=&4\mp1-2V^{(c_3)}(\gamma)-2V^{(\phi\pi)}(\gamma)
	                 +2I_{hh}(\gamma)+2I_{c\bar{c}}(\gamma)\\
                &&-\sum_i (4-2V^{(c_3)}(\gamma_i)
     -2V^{(\phi\pi)}(\gamma_i)+2I_{hh}(\gamma_i)+2I_{c\bar{c}}(\gamma_i))\\
r(\bar{\gamma})+\sum_i\rho(\gamma_i)&=&
	4\mp1-2V^{(c_3)}(\bar{\gamma})-2V^{(\phi\pi)}(\bar{\gamma})
	+2I_{hh}(\bar{\gamma})+2I_{c\bar{c}}(\bar{\gamma}) \stackrel{?}{\ge} 5\,.\nonumber
\end{eqnarray}
The estimates for $c(\gamma)$ are also valid for $c(\bar{\gamma})$, hence this 
inequality is satisfied.

For the case $V^{(-)}\in \gamma_{i_0}$ the following equations are
relevant 
\be\label{stC3i0}
\rho(\gamma)=5 \qquad \rho(\gamma_i)=4 \quad (i\not=i_0) \qquad \rho(\gamma_0)=5 .
\ee
The equation for $r(\bar{\gamma})$ is unchanged, but due to the presence of
$V^{(-)}$ in $\gamma_{i_0}$ the final equation reads
\be\label{stC3i0f}
r(\bar{\gamma})+\sum_i\rho(\gamma_i)=
                    5\mp1-2V^{(c_3)}(\bar{\gamma})-2V^{(\phi\pi)}(\bar{\gamma})
	+2I_{hh}(\bar{\gamma})+2I_{c\bar{c}}(\bar{\gamma}) \stackrel{?}{\ge} 5\,.
\ee
The question then is, whether
$\tilde{c}(\bar{\gamma})\equiv -2V^{(c_3)}(\bar{\gamma})
-2V^{(\phi\pi)}(\bar{\gamma})
	+2I_{hh}(\bar{\gamma})+2I_{c\bar{c}}(\bar{\gamma}) \ge 1$.
As in (\ref{cgaes}) we rewrite this expression explicitly in sums over vertices 
and their line ``occupation''
\begin{eqnarray}\label{vlo}
\tilde{c}(\bar{\gamma})&=&\sum_{i\in c_1,c_2}(c_{h,i}-n_{h,i})(\bar{\gamma})
	 +\sum_{i\in c_3}(c_{h,i}-n_{h,i}-2)(\bar{\gamma})\nonumber \\
	&&+\sum_{i\in\phi\pi}((1-n_{h,i}-2)+(2-n_{\tilde{c},i}))(\bar{\gamma})\\
	&&+(c_{h,V^{(-)}}-n_{h,V^{(-)}})+(c_{c,V^{(-)}}-n_{c,V^{(-)}})-2\ge 0 \nonumber
\end{eqnarray}
The first two lines represent a situation without $V^{(-)}$ hence the estimates
as before apply, these contributions are non-negative. For the third line we
distinguish two cases:\\
(1) $(n_{c,V^{(-)}})_\gamma=(n_{c,V^{(-)}})_{\bar{\gamma}_{i_0}}=1$ 
(notation: $\bar{\gamma}_{i_0}\equiv\gamma/\gamma_{i_0}$)\\
Here the bracket $c_{c,V^{(-)}}-n_{c,V^{(-)}}$ vanishes. However the first 
bracket (referring to the
$h$-lines) contributes at least 2. Hence the total sum is non-negative.\\
(2) $(n_{c,V^{(-)}})_\gamma=(n_{c,V^{(-)}})_{\bar{\gamma}_{i_0}}=0$\\
Now since the $c\bar{c}$-line starting at $V^{(-)}$ goes straight through the 
whole diagram $\gamma$, it can not form a $c\bar{c}$-loop (it carries a FP-charge).
It must meet at least one $\phi\pi$-vertex $V^{(\phi\pi)}_{*}$. If this vertex
belongs to $\gamma_{i_0}$, it is contracted with $V^{(-)}$ to form a new vertex
in $\bar{\gamma}_{i_0}$ which has one negative FP-charge. Then this is the previous
case. If it does not belong to $\gamma_{i_0}$ then this $V^{(\phi\pi)}_{*}$ appears
as an ordinary FP-vertex in $\bar{\gamma}_{i_0}$ and its contribution is
covered by the second line in (\ref{vlo}). 
Hence the overall estimate holds true and condition \eqref{c3} is satisfied.\\
The condition \eqref{c4}: $\rho(\gamma)=5 \le \delta(\gamma)+1 =5+1$ is satisfied by 
the definition of the subtraction degrees.
In the context of condition \eqref{c4} it is of quite some interest to investigate,
whether the upper limit $\rho(\gamma)=\delta (\gamma)+1$ is consistent with
all the other conditions.
We start with condition \eqref{c1} $\rho(\gamma)\le r(\gamma)$.
For 1PI diagrams diagrams $\gamma$ containing the vertex $V^{(-)}$ this 
means to check, whether 
\be
\delta(\gamma)+1=6\le r(\gamma)= 4-2V^{(c_3)}-2V^{(\phi\pi)}+2I_{hh}+2I_{c\bar{c}}
                    +\left\{\begin{array}{l}
		                           -1\\
		                           +1 .
			\end{array}\right. 
\ee
Rewritten in terms of line balances this means (see (\ref{cgaes}))
\begin{eqnarray}\label{ulrh}
	0&\le&-2+\sum_{i\in c_1,c_2}(c_{h,i}-n_{h,i})
	 +\sum_{i\in c_3}(c_{h,i}-n_{h,i}-2) \\
	&&+\sum_{i\in\phi\pi}((1-n_{h,i}-2)+(2-n_{\tilde{c},i}))\nonumber\\
	&&+(c_{h,V^{(-)}}-n_{h,V^{(-)}})+(1-n_{c,V^{(-)}})
	                    +\left\{\begin{array}{l}
		                           -1\\
		                           +1
			\end{array}\right.\nonumber
\end{eqnarray}
Since the sums in the first and second line are non-negative (s. discussions
above), this boils down to 
\be
(c_{h,V^{(-)}}-n_{h,V^{(-)}})+(1-n_{c,V^{(-)}})
	                    +\left\{\begin{array}{l}
		                           -3\\
		                           -1
			\end{array}\right.\nonumber
\ge0
\ee
(Let us recall: upper entry $-3$ stands for contributions $\int c(\partial)^3 h\cdots h$,
lower entry $-1$ for $\int c((\partial)^5 h\cdots h$ to $V^{(-)}$.)
But we only know for sure that 
$(c_{h,V^{(-)}}-n_{h,V^{(-)}})+(1-n_{c,V^{(-)}})\ge 2$.
Hence, if this lower bound can indeed be realized, the upper limit for 
$\rho(\gamma)$ would not be allowed in the derivation of the ST. It would however
be allowed for the Green functions constructed out of $\lbrack NP \rbrack^4_4\,$ 
normal products.
If indeed  $\rho(\gamma)=\delta(\gamma)+1$ can not be used then the
IR-subtractions within $\tau(\gamma)$ (\ref{sbtr}) are active i.e.\
UV-subtractions alone would not guarantee convergence. In QED 
$\rho(\gamma)=\delta(\gamma)+1$ is allowed, hence by (\ref{rmss}) only 
UV-subtractions are active. To the contrary, as here, in Yang-Mills (YM) it is not. 
Of course, at $s=1$ the dependence on $M$ disappears if the LZ-equation 
holds (cf. (\ref{lzv})).
Again, as for Lagrangian vertices we can refer also in the present case to
Lowenstein's theorem for convergence in the same sense as above.

\subsection{Slavnov-Taylor identity\label{se:stidentity}}
The ST identity which we have to establish to higher orders takes the same form 
as in tree approximation, (\ref{fbrst}), supplemented however
by the $m^2$-dependent gauge fixing, (\ref{cmgf}), and Faddeev-Popov-terms, 
(\ref{FaPo}), i.e.
\be\label{2fbrst}
\mathcal{S}(\Gamma)\equiv
\int\Big(\frac{\delta\Gamma}{\delta{K}}\frac{\delta\Gamma}{\delta h}
+\frac{\delta\Gamma}{\delta L}\frac{\delta\Gamma}{\delta c}
+b\frac{\delta\Gamma}{\delta\bar{c} }\Big)=0
\ee
\begin{eqnarray}
\label{2cmgf}
	\Gamma_{\rm gf}&=&-\frac{1}{2\kappa}\int dxdy\, h^{\mu\nu}(x)
	(\partial_\mu b_\nu+\partial_\nu b_\mu)(y)\Big\lbrace\big( \frac{\Box}{4\pi^2} + m^2 \big)
\frac{1}{(x-y)^2}\Big\rbrace \\
 &&-\int\frac{\alpha_0}{2}\eta^{\mu\nu} b_\mu b_\nu\\
\label{2FaPo}
	\Gamma_{\phi\pi}&=&-\frac{1}{2}\int dxdy\, \brsts\,h^{\mu\nu}(x)
(\partial_\mu \bar{c}_\nu
	+\partial_\nu \bar{c}_\mu)(y)\Big\lbrace\big( \frac{\Box}{4\pi^2} + m^2 \big)\frac{1}{(x-y)^2}\Big\rbrace .
\end{eqnarray}
The $b,\bar{c}$-field equations of motion take now the form 
\begin{eqnarray}
\label{2beq}
\frac{\delta \Gamma}{\delta b^\rho}&=&\kappa^{-1}\int 
                              dy\,\partial^\mu h_{\mu\rho}(y)
             \Big\lbrace\big( \frac{\Box}{4\pi^2} + m^2 \big)\frac{1}{(x-y)^2}\Big\rbrace-\alpha_0 b_\rho\\
\label{2ghe}
	\frac{\delta\Gamma}{\delta \bar{c}_\rho(x)}&=&
	-\int dy\, \kappa^{-1} \partial_\lambda\frac{\delta\Gamma}{\delta K_{\lambda\rho}(y)} 
             \Big\lbrace\big( \frac{\Box}{4\pi^2} + m^2 \big)\frac{1}{(x-y)^2}\Big\rbrace .
\end{eqnarray}
Again the $b$-field equation can be integrated trivially back to (\ref{2cmgf}) and 
therefor the functional $\bar{\Gamma}$ be introduced as in the tree approximation
\be\label{2Gmmbr}
\Gamma = \Gamma_{\rm gf}+\bar{\Gamma} .
\ee
(\ref{rstc}) is changed into
\be\label{2rstc}
\kappa^{-1}\int dy\,
           \partial_\lambda\frac{\delta\bar{\Gamma}}{\delta K_{\mu\lambda}(y)}
			      \Big\lbrace\big( \frac{\Box}{4\pi^2} + m^2 \big)\frac{1}{(x-y)^2}\Big\rbrace
+\frac{\delta\bar{\Gamma}}{\delta\bar{c}_\mu} =0,
\ee
whereas (\ref{sceH}) becomes
\be\label{2sceH}
H_{\mu\nu}(x)=K_{\mu\nu}(x)
+\frac{1}{2}\int dy\,(\partial_\mu\bar{c}_\nu+\partial_\nu\bar{c}_\mu)(y)
   \Big\lbrace\big( \frac{\Box}{4\pi^2} + m^2 \big)\frac{1}{(x-y)^2}\Big\rbrace .
\ee
The relations (\ref{brGm}) are unchanged:
\begin{eqnarray}\label{2brGm}
\mathcal{S}(\Gamma)&=&\frac{1}{2}\mathcal{B}_{\bar{\Gamma}}\bar{\Gamma}=0\\
	\mathcal{B}_{\bar{\Gamma}}&\equiv& 
	\int\Big(
  \frac{\delta\bar{\Gamma}}{\delta H}\frac{\delta}{\delta h}
+ \frac{\delta\bar{\Gamma}}{\delta h}\frac{\delta}{\delta H} 
+ \frac{\delta\bar{\Gamma}}{\delta L}\frac{\delta}{\delta c}
+ \frac{\delta\bar{\Gamma}}{\delta c}\frac{\delta}{\delta L}\Big) .
\end{eqnarray}

In the BPHZL renormalization scheme the starting point for establishing equations
like the above one's to all orders is a $\Gamma_{\rm eff}$ with which one calculates accordingly subtracted Feynman diagrams. 
Here we choose
\be\label{Gmmff}
\Gamma_{\rm eff}=\Gamma^{\rm class}_{\rm inv} +\Gamma_{\rm gf}+\Gamma_{\phi\pi}
                                 +\Gamma_{\rm e.f.}+\Gamma_{\rm ct} .
\ee
In addition to (\ref{ivc}),(\ref{clssct}),(\ref{2cmgf}), and (\ref{2FaPo})
one has to take into account the changes caused by the auxiliary mass term
in (\ref{gmps1}) and (\ref{gmps2}).
$\Gamma_{\rm ct}$ will collect counterterms as needed. All these
expressions are to be understood as normal products, i.e.\ insertions into Green
functions with power counting degrees $\delta=\rho=4$.

Starting from $Z$, the generating functional for general Green functions, 
and from the definition of $\mathcal{S}$ in (\ref{sma}) we
postulate 
\be\label{Zbrst}
\mathcal{S}Z=0.
\ee
Then the action principle yields
\be\label{acZbrst}
\mathcal{S}Z=\Delta_Z\cdot Z= \Delta_Z +O(\hbar \Delta_Z),
\ee
where $\Delta_Z\equiv[\Delta_Z]^5_5$ is an integrated insertion with 
$Q_{\phi\pi}(\Delta_Z)=+1$.
Again, by invoking the action principle one can realize the $b$-field
equation of motion (\ref{2beq}), with (\ref{2rstc}), now on the renormalized
level, as a consequence of (\ref{2fbrst}). This admits (\ref{2brGm}) as a 
postulate and results into
\begin{eqnarray}\label{rgheq}
   \mathcal{S}(\Gamma)&=&\Delta\cdot\Gamma\\
\frac{1}{2}\mathcal{B}_{\bar{\Gamma}}\bar{\Gamma}&=&\Delta+O(\hbar\Delta) .
\end{eqnarray}
Here $\Delta\equiv [\Delta]_5^5$ with $Q_{\phi\pi}(\Delta)=+1$
does not dependent on $b$ and $\bar{c}$. These relations admit a cohomological 
treatment, since 
\be\label{cstc}
\mathcal{B}_{\bar{\Gamma}}\mathcal{B}_{\bar{\Gamma}}\bar{\Gamma} =0, \qquad
\mathcal{B}_{\bar{\Gamma}}\mathcal{B}_{\bar{\Gamma}}=0,
\ee
the latter being true as a necessary condition, if (\ref{2brGm}) is to be 
satisfied.
Since in the tree approximation (\ref{2brGm}) holds one has
\be\label{2cstc}
\brstb\Delta=0 \quad {\rm for} \quad \brstb\equiv \mathcal{B}_{\bar{\Gamma}_{\rm class}}
\qquad {\rm with} \quad \brstb^2=0
\ee
as the final consistency condition to be solved. 
The standard way to solve this cohomology problem is to list contributions to $\Delta$ by starting with terms depending on external fields and then those consisting of elementary fields only, i.e.
\be\label{chmlg}
\Delta= \int(K_{\mu\nu}\Delta^{\mu\nu}(h,c)+L_\rho\Delta^\rho(h,c))
               +\Lambda(h,c) .
\ee
All terms are insertions compatible with $[...]^5_5$
and $Q^{\phi\pi}=+1$. (Recall that $Q^{\phi\pi}(K)=-1$ and 
$Q^{\phi\pi}(L)=-2$.)
In \cite{Barnich:1994kj,Barnich:1995ap} it is shown, that all these contributions eventually are $\brstb$-variations. 
This is true even for the $\Lambda$-term. 
This means that pure gravity has no anomalies, the solution reads:
\be\label{fchgrv}
\Delta=\brstb \hat{\Delta}
\ee
with a $\hat{\Delta}$ which can be absorbed into $\Gamma_{\rm eff}$. 
In the quoted references the algebra leading to this result has been performed by using cohomological methods.
Without power counting and convergence and not within a concrete renormalization scheme, this represents a classical consideration.
In the present context we have, however, supplied it with ``analytic''
information, i.e.\ assured the existence of the relevant quantities as
insertions into existing Green functions. 
The result is thus that we have indeed a ST-identity which holds as inserted 
into general Green's functions of elementary fields, at non-exceptional momenta 
and $s=1$. 

Along the lines given in the tree approximation one can now establish the unitarity of the $S$-matrix. It is however clear
that such a construction is to a large extent purely formal, because one has to go on-shell and hits physical IR divergences there in many configurations of incoming and outgoing particles.\\
Let us nevertheless sketch some of the required steps. First of all the matrix of residua $z^{-1}$ becomes relevant. Then like in the tree approximation the state space operator $Q^{\rm BRST}$ can be calculated with the same arguments as there: only linear
terms in the functional transformation contribute. They appear however with factors which have to be shown via some tests on the ST to permit a multiplicative renormalization of the tree approximation charge. With this result one can deduce that the $S$-matrix maps physical states onto physical states. 
These physical states have to be constructed in two steps: In the first one a state $|{\rm phys\rangle}$ is called ``physical'' if it is annihilated by $Q^{\rm BRST}$, i.e.
\be\label{phsst} 
Q^{\rm BRST}|\rm{phys}\rangle=0
\ee 
This requirement defines a linear subspace in the full indefinite metric Fock space and eliminates states with negative norm.
In the second step one forms equivalence classes of physical states which differ only by the number of particles which generate vanishing norm. The completion of this state of equivalence classes contains then only states with non-zero norm.
On this physical Hilbert space the $S$-matrix is unitary.
It is worthwhile to mention that this construction has been shown to exist rigorously e.g.\  in the context of Yang-Mills theory with complete breakdown of internal symmetry to a completely massive theory \cite{Becchi:1985bd}.
Due to on-shell IR-divergences it is only formally valid in the present case. One can however expect that scattering amplitudes which are not affected by IR-divergences are physically meaningful.

Based on the ST one may construct Green functions of BRST-covariant operators which are independent of gauge parameters
and could then serve as building blocks for observables. 
But this will not be covered in this work and is left for future research.

\subsection{Normalization conditions II\label{se:nc2}}
The normalization conditions (\ref{trnorm1})-(\ref{trnorm5}) have to be 
modified such 
that they are compatible with higher orders of perturbation theory: they
have to be taken at values in momentum space which are consistent with
the subtraction procedure.
They read
\begin{eqnarray}\label{highnorm}
\frac{\partial}{\partial p^2}\,\gamma^{(2)}_{\rm TT\,|{\substack{p=0 \\ s=1} }}&=
				   &c_3\kappa^{-2}\\
\frac{\partial}{\partial p^2}\frac{\partial}{\partial p^2}\,
	\gamma^{(2)}_{\rm TT\,|{\substack{p^2=-\mu^2\\ s=1} }}&=&-2c_1\\
\frac{\partial}{\partial p^2}\frac{\partial}{\partial p^2}\,
	\gamma^{(0)}_{\rm TT\,|{\substack{p^2=-\mu^2 \\ s=1} }}
                                        	&=&2(3c_2+c_1)\\
	\Gamma_{h^{\mu\nu}} &=&-\eta_{\mu\nu}c_0=0\\			
\frac{\partial}{\partial p_\sigma}
	\Gamma_{K^{\mu\nu}c_\rho|{\substack{p^2=-\mu^2 \\ s=1} }}&=&
                                -i\kappa(\eta^{\mu\sigma}\delta^\nu_\rho
	                          +\eta^{\nu\sigma}\delta^\mu_\rho
		  -\eta^{\mu\nu}   \delta^\sigma_\rho)\label{highnorm1} \\
\frac{\partial}{\partial p^\lambda}
	\Gamma_{{L_\rho}c^\sigma c^\tau|{\substack{p^2=-\mu^2 \\ s=1} }}&=&
			-i\kappa(\delta^\rho_\sigma\eta_{\lambda\tau}
				  -\delta^\rho_\tau\eta_{\lambda\sigma}).
\end{eqnarray}
Imposing the $b$-equation of motion (\ref{beq}) still fixes $\alpha_0$ 
and the $b$-amplitude, whereas (\ref{highnorm1}) again
fixes the $h$- and $K$-amplitudes. 

\section{Invariant differential operators and invariant insertions\label{se:invdiffop}}
Here we develop the concept of BRST-invariant differential operators
and their one-to-one counterparts, BRST-invariant insertions.
One can essentially follow the paper \cite{Piguet:1984js} and translate from YM to gravity.

Suppose a model satisfies the WI of a linear transformation
\be\label{ltWI}
W^a\Gamma\equiv\int\delta^a\phi\frac{\delta\Gamma}{\delta\phi}=0
\ee
and $\lambda$ is a parameter of the theory (e.g.\ coupling, mass, 
normalization parameter) of which the WI-operator $W^a$ does not
depend. Then $\lambda\partial_\lambda$ commutes with $W^a$, i.e.
\be\label{sc0}
[\lambda\partial_\lambda,W^a]=0.
\ee
Then the action principle tells us that 
\be\label{srt}
\lambda\partial_\lambda\Gamma=\Delta_\lambda\cdot\Gamma .
\ee
Applying $W^a$ to (\ref{srt}) and using (\ref{sc0}) we find
\be\label{sct}
W^a(\Delta_\lambda\cdot\Gamma)= W^a\Delta_\lambda+O(\hbar\Delta_\lambda)=0,
\ee
which expresses the invariance of $\Delta_\lambda$ under the symmetry 
transformation $W^a$: $\lambda\partial_\lambda$ and $\Delta_\lambda$ are 
called symmetric with respect to the symmetry $W^a$.

For the $\Gamma$-non-linear BRST-symmetry one has to proceed slightly 
differently. We shall call an insertion $\Delta$ BRST-symmetric if to first 
order in $\epsilon$
\begin{eqnarray}
	\mathcal{S} (\Gamma_\epsilon)&=&O(\epsilon^2) \label{scbrst}\\
	{\rm for} \qquad \Gamma_\epsilon&=&\Gamma+\epsilon\Delta\cdot\Gamma
		   \qquad{\rm with}\qquad \mathcal{S} (\Gamma)=0.
\end{eqnarray}
If $\Delta$ is generated by a differential operator $(\ref{srt})$, this
differential operator will be called BRST-symmetric. Writing (\ref{scbrst})
explicitly we have
\be\label{scbrst2}
\ST(\Gamma)+\epsilon \ST_\Gamma\Delta\cdot\Gamma=O(\epsilon^2)
\ee
\be\label{scbrst3}
\ST_\Gamma\equiv\int\left(
 \frac{\delta\Gamma}{\delta K}\frac{\delta}{\delta h}
+\frac{\delta\Gamma}{\delta h}\frac{\delta}{\delta K}
+\frac{\delta\Gamma}{\delta L}\frac{\delta}{\delta c}
+\frac{\delta\Gamma}{\delta c}\frac{\delta}{\delta L}
+b\frac{\delta}{\delta\bar{c}}\right)
+\chi\frac{\partial}{\delta\alpha_0} ,
\ee
i.e.\ the symmetry condition reads
\be\label{scbrst4}
\ST_\Gamma\Delta\cdot\Gamma=0 .
\ee
A comment is in order. Although later we shall exclusively work in Landau gauge, we carry here the gauge parameter $\alpha_0$ along as
preparation for the general solution with arbitrarily many parameters $z_{nk}$. This 
facilitates the formulation of the general version. Actually relevant at the end
are only the formulae with $\alpha_0=\chi=0$.
The explicit form of $\ST_\Gamma$ precisely defines how to perform the 
variation of the fields. \footnote{This formula shows that it is not the
demand ``linearity in $\Gamma$'' which determines its form, but rather
the demand ``correct  transformation of an insertion ''.}
The operator $\ST_\Gamma$ is helpful for rewriting the gauge fixing and 
$\phi\pi$-contributions to the action \eqref{2cmgf}:
\be\label{vfrmgffp}
\Gamma_{\rm gf}+\Gamma_{\phi\pi}
    = \ST_\Gamma\left(-\frac{1}{2\kappa}\int h^{\mu\nu}(x)
              (\partial_\mu\bar{c}_\nu+\partial_\nu\bar{c}_\mu)(y)
	      \Big\{\big( \frac{\Box}{4\pi^2} + m^2 \big)\frac{1}{(x-y)^2}\Big\}
	       -\int\frac{\alpha_0}{2}\eta^{\mu\nu}\bar{c}_\mu b_\nu\right) .
\ee
(Note: the last term creates a contribution which has not been taken into 
account in (\ref{2cmgf}), however in (\ref{varalph}).)
When going over to $Z$, the generating functional for the general Green 
functions,
it is clear, that gauge fixing and $\phi\pi$-term vanish between physical
states, because they are a BRST-variation.

A necessary condition for insertions to be BRST-symmetric is obtained
by acting with $\delta/\delta b$ on (\ref{scbrst}): 
\be\label{snn}
G\Delta\cdot\Gamma=S_\Gamma\frac{\delta\Delta\cdot\Gamma}{\delta b},
\qquad G^\rho \equiv\frac{\delta}{\delta\bar{c}_\rho(x)}
+\kappa^{-1}\int dy\,\partial_\lambda\frac{\delta\bar{\Gamma}}{\delta K_{\rho\lambda}(y)}
   \Big\{\big( \frac{\Box}{4\pi^2} + m^2 \big)\frac{1}{(x-y)^2}\Big\} .
\ee
For $b$-independent insertions $\Delta$ one must ensure the homogeneous 
ghost equation
\be\label{ghD}
G\Delta\cdot\Gamma = 0 .
\ee
Using the gauge condition
\be\label{gc}
\frac{\delta\Gamma}{\delta b_\rho}=-\alpha_0 \eta^{\rho\lambda}b_\lambda
 +\kappa^{-1}\int dy\,\partial_\mu h^{\mu\rho}(y)\Big\{\big( \frac{\Box}{4\pi^2} + m^2 \big)\frac{1}{(x-y)^2}\Big\} ,
\ee
one can reduce (\ref{snn}) to        
\be\label{rsc}
\mathcal{B}_{\bar{\Gamma}}\Delta\cdot\Gamma=0 .
\ee
In the tree approximation we have called this operator $\brstb$.

Our next task is to construct a {\it basis} for all symmetric insertions 
of dimension 4, $\phi\pi$-charge 0, and independent of $b_\rho$ -- first in the
tree approximation and then to all orders. A systematic way to find them is
to solve the cohomology problem
\be\label{cpr}
\brstb\Delta=0
\ee
for $\Delta$ satisfying
\begin{eqnarray}\label{cpr2}
	\frac{\delta\Delta}{\delta b}=&0,&G\Delta=0\\
	{\rm dim}(\Delta)=&4,&Q_{\phi\pi}(\Delta)=0 .
\end{eqnarray}

Here $\brstb=\mathcal{B}_{\bar{\Gamma}_{\rm class}}$, hence
\begin{eqnarray}\label{lsttrs}
	  \brstb&=& \brsts \qquad{\rm on\, all\, elementary\, fields}\\
\brstb H_{\mu\nu}&=&\frac{\delta\bar{\Gamma}_{\rm cl}}{\delta h^{\mu\nu}}
	     =\frac{\delta\Gamma^{\rm class}_{\rm inv}}{\delta h^{\mu\nu}}
	     -\kappa(H_{\lambda\mu}\partial_\nu c^\lambda
	     +H_{\lambda\nu}\partial_\mu c^\lambda
	     +\partial_\lambda(H_{\mu\nu}c^\lambda))\label{lsttrsH}\\
   \brstb L_\rho&=&\frac{\delta{\bar\Gamma}_{\rm cl}}{\delta c^\rho}=
	    \kappa(2\partial^\lambda H_{\lambda\rho}
	    +2\partial_{\lambda'}(H_{\rho\lambda}h^{\lambda'\lambda}
	    +H_{\lambda'\lambda}\partial_\rho h^{\lambda\lambda'}))\\
	   &&\qquad\quad  -\kappa(L_\lambda\partial_\rho c^\lambda
		    +\partial_\lambda(L_\rho c^\lambda)) . \label{lsttrsL}
\end{eqnarray}
In order to proceed we first separate the $\alpha_0$-dependence
\be\label{s0d}
\Delta=\chi \Delta_- +\Delta_0 .
\ee
We now define
\be\label{bbr}
\bar{\brstb}=\left\{\begin{array}{l}
	      \brstb\qquad{\rm on}\quad h,c,H,L\\
              0\qquad{\rm on}\quad \alpha_0
              \end{array}
       \right.
\ee
and note that
\be\label{bbra}
\partial_{\alpha_0}(\brstb\psi)=0\qquad{\rm for}\quad \psi=h,c,H,L
\ee
with $\bar{\brstb}^2$=0, since $\bar{\Gamma}_{\rm cl}$ is independent of $\alpha_0$.
(\ref{cpr}) implies
\be\label{spr}
\bar{\brstb}\Delta_- -\partial_{\alpha_0}\Delta_0=0 \qquad \bar{\brstb}\Delta_0=0,
\ee
hence
\be\label{spr2}
\Delta=\brstb\hat{\Delta}_- +\hat{\Delta}_0.
\ee
Here $\hat{\Delta}_0$ is $\alpha_0$-independent and $\bar{\brstb}$-invariant.
Since $\bar{c}$ does not occur, a negative $\phi\pi$-charge can only be
generated by external fields, hence
\be\label{lfe}
\hat{\Delta}_- = \int(f_H(\alpha_0)H_{\mu\nu}h^{\mu\nu}
		      +f_L(\alpha_0)L_\rho c^\rho)
\ee
which is the precise analogue of \cite[(4.19)]{Piguet:1984js},
is certainly a solution. However in the present case the field $h^{\mu\nu}$
has canonical dimension zero, whereas its counterpart in Yang-Mills theory, 
the vector field $A_\mu$ has dimension one. So every function 
$\mathcal{F}^{\mu\nu}(h)$ is also a solution. For the time being we continue
with (\ref{lfe}) and discuss the general solution 
at a later stage (cf. Sect. \ref{se:generalsolutionSTI}).
It is worth solving the subproblem
\be\label{spr3}
\partial_{\alpha_0}\hat{\Delta}_0=0 \qquad \bar{\brstb}\hat{\Delta}_0=0
\ee
explicitly. 
We start listing the contributions to $\hat{\Delta}_0$ ordered by their
external field dependence, i.e.
\be\label{Ld}
\hat{\Delta}_0=-f_L(0)\kappa\int L_\rho c^\lambda\partial_\lambda c^\rho
			 +\cdots({\rm indep.\, of}\, L),
\ee
where $f_L(0)$ is an arbitrary number independent of $\alpha_0$. 
With (\ref{lsttrsL}) this term can be rewritten as
\be\label{Ld2}
\hat{\Delta}_0=f_L(0)\bar{\brstb}(\int L_\rho c^\rho)
                                +\cdots({\rm indep.\, of}\, L)
\ee
\be\label{Ld3}
\hat{\Delta}_0= \brstb \int(f_L(0)L_\rho c^\rho)
                                +\cdots({\rm indep.\, of}\, L) .
\ee
We next make explicit the $H$-dependence
\be\label{Hd}
\hat{\Delta}_0= \brstb \int(f_L(0) L_\rho c^\rho)
	  +\int H_{\mu\nu}F_{(+)}^{\mu\nu}(h,c)+\cdots(L,H)-{\rm indep.}
\ee
The postulate (\ref{spr3}) reads
\begin{eqnarray}\label{Hd2}
	0&=&\bar{\brstb}\hat{\Delta}_0=
       \int\Big(\frac{\delta\bar{\Gamma}_{\rm cl}}{\delta h}F^{(+)}
-H\bar{\brstb}F^{(+)}\Big)+(L,H)-{\rm indep.}\\
	&=:&-\int H\mathcal{C}F^{(+)}+(L,H)-{\rm indep.}
\end{eqnarray}
and defines a transformation $\mathcal{C}$ as the coefficient of $H$ in
(\ref{Hd}):
\be\label{mthcC}
\mathcal{C}F_{(+)}=\bar{\brstb}F^{(+)}
      +\kappa(\partial_\lambda c^\mu F^{\nu\lambda}_{(+)}
             +\partial_\lambda c^\nu F^{\mu\lambda}_{(+)}
	     -c^\lambda\partial_\lambda F^{\mu\nu}_{(+)}).
\ee
This transformation is nilpotent and satisfies, due to (\ref{Hd2}),
\be\label{mthcC2}
\mathcal{C}F_{(+)}=0
\ee
One solution is 
\be\label{sF}
F^{\mu\nu}_{(+)}=\mathcal{C}(f_H(0)h^{\mu\nu}).
\ee
Since
\be\label{sF2}
\mathcal{C}(h^{\mu\nu})=
              \kappa(-\partial^\mu c^\nu-\partial^\nu c^\mu) ,
\ee
it fits correctly to the $H$-dependent part of $(\ref{lsttrsH})$ in
(\ref{Hd2}).
One thus arrives for this solution at
\be\label{sHd3}
\bar{\brstb}\int f_H(0)H_{\mu\nu}h^{\mu\nu}
		       =\int H_{\mu\nu}\mathcal{C}(f_H(0)h^{\mu\nu}) ,
\ee
i.e.\ the $H$-dependent part in $\hat{\Delta}_0$ is also a variation.
As mentioned above this is not the most general solution, but that will
be treated later with the analogous outcome.

The remaining contributions to $\hat{\Delta}_0$ depend only on $h$ and must not depend on $\alpha_0$. 
The only invariants are the terms appearing in $\Gamma^{\rm class}_{\rm inv}$. 
They are not variations, but constitute obstruction terms to the $\bar{\brstb}$-cohomology. 
Altogether we thus have
\be
\Delta_0= \brstb \int(f_L(0)L_\rho c^\rho+f_H(0)H_{\mu\nu}h^{\mu\nu})
	    +\int\,\sqrt{-g}(\hat{c}_3R+\hat{c}_1R^{\mu\nu}+\hat{c}_2R^2) .
\ee
(The factors $\hat{c}$ are independent of $\alpha_0$.)
In tree approximation we end up with five invariant insertions of dimension 4
and $\phi\pi$-charge 0, which are independent of $b_\rho$ and satisfy the 
ghost equation:
\begin{eqnarray}\label{sns5}
	\Delta'_L&=&\brstb\left(f_L(\alpha_0)\int L_\rho c^\rho\right)\\
	\Delta'_H&=&\brstb\left(f_H(\alpha_0)\int H_{\mu\nu} h^{\mu\nu}\right)\\
	\Delta_{c_3}&=&c_3\kappa^{-2}\int\sqrt{-g}\kappa^{-2}R
  \quad\,	\Delta_{c_1}=c_1\int\sqrt{-g}R^{\mu\nu}R_{\mu\nu}
  \quad\, \Delta_{c_2}=c_2\int\sqrt{-g}R^2 .
\end{eqnarray}
(Here we 
renamed the couplings of the non-variations.)
In higher orders we may define easily invariant insertions for those which are not variations:
\be\label{hghcpls}
\Delta_{c_i}:=c_i\frac{\partial}{\partial c_i}\Gamma\quad 
                                   (i=1,2,3 \quad{\rm no\,\, sum}),
\ee
however it is clear that the $(s-1)$-dependent normal products
$c_{31}[\kappa^{-1}m\int\sqrt{-g}R\,]^4_4$ and 
$c_{32}1/2[m^2\int\sqrt{-g}R\,]^4_4$
also belong to the basis in higher orders and make part of $\Gamma_{\rm eff}$.
Hence we define them also as invariant by the respective derivation with
respect to their coupling
\be\label{gnbss}
\Delta_{c_{31}}:=c_{31}\frac{\partial}{\partial c_{31}}\Gamma \qquad
\Delta_{c_{32}}:=c_{32}\frac{\partial}{\partial c_{32}}\Gamma .
\ee
Accordingly we change the notation $c_3\rightarrow c_{30}$.
The other terms we also try to represent as symmetric {\sl differential}
operators acting on $\Gamma$.\\
We rewrite $\Delta'_L$:
\begin{eqnarray}\label{Ldo}
	\Delta'_L= \brstb\left(f_L(\alpha_0)(\alpha_0)\int L_\rho c^\rho\right)
  &=&\chi f'_L\int Lc
     +f_L\int\left(\frac{\delta\bar{\Gamma}_{\rm cl}}{\delta c}
     +L\frac{\delta\bar{\Gamma}_{\rm cl}}{\delta L}\right)\\
  &=&\chi f'_L\int Lc
     +f_L\int\left(-c\frac{\delta\bar{\Gamma}_{\rm cl}}{\delta c}
     +L\frac{\delta\bar{\Gamma}_{\rm cl}}{\delta L}\right)\\
	&=&-f_L\mathcal{N}_L\Gamma_{\rm cl}
     +\chi f'_L\int Lc,
\end{eqnarray}
where $\mathcal{N}$ denote a leg-counting operator. This suggests defining
$\Delta_L$ to all orders by
\begin{eqnarray}\label{Ldoh}
\Delta_L\cdot\Gamma&=&f_L(\alpha_0)\mathcal{N}_L\Gamma-\chi f'_L\int Lc,\\
\mathcal{N}_L&\equiv&\int\Big(c\frac{\delta}{\delta c}-L\frac{\delta}{\delta L}\Big)
    =N_c-N_L .
\end{eqnarray}
It is to be noted that the $\chi$-dependent term in (\ref{Ldoh}) is well 
defined since $L$ is an external field, hence the expression is linear in the
quantized field (c). $\Delta_L$ does obviously not depend on $b_\rho$, it
satisfies the ghost equation and it fulfills (\ref{rsc}), since it can be 
written as
\be\label{fdL}
\Delta_L\cdot\Gamma=-\mathcal{B}_{\bar{\Gamma}}\left(f_L\int Lc\right),
\ee
and since $\mathcal{B}_{\bar{\Gamma}}$ is nilpotent. Hence it is a 
BRST-symmetric operator to all orders.

Finally we have to extend $\Delta_H'$. We first rewrite it in the form
\be\label{Hdo}
\Delta_H'= \brstb\left(f_H(\alpha_0)\int H_{\mu\nu}h^{\mu\nu}\right)
	=f_H N_H\bar{\Gamma}_{\rm cl}-f_HN_H\Gamma_{\rm cl}
		+\chi f'_H\int H_{\mu\nu}h^{\mu\nu}.
\ee
Next we go over to $\Gamma_{\rm cl}$ in the variables $K$ and $\bar{c}$:
\begin{eqnarray}\label{Hon}
	\Delta_H'&=&f_H(N_h-N_K-N_b-N_{\bar{c}}
	+2\alpha_0\partial_{\alpha_0}+2\chi\partial_\chi)\Gamma_{\rm cl}\\
	&&+\chi f'_H\Big(\int \Big( Kh-\bar{c}\frac{\delta\Gamma_{\rm cl}}{\delta b}\Big)
	   +2\alpha_0\frac{\partial}{\partial \chi}\Gamma_{\rm cl}\Big) .
\end{eqnarray}
This suggests as definition of $\Delta_H$ to all orders
\begin{eqnarray}\label{fdH}
\Delta_H\cdot\Gamma:&=&f_H\mathcal{N}_K\Gamma
	+\chi f'_H \Big(\int \Big( Kh-\bar{c}\frac{\delta\Gamma_{\rm cl}}{\delta b}\Big)
	                 +2\alpha_0\frac{\partial}{\partial \chi}\Gamma\Big)\\
    \mathcal{N}_H&\equiv& N_h-N_K-N_b-N_{\bar{c}}
	                 +2\alpha_0\partial_{\alpha_0}+2\chi\partial_\chi.
\end{eqnarray}
Or else
\be\label{fdHa}
\Delta_H\cdot\Gamma :=\ST_\Gamma\left(f_H(\alpha_0)
       \Big(\int \Big(Kh-\bar{c}\frac{\delta\Gamma}{\delta b} \Big)
                        +2\alpha_0\frac{\partial\Gamma}{\partial\chi}\Big)\right).
\ee
In view of 
\be\label{npS}
\ST_\Gamma \ST_\Gamma=0
\ee
for all $\Gamma$ with $\ST(\Gamma)=0$, $\Delta_H$ is BRST-symmetric once we 
have verified that it is independent of $b_\rho$ and satisfies the ghost 
equation.\\
\be\label{chbi}
\frac{\delta}{\delta b}(\Delta_H\cdot\Gamma)=0
\ee
is readily checked in the form (\ref{fdH}). 
\be\label{chgh}
G(\Delta_H\cdot\Gamma)=0
\ee
is best checked in the form (\ref{fdHa}) by observing that
\be\label{hpch}
G\left(\int\Big(Kh-\bar{c}\frac{\delta\Gamma}{\delta b}\Big)
              +2\alpha_0\frac{\partial\Gamma}{\partial\chi}\right)=0,
\ee
and
\be\label{ghsg}
\{G,\ST_\Gamma\}=0
\ee
(this latter property being due to $G\Gamma=-1/2 \chi b$).

To summarize in compact notation we denote the above symmetric differential
operators by
\be\label{tns}
\nabla_i\in \{c_1\partial_{c_1}, c_2\partial_{c_2}, 
c_{30}\partial_{c_{30}}, c_{31}\partial_{c_{31}}, c_{32}\partial_{c_{32}}, 
         \mathcal{N}_H, \mathcal{N}_L \}
\ee
and have with (\ref{hghcpls}),(\ref{gnbss}), (\ref{fdL}), and (\ref{fdHa})
defined a basis of symmetric insertions to all orders by
\be\label{gnp}
\nabla_i\Gamma \stackrel{.}{=} \Delta_i\cdot\Gamma.
\ee
The fact that symmetric differential operators and symmetric insertions
are in one-to-one correspondence just means that adding symmetric 
counterterms $\Delta_i$ to $\Gamma$ is renormalizing the corresponding
quantity $i$ indicated by $\nabla_i$ of the theory. Fixing the arbitrary 
parameters in the symmetric insertions (\ref{sns5}) is again performed by
satisfying normalization conditions and the present analysis shows
that the conditions (\ref{highnorm}) are appropriate. In higher orders
the Euclidean point $-\mu^2$ is relevant. $\alpha_0=0$ and $\chi=0$
are to be chosen now. 
Once one has satisfied these normalization conditions the theory is completely fixed.

\section{Removing auxiliary mass dependence via Zimmermann Identities\label{se:removingauxmassZI}}
Above we have introduced among the symmetric insertions several which depend
on the auxiliary mass. Here we study to which extent they can be effectively removed by using ZI's \cite{Zimmermann:1972te}.

\subsection{Shift\label{se:shift}} 
In (\ref{ivce}) we replaced
$c_3\kappa^{-2}
{\rm within}\quad\gamma^{(r)}_{KL},\, r=2,K=L=T$ by 
$c_3\kappa^{-2} \rightarrow c_{30}\kappa^{-2}+m\kappa^{-1}c_{31}
+\frac{1}{2}m^2 c_{32},
$
where $m\equiv M(s-1)$.
On the level of symmetric insertions this replacement corresponds to enlarging the basis of naively BRST-invariant insertions with $\rho=\delta=4$  
by $c_{31}m\kappa^{-1}\int\sqrt{-g}R$ and
    $c_{32}\frac{1}{2} m^2\int\sqrt{-g}R$, which are to be taken into account in 
$\Gamma_{\rm eff}$ .\\
Then the question is, whether one can via ZI's eliminate the $m$-terms and
maintain invariance. The sought invariant $[...]^4_4$ insertions are 
defined to all orders as symmetric insertions via the invariant derivatives
\begin{eqnarray}\label{smns}
\big[\kappa^{-2}\int\sqrt{-g}R\,\big]^{4}_{4}&=&
	                 \frac{\partial}{\partial c_{30}}\Gamma\\
\big[\kappa^{-1}\int\sqrt{-g}m R\,\big]^{4}_{4}&=&
	                 \frac{\partial}{\partial c_{31}}\Gamma\\
\big[\int\sqrt{-g}\frac{1}{2}m^2R\,\big]^4_4&=&
	                 \frac{\partial}{\partial c_{32}}\Gamma
\end{eqnarray}
and the symmetric counting operators $\mathcal{N}_{H,L}$.
The relevant ZI's have the form
\begin{eqnarray}\label{sZ}
\big[\kappa^{-2}\int\sqrt{-g}R\,\big]^{4}_{4}&=&
      \big[\kappa^{-2}\int\sqrt{-g}R\,\big]^{3}_{3}+[...]^4_4\label{sZ0}\\
	{\rm with}\quad [...]^4_4&=&[\int\,\sqrt{-g}(\kappa^{-2}u_{0}R
  +u_{31}m\kappa^{-1}R+u_{32}\frac{1}{2}m^2R\nonumber\\
	&& + u_1 R^{\mu\nu}R_{\mu\nu}+u_2R^2)+u_h\,\mathcal{N}_H+u_c\,\mathcal{N}_L]^4_4\\
\big[\kappa^{-1}\int\sqrt{-g}mR\,\big]^{4}_{4}&=&
     m\big[\kappa^{-1}\int\sqrt{-g}\kappa^{-1}R\,\big]^{3}_{3}+[...]^4_4
	                                                \label{sZ1}\\
	{\rm with}\quad [...]^4_4&=&[\int\,\sqrt{-g}(\kappa^{-2}v_{30}R
	+v_0m\kappa^{-1}R+v_{31}\frac{1}{2}m^2R\nonumber\\
		       && + v_1 R^{\mu\nu}R_{\mu\nu}+v_2R^2)
	                +v_h\,\mathcal{N}_H+v_c\,\mathcal{N}_L]^4_4
\end{eqnarray}
and
\begin{eqnarray}	                
	\big[\int\sqrt{-g}\frac{1}{2}m^2 R\,\big]^4_4&=&
     m\big[\int\sqrt{-g}\frac{1}{2}mR\,\big]^3_3+[...]^4_4
	                                                \label{sZ2}\\
	{\rm with}\quad [...]^4_4&=&[\int\,\sqrt{-g}(\kappa^{-2}w_{30}R
	+w_{31}m\kappa^{-1}R+w_0\frac{1}{2}m^2R\nonumber\\
		       && + w_1 R^{\mu\nu}R_{\mu\nu}+w_2R^2)
	                +w_h\,\mathcal{N}_H+w_c\,\mathcal{N}_L]^4_4 .
\end{eqnarray}
All coefficients $u,v,w$ are of order $\hbar$. The terms multiplied by 
$u_0,v_0,w_0$ resp.\ will be absorbed on the resp.\ lhs
and then the resp.\ line  divided by $1-u_0,1-v_0,1-w_0$, such that the
normal products on the rhs have the factors $(1-u_0)^{-1}, (1-v_0)^{-1}, (1-w_0)^{-1}$
in the resp.\ line. From this representation it is then obvious that all 
$[...]^3_3$ insertions on the rhs are symmetric, because all other insertions
are symmetric. Since the relevant determinant in this linear system of 
equations is clearly non-vanishing, one can solve for all hard insertions
$[\int\sqrt{-g}R(\kappa^{-2}, m\kappa^{-1},\frac{1}{2}m^2)]^4_4$ in terms of 
the soft one's together with $(c_1,c_2,\mathcal{N}_{H,L})$-terms. But those soft insertions
which contain the factor $m$ vanish at $s=1$, hence all hard $m$-dependent
insertions have been eliminated. And the hard insertion 
$[\kappa^{-2}\int\sqrt{-g}R]^4_4$ has been effectively replaced by its soft 
counterpart.
These considerations are crucial for deriving the parametric
differential equations in symmetric form and without $m$-dependence at $s=1$.

\subsection{Push\label{se:push}}
Next we consider the problem of removing Push by using appropriate ZI's. 
First we treat the contributions of Push to $\Gamma^{\rm class}_{\rm inv}$ (cf. (\ref{mcoffs})).
They occur in the second power of $h$ and have the form (see (\ref{gmps1})),(\ref{gmps2}))
\be\label{pshnv}
\Gamma_{(hh)}(m^2)=
\int\,h^{\mu\nu}(m^2\hat{\gamma}^{(2)}_{\rm TT}P^{(2)}_{\rm TT}
           +m^2\hat{\gamma}^{(0)}_{\rm TT}P^{(0)}_{\rm TT})_{\mu\nu\rho\sigma}
                                h^{\rho\sigma}.
\ee
In higher orders we have just the same terms, but now to be understood as
normal products $[...]^4_4$ in $\Gamma_{\rm eff}$. We use the ZI
\begin{multline}\label{pZi}
	[\int\,h^{\mu\nu}(m^2\hat{\gamma}^{(2)}_{\rm TT}P^{(2)}_{\rm TT}
           +m^2\hat{\gamma}^{(0)}_{\rm TT}P^{(0)}_{\rm TT})_{\mu\nu\rho\sigma}
				h^{\rho\sigma}]^4_4\cdot\Gamma_{(hh)}
	\phantom{M(s-1)}			\\
	=M(s-1)[\int\,h^{\mu\nu}(m\hat{\gamma}^{(2)}_{\rm TT}P^{(2)}_{\rm TT}
           +m\hat{\gamma}^{(0)}_{\rm TT}P^{(0)}_{\rm TT})_{\mu\nu\rho\sigma}
		h^{\rho\sigma}]^3_3\cdot\Gamma_{(hh)}\\
	      +[{\rm corr.s}]^4_4\cdot\Gamma_{(hh)}.
\end{multline}
Here the $\hat{\gamma}$'s are interpreted as differential operators and
$m\equiv M(s-1)$ is to be recalled.
The corrections comprise first of all the starting term from the lhs with
a coefficient $q=O(\hbar)$. We bring it to the lhs and divide by $1-q$.
This yields
\begin{multline}\label{pZi2}
	[\int\,h^{\mu\nu}(m^2\hat{\gamma}^{(2)}_{\rm TT}P^{(2)}_{\rm TT}
           +m^2\hat{\gamma}^{(0)}_{\rm TT}P^{(0)}_{\rm TT})_{\mu\nu\rho\sigma}
				h^{\rho\sigma}]^4_4\cdot\Gamma_{(hh)}
	\phantom{M(s-1)}\\
           = \frac{M(s-1)}{1-q}
 [\int\,h^{\mu\nu}(m\hat{\gamma}^{(2)}_{\rm TT}P^{(2)}_{\rm TT}
           +m\hat{\gamma}^{(0)}_{\rm TT}P^{(0)}_{\rm TT})_{\mu\nu\rho\sigma}
		h^{\rho\sigma}]^3_3\cdot\Gamma_{(hh)}\\
	      +\frac{1}{1-q}[{\rm corr.s}]^4_4\cdot\Gamma_{(hh)}.
\end{multline}
As correction terms appear the $hh$-vertex functions with all 
$[...]^4_4$-insertions. We now can demand $\brsts_0$-invariance because this
is a linear transformation. Among the $\hat{\gamma}^{(r)}_{\rm K,L}$-
contributions precisely those with $r=2,0; K=L=T$ are $\brsts_0$-invariant (see App.\ B),
hence they have been absorbed already. The other contributions go with the
symmetric differential operators $\mathcal{N}_{\rm H,L}$. These are however
BRST-variations and thus vanish between physical states. Therefore this
part of Push does at $s=1$ not contribute to physical quantities. 

The second (and last) appearance of Push is within gauge fixing and 
$\phi\pi$-terms. 
\begin{eqnarray}\label{pshgf}
(\Gamma_{\rm gf}+\Gamma_{\phi\pi})(m^2))&=&
	  -\frac{1}{2}\int \Big( \frac{1}{\kappa}h^{\mu\nu}(x)
    (\partial_\mu b_\nu+\partial_\nu b_\mu)(y)\frac{m^2}{(x-y)^2}\nonumber\\
&&\qquad+ D^{\mu\nu}_\rho c^\rho(x)(\partial_\mu\bar{c}_\nu
	 +\partial_\nu\bar{c}_\mu)(y)\frac{m^2}{(x-y)^2} \Big)\nonumber\\
	&=&-\frac{1}{2}\int\, \brsts_\Gamma \Big(h^{\mu\nu}(x)(\partial_\mu\bar{c}_\nu
			     +\partial_\nu\bar{c}_\mu)(y)\frac{m^2}{(x-y)^2} \Big).
\end{eqnarray}
The product in the last line is point split in $(x\leftrightarrow y)$. 
Divergences can be 
developed at coinciding points in such a way that they can be controlled by a ZI
\begin{eqnarray}\label{dcs}
[h^{\mu\nu}(x)(\partial_\mu\bar{c}_\nu
		       +\partial_\nu\bar{c}_\mu)(y)m^2]^4_4\cdot\Gamma&=&
m[h^{\mu\nu}(x)(\partial_\mu\bar{c}_\nu
               +\partial_\nu\bar{c}_\mu)(y)m]^3_3\cdot\Gamma\nonumber\\
	&&+\,[{\rm corr.s}]^4_4\cdot\Gamma
\end{eqnarray}
Among the corrections, again, appears the normal product of the lhs, which can 
be absorbed there, such that on the rhs only all other insertions
of dimension 4 and $\phi\pi$-charge $-1$ show up. These are 
$K_{\mu\nu}h^{\mu\nu},L_\rho c^\rho$ which are both naively defined because
they are linear in the quantized fields. At $s=1$ they are the only surviving
terms which contribute in (\ref{pshgf}) and then eventually vanish after
integration between physical states.

\section{The invariant parametric differential equations\label{se:invparadiffeq}}

\subsection{The Lowenstein-Zimmermann equation\label{se:LZeq}}
Green functions must be independent of the auxiliary mass $M$ at $s=1$, so
one has to know the action of $M\partial_M$ on them. Since the ST-identity
does not depend on $M$, $M\partial_M$ is a BRST-invariant differential 
operator and can be expanded in the basis provided by $(\ref{tns})$.
In fact with the ZI's (\ref{sZ1}) and (\ref{sZ2}) and the discussion
there we can consider the basis of symmetric differential operators to
be given by $c_{30} \partial_{c_{30}}, c_1 \partial_{c_1}, c_2\partial_{c_2}$ complemented
with the symmetric counting operators $\mathcal{N}_{H,L}$.
Furthermore we have shown that the contributions
coming from Push (\ref{pZi2}) and the contributions from Shift
go at most into the symmetric counting operators. Hence
\be\label{sMdM}
M\partial_M\Gamma=(-\beta^{\rm LZ}_{30} c_{30} \partial_{c_{30}}
-\beta^{\rm LZ}_{1} c_1 \partial_{c_{1}}
-\beta^{\rm LZ}_{2} c_2 \partial_{c_{2}}
   +\gamma^{\rm LZ}_h\mathcal{N}_H+\gamma^{\rm LZ}_c\mathcal{N}_L)\Gamma .
\ee
The coefficient functions $\beta,\gamma$ can be determined by
testing on the normalization conditions. 
The test on \eqref{sMdM} involving external fields
\be\label{Lcc1}
\frac{\partial}{\partial p^\lambda}\Gamma_{L_\rho c^\sigma c^\tau}
	       \,|_{\substack{ p^2=-\mu^2 \\ s=1} }	
      =-i\kappa(\delta^\rho_\sigma\eta_{\lambda\tau}
               -\delta^\rho_\tau\eta_{\lambda\sigma})
\ee
implies
\be\label{Lccp}
M\partial_M\,\partial_p \Gamma_{Lcc}\,|_{\substack{ p^2=-\mu^2 \\ s=1}}
-\gamma^{\rm LZ}_c(\partial_p \Gamma_{Lcc}\,|_{\substack{ p^2=-\mu^2 \\ s=1}})=0 .
\ee
Since the $M$-derivative in the first term is not in conflict with going
to the argument of $\Gamma$ the first term vanishes and hence
$\gamma^{\rm LZ}_c=0$.
Quite analogously we may proceed for
\be\label{Kc1}
\frac{\partial}{\partial p^\sigma}\Gamma_{K^{\mu\nu}c_\rho}
	                              \,|_{\substack{ p^2=-\mu^2 \\ s=1}}	
      =-i\kappa(\eta^{\mu\sigma}\delta_\rho^\nu+\eta^{\nu\sigma}\delta_\rho^\mu
                              -\eta^{\mu\nu}\delta_\rho^\sigma).
\ee
Here this test on (\ref{sMdM}) yields
\be\label{Kcp}
M\partial_M\,\partial_p \Gamma_{Kc}\,|_{\substack{ p^2=-\mu^2 \\ s=1}}
-\gamma^{\rm LZ}_h (-\partial_p \Gamma_{Kc}\,|_{\substack{ p^2=-\mu^2 \\ s=1}})
-\gamma^{\rm LZ}_c (-\partial_p \Gamma_{Kc}\,|_{\substack{ p^2=-\mu^2 \\ s=1}})=0 .
\ee
With the same argument as before, $\gamma^{\rm LZ}_c=0$ and 
$\gamma^{\rm LZ}_h=0$ follows.

For obtaining the $\beta$-functions we use the normalization conditions
(\ref{highnorm}) for $\gamma^{(2)}_{\rm TT}$ and $\gamma^{(0)}_{\rm TT}$. 
The test
\be\label{hTT3}
\frac{\partial}{\partial p^2}\gamma^{(2)}_{\rm TT}\,|_{\substack{ p^2=0 \\ s=1}}
=c_{30}\kappa^{-2}
\ee
implies
\be\label{hTT3p}
M\partial_M\frac{\partial}{\partial p^2}\gamma^{(2)}_{\rm TT}\,|_{\substack{ p^2=0 \\ s=1}}
+c_{30}\kappa^{-2}\beta^{\rm LZ}_{c_{30}}=0 .
\ee
Since the normalization does not involve $M$, the first term is zero, hence 
$\beta^{\rm LZ}_{c_{30}}=0$.
It is clear that the other $\beta$-functions vanish too.
Hence at $s=1$ the LZ-equation 
\be\label{lzv}
M\partial_M \Gamma|_{s=1} =0
\ee
holds and reveals that the vertex functions are independent of $M$ at $s=1$.

\subsection{The Renormalization Group equation\label{se:RGeq}}
The RG-equation formulates the response of the system to the variation of
the normalization parameter $\mu$, (see (\ref{highnorm})),
where e.g.\ couplings or field amplitudes are defined. Since the ST-operator does not depend on $\mu$ the partial differential operator
$\mu\partial_\mu$ is symmetric and can be expanded in the basis 
(\ref{tns}). Quite analogously to the LZ-equation (by removing Push and Shift) we end up with 
\be\label{rg1}
	\mu\partial_\mu\Gamma_{|s=1}=
	(-\beta^{\rm RG}_{30} c_{30} \partial_{c_{30}}		      
	-\beta^{\rm RG}_{c_1} c_1 \partial_{c_1}  
	-\beta^{\rm RG}_{c_2} c_2 \partial_{c_2}	
	+\gamma^{\rm RG}_h\,\mathcal{N}_H
	+\gamma^{\rm RG}_c\,\mathcal{N}_L)\Gamma_{|s=1} \, .
\ee
We observe that some normalization conditions involve $\mu$,
hence performing derivatives wrt $\mu$ does not commute with choosing
arguments for the relevant vertex functions and we expect non-trivial 
coefficient functions. 
Again we start with those tests which involve external fields, i.e.
\be\label{Lcc2}
\frac{\partial}{\partial p^\lambda}\Gamma_{L_\rho c^\sigma c^\tau}
	       \,|_{\substack{ p^2=-\mu^2 \\ s=1}}	
      =-i\kappa(\delta^\rho_\sigma\eta_{\lambda\tau}
               -\delta^\rho_\tau\eta_{\lambda\sigma}).
\ee
Now $\mu\partial_\mu$ does not commute with choosing a $\mu$-dependent
argument, hence
\be
\mu\partial_\mu\frac{\partial}{\partial p^\lambda}
     \Gamma_{L_\rho c^\sigma c^\tau}
	       \,|_{\substack{ p^2=-\mu^2 \\ s=1}} 
      +i\gamma^{\rm RG}_c\kappa(\delta^\rho_\sigma\eta_{\lambda\tau}
               -\delta^\rho_\tau\eta_{\lambda\sigma})=0
\ee
which determines $\gamma^{\rm RG}_c$.
For the normalization condition
\be\label{Kc2}
\frac{\partial}{\partial p^\sigma}\Gamma_{K^{\mu\nu}c_\rho}
	                              \,|_{\substack{ p^2=-\mu^2 \\ s=1}}	
      =-i\kappa(\eta^{\mu\sigma}\delta_\rho^\nu+\eta^{\nu\sigma}\delta_\rho^\mu
                              -\eta^{\mu\nu}\delta_\rho^\sigma)
\ee
the structure is exactly the same as in the preceding
example such that the result is
\be
\mu\partial_\mu\frac{\partial}{\partial p^\sigma}\Gamma_{K^{\mu\nu}c_\rho}
	   \,|_{\substack{ p^2=-\mu^2 \\ s=1}}
+(\gamma^{\rm RG}_c-\gamma^{RG}_h)i\kappa(\eta^{\mu\sigma}\delta_\rho^\nu
+\eta^{\nu\sigma}\delta_\rho^\mu-\eta^{\mu\nu}\delta_\rho^\sigma)=0.
\ee
This equation gives $\gamma^{\rm RG}_h$.
The $\beta$-functions will be determined by the normalization conditions
for the couplings.
The normalization condition
\be\label{c3n}
\partial_{p^2}\gamma^{(2)}_{\rm TT}\,|_{\substack{ p^2=0 \\ s=1}} 
                             =c_{30}\kappa^{-2}
\ee
is independent from $\mu$ hence it implies
\be
\mu\partial_\mu \partial_{p^2}\gamma^{(2)}_{\rm TT}\,|_{\substack{ p^2=0 \\ s=1}}
              =0=-\beta^{\rm RG}_{30}c_{30}\kappa^{-2} 
      +2c_{30}\kappa^{-2}\gamma^{\rm RG}_h.
      \ee
This determines $\beta^{\rm RG}_{c_{30}}$.
The other normalization conditions, however depend on $\mu$ and thus result into 
\begin{eqnarray}
\mu\partial_\mu 
\partial_{p^2}\partial_{p^2}\gamma^{(2)}_{\rm TT}\,|_{\substack{ p^2=-\mu^2 \\ s=1}}
	&=&2c_1\beta^{\rm RG}_1-2c_1\gamma^{\rm RG}_h\\
\mu\partial_\mu
\partial_{p^2}\partial_{p^2}\gamma^{(0)}_{\rm TT}\,|_{\substack{ p^2=-\mu^2 \\ s=1}}
	&=&-6c_2\beta^{\rm RG}_{c_2}+2c_1\beta^{\rm RG}_1 
             +2(3c_2-c_1)\gamma^{\rm RG}_h.
\end{eqnarray}
These equations determine $\beta^{\rm RG}_1,\beta^{\rm RG}_2$. These coefficient functions depend on the product $\mu\kappa$. Since we work in Landau gauge, they do not depend on a gauge parameter. 

\subsection{The Callan-Symanzik equation\label{se:CSeq}}
The CS-equation describes the response of the system to 
the variation of all parameters carrying the dimension of mass.
Here $M$, $\mu$, and $\kappa$. The variation
of $M$ has been covered by the LZ-equation with the result that Green
functions do not depend on it at $s=1$. The variation of $\mu$ has been
treated as well. As far as $\kappa$ is concerned
it is crucial to observe that the ST-identity depends on it, hence it
does not per se give rise to a symmetric differential operator. However
acting with $-\kappa\partial_\kappa$ on $\Gamma^{\rm class}$ we find 
\be\label{smsch}
	-\kappa\partial_\kappa \Gamma^{\rm class}
= (2c_3\partial_{c_3} +(N_b-2\alpha_0\partial_{\alpha_0})
	      -N_K-N_L)\Gamma^{\rm class}.
\ee
Hence the combination
\be\label{skpp}
-\kappa\partial_\kappa-2c_{30}\partial_{c_{30}}
                      -(N_b-2\alpha_0\partial_{\alpha_0})
                      +N_K+N_L
\ee
is independent of $\kappa$ on $\Gamma^{\rm class}$, the variation of
$\kappa$ is just balanced by the other derivatives, this combination forms a 
differential operator which commutes with the ST-identity, and thus is
symmetric.

In higher orders we can therefore expand this operator in the basis 
$(\ref{tns})$ and obtain
\begin{multline}\label{cs1}
	(-\kappa\partial_\kappa-2c_{30}\partial_{c_{30}}
                      -(N_b-2\alpha_0\partial_{\alpha_0})
		      +N_K+N_L)\Gamma_{|s=1}=\\
	(-\beta_{30} c_{30} \partial_{c_{30}}		      
	-\beta_{c_1} c_1 \partial_{c_1}  
	-\beta_{c_2} c_2 \partial_{c_2}	
	+\gamma_h\,\mathcal{N}_H
	+\gamma_c\,\mathcal{N}_L)\Gamma_{|s=1},
\end{multline}
where the contributions going with the variation of $c_{31},c_{32}$ 
have been eliminated with the ZI's (\ref{sZ1}) and (\ref{sZ2}). 
Like for the LZ equation \eqref{sMdM} the coefficient functions vanish, since the normalization conditions and the differential operator are not in conflict with each other, i.e.\
\begin{align}\label{cs2}
	(-\kappa\partial_\kappa-2c_{30}\partial_{c_{30}}
                      -(N_b-2\alpha_0\partial_{\alpha_0})
		      +N_K+N_L)\Gamma_{|s=1}= 0 
\end{align}
We eliminate in the RG-equation (\ref{rg1}) the hard insertion 
$c_{30}\partial_{c_{30}}$ and add the result to (\ref{cs2}) obtaining the
CS-equation in its conventional form
\begin{multline}\label{cs3}
(\mu\partial_\mu-\kappa\partial_\kappa-2c_{30}\partial_{c_{30}}
	-(N_b-2\alpha_0\partial_{\alpha_0})+N_K+N_L
	+\beta^{\rm CS}_1 c_1 \partial_{c_1}+\\		      
	+\beta^{\rm CS}_2 c_2 \partial_{c_2}		      
	-\gamma^{\rm CS}_h\mathcal{N}_H
        -\gamma^{\rm CS}_c\mathcal{N}_L)\Gamma_{|s=1}=
\alpha^{\rm CS}[\kappa^{-2}\int\sqrt{-g}R\,]^3_3\cdot\,\Gamma_{|s=1}.
\end{multline}
The coefficient functions are of order $O(\hbar)$. Their values have to be 
determined by testing on the normalization conditions and taking care of 
the soft contribution.
The differential operator can be interpreted as a symmetrized version of the dilatations and the equation then says that in the deep Euclidean region the soft breaking on the rhs becomes negligible and the hard breaking is parametrized by the functions $\beta$ and $\gamma$.
Between physical states only the $\beta$'s would be relevant. 

Before testing on \eqref{cs3}, we have to note that all coefficient functions start with
order $O(\hbar)$. This is clear for $\beta$'s and $\gamma$'s because they
were introduced via the action principle after having applied the
symmetric differential operator to $\Gamma$. But contrary to more conventional models this is here also true for $\alpha^{\rm CS}$, because it was traded against the hard insertion $[\int\sqrt{-g}R]^4_4$.
This has to do with the special character of the symmetric differential
operator and the $\kappa$-dependence of $\Gamma$: The EH action depends on $\kappa$ which carries dimension, but acts as a mass
term only relative to the higher derivative terms. 

We test on
\be\label{Lcc3}
\frac{\partial}{\partial p^\lambda}\Gamma_{L_\rho c^\sigma c^\tau}
	       \,|_{\substack{ p^2=-\mu^2 \\ s=1}}	
      =-i\kappa(\delta^\rho_\sigma\eta_{\lambda\tau}
               -\delta^\rho_\tau\eta_{\lambda\sigma}).
\ee
In order to understand the impact of the symmetric differential operator we start with the tree approximation and find
\be\label{cstr}
(-\kappa\partial_\kappa +1)
\frac{\partial}{\partial p^\lambda}\Gamma^{(0)}_{L_\rho c^\sigma c^\tau}=0 ,
\ee
which is correct, since $\mu\partial_\mu-2c_{30}\partial_{c_{30}}$
does not contribute and from counting operators only $N_L$ does.
In higher orders $\mu\partial_\mu$ no longer commutes with going to the
desired value for $p$, whereas $\kappa\partial_ \kappa-2c_{30}\partial_{c_{30}}+N_L$ does, hence
\begin{eqnarray}\label{Lcch}
\mu\partial_\mu
\frac{\partial}{\partial p^\lambda}\Gamma_{L_\rho c^\sigma c^\tau}
	       \,|_{\substack{ p^2=-\mu^2 \\ s=1}}	
	&-&\gamma^{\rm CS}_c(-i)\kappa(\delta^\rho_\sigma\eta_{\lambda\tau}
	       -\delta^\rho_\tau\eta_{\lambda\sigma})\nonumber\\
	&=&\alpha^{\rm CS}[\kappa^{-2}\int\sqrt{-g}R\,]^3_3\cdot\,
\partial_{p^\lambda}{\Gamma_{L_\rho c^\sigma c^\tau}}_{|{\substack{ p^2=-\mu^2 \\ s=1}}}.
\end{eqnarray}
Herewith $\gamma^{\rm CS}_c$ is determined. (The $\alpha$-term contributes not earlier than in two loops, since we are concerned with
1PI diagrams.) 

We test on
\be\label{Kc3}
\frac{\partial}{\partial p^\sigma}\Gamma_{K^{\mu\nu}c_\rho}
	                              \,|_{\substack{ p^2=-\mu^2 \\ s=1}}	
      =-i\kappa(\eta^{\mu\sigma}\delta_\rho^\nu+\eta^{\nu\sigma}\delta_\rho^\mu
                              -\eta^{\mu\nu}\delta_\rho^\sigma)
\ee
and, again, because also the term $-\kappa\partial_\kappa-2c_{30}\partial_{c_{30}}+N_K$ commutes with going to a specific value of $p$, we find in higher orders
\begin{eqnarray}\label{cscffK}
\mu\partial_\mu
\frac{\partial}{\partial p^\sigma}\Gamma_{K^{\mu\nu}c_\rho}
	                              \,|_{\substack{ p^2=-\mu^2 \\ s=1}}	
&-&(\gamma^{\rm CS}_h-\gamma^{\rm CS}_c)i\kappa(\eta^{\mu\sigma}\delta_\rho^\nu
+\eta^{\nu\sigma}\delta_\rho^\mu-\eta^{\mu\nu}\delta_\rho^\sigma)\nonumber\\
	&=&\alpha^{\rm CS}[\kappa^{-2}\int\sqrt{-g}R\,]^3_3\cdot\,
\frac{\partial}{\partial p^\sigma}{\Gamma_{K^{\mu\nu}c^\rho}}_{|s=1} .
\end{eqnarray}
This yields eventually $\gamma^{\rm CS}_h$.
With the same argument $\alpha^{\rm CS}$ and the $\beta^{\rm CS}_{1,2}$ are given by
\begin{equation}
\mu\partial_\mu
     \partial_{p^2}\gamma^{(2)}_{\rm TT}\,|_{\substack{ p^2=-\mu^2 \\ s=1}}
	-2c_{30}\kappa^{-2}\gamma^{\rm CS}_h
	=\alpha^{\rm CS}[\kappa^{-2}\int\sqrt{-g}R\,]^3_3
		       \cdot\,\mathbb{P}^{(2)}_{30}\Gamma_{|s=1}\label{phccs}
\end{equation}
\begin{multline}		       
\mu\partial_\mu
\partial_{p^2}\partial_{p^2}\gamma^{(2)}_{\rm TT}\,|_{\substack{ p^2=-\mu^2 \\ s=1}}
	-2c_1\beta^{\rm CS}_1-2c_1\gamma^{\rm CS}_h
	=\alpha^{\rm CS}[\kappa^{-2}\int\sqrt{-g}R\,]^3_3
		       \cdot\,\mathbb{P}^{(2)}_1\Gamma_{|s=1}\label{bc1}
\end{multline}
\begin{multline}		       
\mu\partial_\mu
\partial_{p^2}\partial_{p^2}\gamma^{(0)}_{\rm TT}\,|_{\substack{ p^2=-\mu^2 \\ s=1}}
	+6c_2\beta^{\rm CS}_2-2c_1\beta^{\rm CS}_1 
             +2(3c_2-c_1)\gamma^{\rm CS}_h\\
	=\alpha^{\rm CS}[\kappa^{-2}\int\sqrt{-g}R\,]^3_3
			\cdot\,\mathbb{P}^{(0)}_2\Gamma_{|s=1}.\label{bc2}
\end{multline}
(\ref{phccs}) determines $\alpha^{\rm CS}$ and (\ref{bc1}), (\ref{bc2}) determine
$\beta^{\rm CS}_{1,2}$, resp.
The symbols $\mathbb{P}$ stand for projectors of $\Gamma_{hh}$ into the components
\be
\mathbb{P}_{30}\to\partial_{p^2} \gamma^{(2)}_{\rm TT}(p^2)\qquad
\mathbb{P}_1\to\partial_{p^2}\partial_{p^2} \gamma^{(2)}_{\rm TT}(p^2)\qquad
\mathbb{P}_2\to\partial_{p^2}\partial_{p^2} \gamma^{(0)}_{\rm TT}(p^2).
\ee
These are part of the full vertex functions of higher orders. Clearly
those admit also the expansion in the projector basis as in the
classical approximation.
Also the coefficient functions of the CS-equation depend only on 
$\mu\kappa$, besides the parameters $c_1,c_2,c_{30}$. 

\section{Traces of the Einstein-Hilbert Theory\label{se:removingregulators}}

It has already been observed by \cite{Stelle} that the introduction of $R^{\mu\nu} R_{\mu\nu}$ and $R^2$ in the classical action leads to a regularization of the $h$-field propagator analogously to the Pauli-Villars regularization (cf.\ \cite{Bogolyubov:1980nc}). This regularization is not sufficient to render
the model finite, but it becomes power counting renormalizable. This implies
that all standard tools of, say, BPHZL renormalization become available. 
Furthermore, the BPHZL renormalization scheme may be formulated with such regularization, but has been shown to be independent of it \cite{Zimmermann:1975gk,Clark:1976ym} provided the regulator-free model is finite.
Unsurprisingly, it can be shown that in our construction the limit $c_1, c_2 \rightarrow 0$ exists up to one-loop diagrams so that the result of \cite{tHooft:1974toh} can be recovered.
For diagrams of higher loop order, new divergencies occur which are not treated by the subtractions in the BPHZL scheme.
Those additional divergencies can be verified by setting the UV-degrees in \eqref{dgpr} and \eqref{dgpr2} equal to $-2$ and subsequently following the argument in Sect. \ref{se:powercounting} with these new degrees.
This just means for our work that beyond one loop we have to take non-vanishing parameters $c_1$ and $c_2$ and have to examine in which sense we find the EH theory in our model.

\subsection{Projection to Einstein-Hilbert\label{se:projEH}}

We still have to check in some detail how the $S$-matrix (\ref{sma}) is affected by this limit. The factor 
$K(x-y)$ is the wave operator of the free theory, hence given by 
$\Gamma^{(0)}_{\Phi_i\Phi_j}$ (recall that the fields $\Phi$ are the free 
$\Phi_{\rm in}$ fields). At $c_1=c_2=0$ the $hh$-submatrix has only $p^2$-contributions, no $(p^2)^2$, hence projects to the pole at $p^2=0$ (for $s=1$), as desired.
The matrix $z^{-1}$, commonly the wavefunction renormalization matrix, is here in fact the matrix $r$ of the residues of the poles, since the $h$-wave function has been fixed in $(\ref{highnormG1})$ (and the others by the $b$-equation of 
motion).
Contributions of the possible second singularity of the propagator is projected to zero because no respective factor in the numerator, coming from $\Phi_{\mathrm{in}}$, is available.
Hence for physical quantities they are always projected to 0, as we have seen for the $S$-matrix.\\
Before the fields $\Phi_{\rm in}$ project to the mass shells one can introduce a 
$\Phi_{\underbar{$\scriptstyle \mathrm{in}$}}=z\Phi_{\rm in}$ with the implication 
$z^{-1T}K(x-y)z^{-1}=\Gamma_{\Phi(x)\Phi(y)}$ -- here the {\bf full} $\Gamma_{\Phi\Phi}$. Then one can use the results of ST etc.\ and derive in analogy to the tree approximation that the commutator of $:\Sigma:$ with ST generates again $Q^{\rm BRST}$ as needed.\\
A comment is in order. 
The reason for going via $c_1,c_2$ from the very beginning can be understood just as a means to avoid ``unnecessary'' even higher derivative counterterms (conf.\ \cite{Goroff:1985th}). 
This can be seen as follows:
Starting with $c_3$-terms alone, one realizes in one-loop that higher derivative counterterms are required. 
Absorbing these and transitioning into a new propagator, the power counting becomes the same as in the $(c_1,c_2,c_3)-$model.
This round-about procedure has been circumvented by starting immediately with all terms guaranteeing power counting renormalizability.
In this context, it is quite natural to consider even higher orders of derivatives of the metric in the classical action, which would render the model super-renormalizable (conf.\ \cite{Asorey:1996hz}). 
However, these higher orders do not have a regularizing effect at the order $\hbar$ so that the occurring divergencies have to be treated separately.
Thus the analytic structure of such models is obscured to a certain extent.

\subsection{Parametric differential equations of the $S$-matrix\label{se:paradiffeqS}}
It is of quite some interest to investigate how the $S$-matrix behaves under
RG transformation and under scaling, i.e.\ under action of the CS-operator.\\
First we need the expressions of the symmetric differential operators 
$\mathcal{N}_{\rm H,L}$ (cf. (\ref{fdHa}) and (\ref{fdL})) when they act on $Z$:
\be\label{fdaHz}
\mathcal{N}_{\rm H}Z\equiv i\int\Big(-J_h\frac{\delta}{\delta J_h}-K\frac{\delta}
                {\delta K}+j_{\bar{c}}\frac{\delta }{\delta j_{\bar{c}}}\Big)Z
                \qquad
\mathcal{N}_{\rm L}Z\equiv i \int \Big(-j_c\frac{\delta}{\delta j_c}
                  -L\frac{\delta}{\delta L}\Big)Z    .            
\ee 
Next we introduce 
\be\label{ffS}
\hat{S}(\uJ)\equiv :\Sigma:Z(\uJ),
\ee
a kind of off-shell $S$-matrix.
In order to see how the $S$-matrix transforms under the RG we look at 
\begin{eqnarray}\label{rgsma}
\mu\partial_\mu\hat{S}(\uJ)&=&:(\mu\partial_\mu Y)e^Y:Z(\uJ)
                                    +:\Sigma:\mu\partial_\mu Z(\uJ)\\
Y&\equiv&\int dxdy\Phi_{\rm in}(x)K(x-y)z^{-1}\frac{\delta}{\delta \uJ} \,,
 \qquad      K(x-y)=\Gamma^{(0)}_{\Phi\Phi}\nonumber\\
 \mu\partial_\mu Y&=&\int dxdy\, \Phi_{\rm in}(x)K(x-y)(\mu\partial_\mu z^{-1})
  \frac{\delta}{\delta \uJ} = 0   \nonumber    
\end{eqnarray}
with $z^{-1}$ being the residue matrix of the poles at $p^2=0$. 
In the $hh$-sector these residues are independent of $\mu$: for the spin two part directly 
as guaranteed by the subtraction scheme (\ref{highnorm}); in the spin zero part then indirectly via ST. 
In the $bh$-mixed sector they are 
$\mu$-independent because they are directly determined by the gauge fixing which is independent of it.\\
In the second term of $(\ref{rgsma})$ the operators $\mathcal{N}$ do not contribute,
because they are BRST-variations and therefore mapped to zero by $:\Sigma:$.
The final outcome is 
\be\label{rgsma2}
\mu\partial_\mu S=(-\beta^{\rm RG}_{30} c_{30} \partial_{c_{30}} - \beta^{\rm RG}_1 c_1 \partial_{c_1} - \beta^{\rm RG}_2 c_2 \partial_{c_2})  S.
\ee
For $S$-matrix elements which exist, regarding the infrared, this relation applies. It is remarkable that (although here it is formal in many cases) this is the analogue to the result which Zimmermann has derived axiomatically for massless $\phi^4$-theory \cite{Zimmermann:1979fd}.

With completely analogous arguments one can derive the CS equation for the $S$-operator, i.e.
\begin{align}
(\mu\partial_\mu-\kappa\partial_\kappa-2c_{30}\partial_{c_{30}} + \beta^{\rm CS}_1 c_1 \partial_{c_1} + \beta^{\rm CS}_2 c_2 \partial_{c_2})S & =
\alpha^{\rm CS}[\kappa^-2\int\sqrt{-g}R]^3_3\cdot S \nonumber \\
& = \alpha^{\rm CS}([\kappa^{-2}\int\sqrt{-g}R]^3_3)^{\rm Op} \,. \label{cssma}
\end{align}
The qualification is as before: the equation is meaningful only for matrix elements which exist regarding the infrared. 
It shows however in those cases how scaling is realized.

\section{General solution of the Slavnov-Taylor identity\label{se:generalsolutionSTI}}
As mentioned at the end of Section \ref{se:propagators} the propagators for the field 
$h^{\mu\nu}$ require to consider it as a field with canonical dimension zero.
It is thus impossible to distinguish via power counting between $h$ and an
arbitrary function $h'(h)$. This is familiar from supersymmetric gauge theories
where in linear realization of supersymmetry the real gauge superfield 
$\phi(x,\theta,\bar{\theta})$, known as ``vector superfield'', also has 
vanishing canonical dimension \cite{Piguet:1984mv}. One 
can take over from there mutatis mutandis the treatment of such fields. 
In the present context this means in particular
that for finding the general solution of the Slavnov-Taylor identity one
just chooses a special one, here $h^{\mu\nu}\equiv h_s^{\mu\nu}$, 
with its transformation law (\ref{brst}) $\brsts h_s^{\mu\nu}\equiv Q_s(h_s)$
and replaces it by a general invertible function $\mathcal{F}(h)$
\be\label{gsst}
\mathcal{F}^{\mu\nu}(h)=z_1 h^{\mu\nu} 
     +\sum_{n,k}z_{nk}F_{n,k}^{\mu\nu}(\underbrace{h...h}_{n}).
\ee
Here $n=2,3,...;k=1,2,...k_{\rm max}(n)$ and $F_{n,k}^{\mu\nu}$ denotes the most 
general contravariant two-tensor
in flat Minkowski space which one can form out of $n$ factors of $h$ and which 
does not contain terms with $\eta^{\mu\nu}$ as factor. The reason for this 
restriction will be explained at the end of this section.\\ 
The coefficients have been denoted
$z_{nk}$ because the redefinition
$h\rightarrow \mathcal{F}$ is just a generalized wave function 
renormalization, the standard one being given by $\mathcal{F}(h)=z_1 h$ leading
to $\hat{H}=z^{-1}_1 H$ in the ST-identity.\\
A remark is in order. That the non-linear redefinition $F^{\mu\nu}_{n,k}(h)$ is not a
formal exercise, but indeed necessary in the course of renormalization,
has been shown explicitly, e.g.\ \cite[formula (1.7)]{vandeVen:1991gw}. It is also to be noted that at every order $n$ in the number of 
field $h$ there are only finitely many free parameters $z_{n,k}$ to be prescribed 
by normalization conditions (s.b.).
\subsection{Tree approximation\label{se:generalsolutiontree}}
On the level of the functional $\Gamma^{\rm class}\equiv \Gamma^s$ this change 
manifests itself in the form
\be\label{gclb}
	\bar{\Gamma}(h,c,H,L)=\bar{\Gamma}^s(\hat{h},\hat{c},\hat{H},\hat{L}),
	\ee
where $\bar{\Gamma}^s(\hat{h},\hat{c},\hat{H},\hat{L})$ is the special 
solution of 
(\ref{brGm}) with $h,c,H,L$ replaced by
\begin{eqnarray}\label{gtrf}
	\hat{h}^{\mu\nu}=\mathcal{F}^{\mu\nu}(h^{\mu\nu}),
	&\hat{H}_{\mu\nu}=\frac{\delta}{\delta \hat{h}^{\mu\nu}}
\int H^{\mu\nu}\mathcal{F}^{-1}_{\mu\nu}(\hat{h})_{|\hat{h}=\mathcal{F}(h)}\\
	\hat{c}^\rho=z_cc^\rho&\hat{L}^\rho= \frac{1}{z_ c}L_\rho .
\end{eqnarray}
Again inspired by the case of supersymmetry \cite[Sect.~5.4, p.~68 ff]{Piguet:1986ug}
we shall now show that the parameters $z_{nk}, n\ge2$, are of  gauge type,
hence unphysical. At the same time this represents a second way to find the
general solution of the ST-identity.
We start from an arbitrary invertible function $M$ and its BRST variation $N$
\be\label{gfct} 
M^{\mu\nu}(h)= a_1 h^{\mu\nu}+\sum_{n,k}a_{n,k}(\underbrace{h\cdots h}_{n})^{\mu\nu}\,,
\qquad \brsts M=N,
\ee
where $n=2,3,...$ and $k=1,...,k_{\rm max}(n)$ being the number of two-tensors 
which can be formed out of $n$ factors $h$ without $\eta_{\mu\nu}$.
($k_{\rm max}(n)$ is finite for every $n$.)
Both are composite hence we couple them to external fields 
$\mathcal{M}$ and $\mathcal{N}$. $M$ will serve as defining a new, non-linear gauge
\be\label{nlg}
\Gamma_{\mathrm{gf}}=\frac{1}{2\kappa}\int
     (\partial_\mu M^{\mu\nu}b_\nu+\partial_\nu M^{\mu\nu}b_\mu)
                 -\frac{1}{2}\int\eta^{\mu\nu}b_\mu b_\nu,
\ee
giving rise to the gauge condition
\begin{eqnarray}\label{ngcd}
\frac{\delta\Gamma_{\mathrm{gf}}}{\delta b_\mu}
	&=&\frac{1}{\kappa}\partial_\lambda M^{\lambda\nu}-b^\mu\\
\frac{\delta\Gamma}{\delta b_\mu}
        &=&\frac{1}{\kappa}\partial_\lambda
             \frac{\delta \Gamma}{\delta\mathcal{M}_{\lambda\mu}}-b^\mu.
\end{eqnarray}
To this gauge fixing the $\phi\pi$-term
\be\label{nlFP}
\Gamma_{\phi\pi}=-\frac{1}{2}\int N^{\mu\nu}(\partial_\mu\bar{c}_\nu
					    +\partial_\nu\bar{c}_\mu)
\ee 
and the ST
\be\label{nlgST}
\ST(\Gamma)\equiv \int \Big( \frac{\delta \Gamma}{\delta K}\frac{\delta\Gamma}{\delta h}
+b\frac{\delta \Gamma}{\delta\bar{c}}
-\mathcal{M}\frac{\delta \Gamma}{\delta\mathcal{N}}
+\frac{\delta \Gamma}{\delta L}\frac{\delta \Gamma}{\delta c} \Big)=0
\ee
are suitable.
Gauge condition (\ref{ngcd}) and ST-identity (\ref{nlgST}) lead to the ghost
equation of motion
\be\label{nlgh}
\frac{\delta\Gamma}{\delta\bar{c}_\mu}
        - \kappa^{-1} \partial_\lambda\frac{\delta \Gamma}{\delta\mathcal{N}_{\lambda\mu}}=0,
\ee
which has the general solution
\begin{eqnarray}\label{gnlsn}
	\Gamma&=&\int(-\frac{1}{2}\eta^{\mu\nu}b_\mu b_\nu)
	          +\bar{\Gamma}(h,c,K,L,\mathcal{M}',\mathcal{N}')\\
&&\mathcal{M}'=\mathcal{M}
	    -\frac{1}{2\kappa}(\partial_\mu b_\nu+\partial_\nu b_\mu)\\
&&\mathcal{N}'=\mathcal{N}
	   -\frac{1}{2}(\partial_\mu\bar{c}_\nu+\partial_\nu\bar{c}_\mu)\\
	\bar{\Gamma}&=&\Lambda(h)+\int(KO(h,c)+\mathcal{M}'M(h)+\mathcal{N}'N(h,c)
	-L_\mu(c^\lambda\partial_\lambda c^\mu)).
 \end{eqnarray}
We now demand BRST invariance, i.e.\ (\ref{nlgST}), providing  
the linearized transformation law
\be\label{trfl}
\mathcal{B}_{\bar{\Gamma}}h^{\mu\nu}= O^{\mu\nu} \qquad
     \mathcal{B}_{\bar{\Gamma}}c^\mu=-\kappa c^\lambda\partial_\lambda c^\mu \qquad
\mathcal{B}_{\bar{\Gamma}}\bar{c}_\mu= b_\mu ,
\ee
calculate the effect on (\ref{gnlsn}) and find the conditions
\begin{eqnarray}
	\mathcal{B}_{\bar{\Gamma}}O&=&0 \label{tr1}\\
	\mathcal{B}_{\bar{\Gamma}}M&=&N \label{tr2}\\
	\mathcal{B}_{\bar{\Gamma}}N&=&0 \label{tr3}\\
	\mathcal{B}_{\bar{\Gamma}}\Lambda&=&0 . \label{tr4}
\end{eqnarray}
The solution of (\ref{tr1}) we know from the first part of this section to be 
\be\label{gnlsh}
O=Q^{\mathcal{F}(h,c)}=
\int\frac{\delta\mathcal{F}^{-1}(\hat{h})}{\delta h} 
					 Q_s(\hat{h},c)|_{\hat{h}=\mathcal{F}(h)}, 
\ee
$\mathcal{F}$ being given from (\ref{gsst}) and $z_1=1$.
Since $\mathcal{B}_{\bar{\Gamma}}$ is nilpotent on functionals $\mathcal{T}(h,c)$
\be\label{nlpt} 
\mathcal{B}_{\bar{\Gamma}}^2 \mathcal{T}=0,
\ee
(\ref{tr3}) follows from (\ref{tr2}) with
\begin{eqnarray}\label{gtrsfl}
N=\bar{\mathcal{B}}_{\bar{\Gamma}}M
	&=&\int dx O(x)\frac{\delta M}{\delta\hat{h}(x)}
        =\int dxdy\frac{\delta\mathcal{F}^{-1}(\hat{h}(x))}{\delta\hat{h}(y)}
			   Q_s(\hat{h},c)(y)\frac{\delta M(h)}{\delta h(x)}\\
	&=&\int dyQ_s(\hat{h},c)(y)\frac{\delta}{\delta\hat{h}(y)}
	    M(\mathcal{F}^{-1}(\hat{h}))|_{\hat{h}=\mathcal{F}(h)}.
\end{eqnarray}
(\ref{tr4}) is solved by
\be\label{gnvt}
\Lambda=\Gamma^{\rm class}_{\rm inv}(\mathcal{F}(h)),
\ee
with $\Gamma^{\rm class}_{\rm inv}$ being given by (\ref{ivc}).
Therefore the general solution of the ST-identity (\ref{nlgST}) is given by
\begin{multline}\label{gsSTc}
\Gamma(h,c,K,L,\mathcal{M}',\mathcal{N}')
	  =	\big[ 
	  \Gamma^{\rm class}_{\rm inv}(\hat{h})
      +\int dxdy K(x)\frac{\delta\mathcal{F}^{-1}(\hat{h}(x))}{\delta\hat{h}(y)}Q_s(\hat{h},c)(y)\\
	  +\int dxdy\mathcal{N}(x)Q_s(\hat{h},c)(y)\frac{\delta}{\delta\hat{h}(y)}M(\mathcal{F}^{-1}(\hat{h}))(x)
	\big]|_{\hat{h}=\mathcal{F}(h)} \\
	+\int (-\kappa L_\mu c^\lambda\partial_\lambda c^\mu +\mathcal{M}'M
	     +\mathcal{N}'N+\eta^{\mu\nu}b_\mu b_\nu).
\end{multline}
In order to compare this general solution with the previous one we define
a new gauge function by
\be
\hat{M}=M(\mathcal{F}^{-1}(\hat{h}))
\ee
with associated 
\begin{multline}\label{qgsST}
\Gamma(h,c,K,L,\mathcal{M}',\mathcal{N}') 
	  = \big[	
	  \Gamma^{\rm class}_{\rm inv}(\hat{h})
		+\int dx\hat{K}(x)Q_s(\hat{h},c)(x) \\
		+\int dxdy\mathcal{N}'(x)Q_s(\hat{h},c)(y)\frac{\delta}{\delta\hat{h}(y)}\hat{M}(\hat{h})(x)\\
		+\int(\mathcal{M}'\hat{M}'(\hat{h})+\frac{1}{4\kappa^2}(\partial_\mu\partial_\nu \hat{h}^{\mu\nu})^2)
	\big]|_{\hat{h}=\mathcal{F}(h)}
	-\kappa\int(L_\mu c^\lambda\partial_\lambda c^\mu) ,
\end{multline}	    
	    where
\be
	\hat{K}=\int dx K(x)\frac{\delta\mathcal{F}^{-1}(\hat{h}(y))}{\delta\hat{h}(x)}|_{\hat{h}=\mathcal{F}(h)}.
\ee
This shows that the solution (\ref{gsSTc}) corresponding to a function 
$\mathcal{F}(h)$ and a gauge function $M(h)$ is modulo the canonical 
transformation 
$h\rightarrow \hat{h}=\mathcal{F}(h)$ and $K\rightarrow \hat{K}(K,h)$ 
equivalent to the solution corresponding to $\mathcal{F}(h)=h$ and gauge 
function $\hat{M}=M(\mathcal{F}^{-1}(h))$. 

At this stage we are able to explain the restrictions on 
$\mathcal{F}(h)$ mentioned
at the beginning of this section. We want the transition 
$h\rightarrow \mathcal{F}(h)$ to be a canonical transformation. But then the
one-particle states associated with the two fields must be the same (up to a
numerical factor). Then $\mathcal{F}$ must start with $z_1 h^{\mu\nu}$ and must
not contain $\eta^{\mu\nu}h^\lambda_{\phantom{\lambda}\lambda}$. \\
In \cite{EKKSI,EKKSII} the conformal transformation properties of the energy-momentum
tensor (EMT) in massless $\phi^4$-theory have been studied. In that context
redefinitions of $h^{\mu\nu}$ \cite{EKKSIII} as here had to be understood because they governed
the renormalization of the EMT. There admitting an $\eta^{\mu\nu}$ would have
mixed renormalization of the EMT as a whole with that of its trace and was 
therefore forbidden altogether. Hence here, too, one does not admit it at any
power of $h$.\\
It is worth mentioning that in the same reference the BRST transformations of
$h^{\mu\nu}$ and their algebra had been derived in form of local Ward identities
for  translations in spacetime. Their explicit solution, i.e.\ representation on 
$h^{\mu\nu}$, turned out to be unstable, namely just admitting the transition
$h^{\mu\nu}\rightarrow \mathcal{F}^{\mu\nu}(h)$. So, that represents a welcome, independent
and explicit proof of the considerations here on the general solution of 
the ST-identity.\\
As a further interesting byproduct of this redefinition question we would like 
to mention that the transition from $h^{\mu\nu}=g^{\mu\nu}-\eta^{\mu\nu}$ to 
the Goldberg variable $\tilde{h}^{\mu\nu}=\sqrt{-g}g^{\mu\nu}-\eta^{\mu\nu}$ 
implies changing one-particle states. This can be seen as follows
\begin{eqnarray}\label{Gldbrgv}
	\sqrt{-g}g^{\mu\nu}&=&\eta^{\mu\nu}+\tilde{h}^{\mu\nu}\\
	         g^{\mu\nu}&=&\eta^{\mu\nu}+h^{\mu\nu}\\
	\tilde{h}^{\mu\nu}-h^{\mu\nu}&=&(\sqrt{-g}-1)(\eta^{\mu\nu}+h^{\mu\nu})\\
	\tilde{h}^{\mu\nu}&=&h^{\mu\nu}
	          -\frac{1}{2}\eta^{\mu\nu}h^\lambda_{\phantom{\lambda}\lambda}
		  +\eta^{\mu\nu} \Big(\frac{1}{8}(h^\alpha_{\phantom{\alpha}\alpha})^2
		  +\frac{1}{4}h^{\alpha\beta}h_{\alpha\beta}\Big)
		  -\frac{1}{2}h^\alpha_{\phantom{\alpha}\alpha}h^{\mu\nu}+O(h^3).
\end{eqnarray}
The $h$-linear term proportional to $\eta^{\mu\nu}$ generates new one-particle
poles relative to the original $h^{\mu\nu}$, as can be seen by comparing the
$\langle hh \rangle$-propagators in our approach with those of \cite{Stelle} and \cite{KuOj}. They belong to 
the spin 0 part of the full field $h^{\mu\nu}$ and will eventually be 
eliminated from the physical spectrum, but they have to be taken care of. Hand 
in hand with this goes a change of the BRST transformation from 
$\brsts h^{\mu\nu}\rightarrow \brsts \tilde{h}^{\mu\nu}$.

\subsection{Gauge parameter independence for the general case\label{se:gpigeneral}}
In the previous subsection we have seen that the field $h^{\mu\nu}$ can be
replaced by a general, invertible function $\mathcal{F}$ of itself, (\ref{gsst}),
and that the parameters $z_{nk}, n=2,3...; k=1,2,...,k_{\rm max}(n)$ are gauge type
parameters. Like for $\alpha_0$ we would like to show that the dependence
of the Green functions from these parameters can be controlled by a suitable
change of the ST-identity (see (\ref{chidouble}) and (\ref{chiZ})).
Hence we introduce anti-commuting parameters $\chi_{nk}$ which form together with
$z_{nk}$ doublets $(z_{nk},\chi_{nk})$ under BRST transformations
\be\label{gd}
\brsts z_{nk}=\chi_{nk} \quad n=2,3,...;k=1,2,...,k_{\rm max}(n) \qquad s\chi_{nk}=0.
\ee
They contribute to the ST-identity
\be\label{gpST}
\ST(\Gamma)+\chi_{nk}\partial_{z_{nk}}\Gamma=0 \,,\qquad 
                             \hat{\ST}Z\equiv \ST Z+\chi_{nk}\partial_{z_{nk}}Z=0.
\ee
If we succeed in proving these generalized ST-identities we know that
the parameters $z_{nk}$ generate unphysical insertions. We just differentiate
(\ref{gpST}) by $\chi_{nk}$ and obtain
\be\label{vch}
\partial_{z_{nk}}Z=-\ST \partial_{\chi_{nk}}Z=i \ST[\Delta^-_{(nk)}Z](J,K,L),
\ee
where $\Delta^-$ is an insertion of dimension 4 and $\phi\pi$-charge -1,
generated by $\partial_{\chi_{nk}}$.
Whereas for the doublet $(\alpha_0,\chi)$ we had to enlarge the gauge fixing
we can proceed here more directly because the parameters $z_{nk}$ show up
only in the redefinition of $h$. It is readily seen that one has to
change only $\bar{\Gamma}$ into
\be\label{pcgb}
\bar{\Gamma}(h,c,H,L,z_{nk},\chi_{nk})
	         =\bar{\Gamma}^s(\hat{h},\hat{c},\hat{H},\hat{L})
		 +\sum_{nk}\chi_{nk}[\int K_{\mu\nu}G^{\mu\nu}_{nk}
		 +r_{nk}[\int L_\mu c^\mu]]
\ee
with
\begin{eqnarray}\label{ctrgm}
        \hat{h}&=&\mathcal{F}(h,z_{nk})
	         \qquad \hat{H}=\frac{\delta}{\delta\hat{h}}
	   \int H\mathcal{F}^{-1}(\hat{h},z_{nk})|_{\hat{h}=\mathcal{F}(h,z_{nk})}\\
	\hat{c}&=&y(z_{nk})c \qquad\quad \hat{L}=\frac{1}{y(z_{nk})}L\\
	G_{nk}(h,z_{nk})&=&-\frac{\partial}{\partial z_{nk}}\mathcal{F}^{-1}(h,z_{nk})|_{\hat{h}=\mathcal{F}(h,z_{nk})} \qquad r_{nk}=-\frac{1}{y(z_{nk})}\frac{\partial}{\partial z_{nk}}y(z_{nk}) 
\end{eqnarray}
and $y(z_{nk})$ is a general function of its arguments. From the preceding subsection
we know that for $\chi_{nk}=0$ this is the general solution of the ST-identity.
For $\chi_{nk}\not=0$ one has to go through (\ref{gpST}) to convince one-self
that this is the case.
The parameters $z_{nk},y(z_{nk})$ will be fixed by normalization conditions. We choose the following one's.\\

The normalization condition (\ref{trnorm5}) fixes $y(z_{nk})=1$, hence
$r(z_{nk})=0$ (note: $n\ge 2$).
In order to fix $z_{nk}$ one has to look in the general solution of the ST-identity
at the term
$
\int H_{\mu\nu}\brsts\mathcal{F}^{\mu\nu}
=\int\sum_{n,k}z_{nk}H_{\mu\nu}\brsts(h...h)^{\mu\nu},
$
where $\brsts$ denotes the standard BRST transformation of $h$, and to project such that e.g.\
\be
\partial_p\Gamma_{Hc\mathcal{P}(\underbrace{h...h}_{n})}|_{p=0}=z_{nk}
\ee
Here $\mathcal{P}$ denotes a suitable projector. We do not work out the details
of its definition.

\subsection{Gauge parameter independence in higher orders\label{se:GPIHO}}
The aim is now to prove (\ref{chiZ}) and (\ref{gpST}) to all orders of 
perturbation theory. Taken together
\be\label{cgpd}
\ST(\Gamma)+(\chi\partial_{\alpha_0}+\sum_{n,k}(\chi_{nk}\partial_{z_{nk}}))\Gamma=0 \,,
\qquad \ST Z+(\chi\partial_{\alpha_0}+\sum_{n,k}(\chi_{nk}\partial_{z_{nk}}))Z=0 .
\ee
We start from
\begin{multline}\label{sggpv}
	\Gamma^s(h,c,\bar{c},b,K,L) = \bar{\Gamma}^s(h,c,\bar{c},K,L)
	-\frac{1}{2\kappa}\int dxdy\, h^{\mu\nu}(x)
(\partial_\mu b_\nu+\partial_\nu b_\mu)(y) \times \\ 
\times\Big\lbrace\big( \frac{\Box}{4\pi^2} + m^2 \big)\frac{1}{(x-y)^2} \Big\rbrace 
    -\frac{1}{2}\alpha_0\int b_\mu b_\nu\eta^{\mu\nu}
\end{multline}
\begin{multline}
\bar{\Gamma}^s(h,c,\bar{c},K,L) = \Gamma^{s\,({\rm class})}_{\rm inv}(h)
	-\frac{1}{2}\int dxdy\, Q^{s\,\mu\nu}(x)(\partial_\mu \bar{c}_\nu+
\partial_\nu \bar{c}_\mu)(y)\times \\
\times \Big\lbrace\big( \frac{\Box}{4\pi^2} + m^2 \big)\frac{1}{(x-y)^2}\Big\rbrace
+\int(K_{\mu\nu}Q^{s\,\mu\nu}(h,c)-\kappa L_\mu c^\lambda\partial_\lambda c^\mu) \\
	-\frac{1}{4}\chi(\bar{c}_\mu b_\nu+\bar{c}_\nu b_\mu)\eta^{\mu\nu}) .
\end{multline}
The $b$-dependent terms can be trivially regained from the gauge condition 
\be\label{3beq}
\frac{\delta \Gamma^s}{\delta b^\rho}=
    \kappa^{-1}\int dy\,\partial^\mu h_{\mu\rho}(y)
         \Big\lbrace\big( \frac{\Box}{4\pi^2} + m^2 \big)\frac{1}{(x-y)^2} \Big\rbrace-\alpha_0 b_\rho 
                  -\frac{1}{2}\chi\bar{c}_\rho,
\ee
whereas the ghost equation of motion reads
\be\label{3ghe}
	\frac{\delta\Gamma^s}{\delta \bar{c}_\rho(x)}=
	-\int dy\,\partial_\lambda\frac{\delta\Gamma^s}{\delta K_{\lambda\rho}(y)} 
     \Big\lbrace\big( \frac{\Box}{4\pi^2} + m^2 \big)\frac{1}{(x-y)^2}\Big\rbrace +\frac{1}{2}\chi b^\rho.
\ee
The general solution has been obtained on the classical level, (\ref{pcgb}), as 
\be\label{2pcgb}
\bar{\Gamma}(h,c,H,L,z_{nk},\chi_{nk})
	         =\bar{\Gamma}^s(\hat{h},\hat{c},\hat{H},\hat{L})
		 +\sum_{nk}\chi_{nk}[\int K_{\mu\nu}G^{\mu\nu}_{nk}
		 +r_{nk}[\int L_\mu c^\mu]]
\ee
with hatted fields given in (\ref{ctrgm}). Due to the presence of the
parameter doublets the ST-identity has the form
\begin{eqnarray}
	\mathcal{S}(\Gamma)&=&\mathcal{B}(\bar{\Gamma})\\
&\equiv&\int\left[
	\frac{\delta\bar{\Gamma}}{\delta H}\frac{\delta\bar{\Gamma}}{\delta h}
       +\frac{\delta\bar{\Gamma}}{\delta L}\frac{\delta\bar{\Gamma}}{\delta c}
	\right]
	+\chi\frac{\partial\bar{\Gamma}}{\partial\alpha_0} 
	+\sum_{n,k}\chi_{n,k}\frac{\partial\bar{\Gamma}}{\partial z_{n,k}}=0 .
\end{eqnarray}
The non-linear operator $\mathcal{B}(\bar{\gamma})$ and the linear operator 
\be
\mathcal{B}_{\bar{\gamma}}\equiv \int\left[
	\frac{\delta\gamma}{\delta H}\frac{\delta}{\delta h} 
       +\frac{\delta\gamma}{\delta h}\frac{\delta}{\delta H}  
       +\frac{\delta\gamma}{\delta L}\frac{\delta}{\delta c}  
       +\frac{\delta\gamma}{\delta c}\frac{\delta}{\delta L}\right]
       +\chi\frac{\partial}{\partial \alpha_0}
      +\sum_{n,k}\chi_{n,k}\frac{\partial}{\partial z_{n,k}}
\ee
satisfy the identities
\begin{eqnarray}
\mathcal{B}_\gamma\mathcal{B}(\gamma)&=&0 \qquad \forall \gamma \label{1lbG}\\
\mathcal{B}_\gamma\mathcal{B}_\gamma&=&0 \qquad {\rm if}\,\, \mathcal{B}(\gamma)=0 .
	\label{2lbG}
\end{eqnarray}
Since the classical action satisfies the ST-identity, we have for the
tree approximation from (\ref{2lbG})
\be\label{ccdt}
\brstb^2=0 \qquad{\rm for} \qquad \brstb\equiv 
                          \mathcal{B}_{\bar{\Gamma}_{\rm class}},
\ee
i.e. $\brstb$ is nilpotent.

The action principle tells now that
\be\label{ctp}
\mathcal{S}\Gamma=\left[\Delta\right]^5_5\cdot \Gamma= \Delta+O(\hbar\Delta) ,
\ee
where $\Delta$ is an insertion with UV=IR-degree=5 and $Q_{\phi\pi}=1$ and we have 
on the rhs separated the trivial diagram contribution (tree diagrams) from higher 
orders (loop diagrams). 
If we do not admit counterterms depending on $\alpha_0$, which is possible
since the $b$-equation of motion can be integrated trivially, we can
discard in the following the contribution of the doublet ($\alpha_0,\chi)$
and have to discuss only the doublets $(z_{nk},\chi_{nk})$.
(\ref{ccdt}) leads then to the consistency condition
\be\label{bccdt}
\brstb \Delta=0,
\ee
which is a classical equation.
Furthermore gauge condition (\ref{3beq}) and ghost equation of motion (\ref{3ghe})
imply that the local functional $\Delta$ only depends on the fields
$h,c,H,L$.

The general solution of (\ref{bccdt}) is given by
\be\label{gsccdt}
\Delta= \brstb\hat{\Delta}+r\mathcal{A}(h,c),
\ee
where $\hat{\Delta}$ is an integrated local insertion (functional of $h,c,H,L$) with
UV=IR-dimension 4 and $Q_{\phi\pi}=0$. $\mathcal{A}$ represents an anomaly, i.e.\ 
has the same properties as $\hat{\Delta}$, but is not a $\brstb$-variation. For 
$z_{nk}=0$ we know already (cf. Sect.\ \ref{se:stidentity}) that the decomposition in 
(\ref{gsccdt}) is valid  and no $\mathcal{A}$(h,c) exists. 
For $z_{nk}\not=0$ no $\mathcal{A}$ can be generated either, but we have to show
that the remaining terms form a $b$-variation.\\
This part of the proof relies only on the doublet structure of $(z_{nk},\chi_{nk})$
and can therefore be taken over literally from \cite[Appendix D, formulae (D.18)--(D.32)]{Piguet:1984mv},
with the result,
that the cohomology is trivial and thus $(\ref{gsccdt})$ verified
with $\mathcal{A}=0$.

In the context of BRST-invariant differential operators we shall need
a corresponding analysis for insertions with the quantum numbers of the action,
i.e.\ UV=IR-dimension=4 and $Q_{\phi\pi}=0$. The field dependent part was
treated above in Sect.\ \ref{se:paraandgpi}, where we constructed the general solution of the
ST-identity. $\Gamma^{\rm class}_{\rm inv}$ turned out to be the only obstruction
to the cohomology,
whereas all external field dependent terms are $\brstb$-variations. The gauge parameter 
dependence is also covered in \cite[Appendix D]{Piguet:1984mv} with the result that
the terms of $\Gamma^{\rm class}_{\rm inv}$ can only have gauge parameter 
independent coefficients, whereas the external field dependent terms are
multiplied with functions of those such that the products are variations under
the general gauge parameter dependent terms. For later use we list them here.
A basis of dimension-4, $\phi\pi$ charge-0 $\brstb$-invariant insertions is provided
by:
\begin{eqnarray}\label{gnvnsrt}
\Gamma_{inv}&=&\int\sqrt{-g}(c_0+c_1R^{\mu\nu}R_{\mu\nu}+c_2R^2+c_3R)(h,z_{nk})\\
\Delta_1(h,c,H,z_{nk},\chi_{nk})&
	=&\brstb\left[d_1(z_1)\int H_{\mu\nu}h^{\mu\nu}  \right] \\
	\Delta_{nk}(h,c,H,z_{nk},\chi_{nk})&
=&\brstb\left[d_{nk}(z_{n,k})\int H_{\mu\nu}(\underbrace{h...h}_{n,k})^{\mu\nu}\right]\\
	\Delta_c(h,c,L)&
=&\brstb\left[e_c\int L_\mu c^\mu\right].
\end{eqnarray}
Recall that counterterms must not depend on 
$\alpha_0$, we work in Landau gauge, $\alpha_0=0$, hence there is also no $\chi$ 
present.

These $\brstb$-invariant insertions are in one-to-one correspondence to $\brstb$-symmetric
differential operators
\begin{eqnarray}\label{bsdffp}
	c_0\partial{c_0}\Gamma&=&\int\sqrt{-g}c_0 \\
	c_1\partial{c_1}\Gamma&=&\int\sqrt{-g}c_1R^{\mu\nu}R_{\mu\nu}\\
	c_2\partial{c_2}\Gamma&=&\int\sqrt{-g}c_2R^2\\
	c_3\partial{c_3}\Gamma&=&\int\sqrt{-g}c_3\kappa^{-2}R\\
\left[d_1(z_1)\mathcal{N}_h+b(d_1)\mathcal{N}^{(-)}_h\right]\Gamma&=&\Delta_1\\
\left[d_{n,k}\partial_{z_{n,k}}+b(d_{n,k})\partial_{\chi_{n,k}}\right]\Gamma&=&
	-\Delta_{n,k}+O(h^{n+1}\\
\left[e_c\mathcal{N}_c+b(e_c)\mathcal{N}^{(-)}_c\right]\Gamma&=&-\Delta_c .
\end{eqnarray}
Here we have defined combinations of counting operators 
\be\label{lcto}
N_\phi\equiv\int\phi\frac{\delta}{\delta\phi} 
\ee
for the fields.\\
\begin{eqnarray}\label{slcto}
	\mathcal{N}_h\Gamma&\equiv&\left[N_h-N_K-N_{\bar{c}}-N_b+2\alpha_0\partial_{\alpha_0}+2\chi\partial_\chi\right]\Gamma\\
	\mathcal{N}^{(-)}_h\Gamma&\equiv&\int Kh-\int\bar{c}\frac{\delta}{\delta b}\Gamma-2\alpha_0\partial_\chi\Gamma\\
\mathcal{N}_c\Gamma&\equiv&\left[N_c-N_L\right]\Gamma\\
	\mathcal{N}^{(-)}_c\Gamma&\equiv&-\int L_\rho c^\rho
\end{eqnarray}
and went back from the variable $H_{\mu\nu}$ in (\ref{gnvnsrt}) to the
variables $K_{\mu\nu},\bar{c}$.

\subsection{Normalization conditions III\label{se:nc3}}
The normalization conditions (\ref{highnorm})--(\ref{highnorm1}) have to be supplemented by those introducing $z_{nk}$ 
and read now
\begin{eqnarray}\label{highnormG}
\frac{\partial}{\partial p^2}\,\gamma^{(2)}_{\rm TT\,|{\substack{ p=0 \\ s=1}}}&=
				   &c_3\kappa^{-2}\\
\frac{\partial}{\partial p^2}\frac{\partial}{\partial p^2}\,
	\gamma^{(2)}_{\rm TT\,|{\substack{ p^2=-\mu^2 \\ s=1}}}&=&-2c_1\\
\frac{\partial}{\partial p^2}\frac{\partial}{\partial p^2}\,
	\gamma^{(0)}_{\rm TT\,|{\substack{ p^2=-\mu^2 \\ s=1}}}
							   &=&2(3c_2+c_1)\\
\Gamma_{h^{\mu\nu}}&=&c_0=0\\							   
\frac{\partial}{\partial p_\sigma}
	\Gamma_{K^{\mu\nu}c_\rho|{\substack{ p^2=-\mu^2 \\ s=1}}}&=&
                                -i\kappa(\eta^{\mu\sigma}\delta^\nu_\rho
	                          +\eta^{\nu\sigma}\delta^\mu_\rho
		  -\eta^{\mu\nu}   \delta^\sigma_\rho)\label{highnormG1} \\
	\partial_p\Gamma_{Kc\mathcal{P}(\underbrace{h...h}_{n})}|_{\substack{ p^2=-\mu^2 \\ s=1}}&=&z_{nk}\\
\frac{\partial}{\partial p^\lambda}
	\Gamma_{{L_\rho}c^\sigma c^\tau|{\substack{ p^2=-\mu^2 \\ s=1}}}&=&
			-i\kappa(\delta^\rho_\sigma\eta_{\lambda\tau}
				  -\delta^\rho_\tau\eta_{\lambda\sigma}).
\end{eqnarray}
Imposing the $b$-equation of motion (\ref{beq}) still fixes $\alpha_0$ 
and the $b$-amplitude, whereas (\ref{highnorm}) again
fixes the $h$-amplitude. $\mathcal{P}$ projects to the $k^{\rm th}$ independent term in $\sum_{n,k}(\underbrace{h...h}_{n})^{\mu\nu}$.

\section{Discussion and conclusions\label{se:DisCon}}
In the present paper we propose the perturbative quantization of classical
Einstein-Hilbert gravity. The version which we discuss has as
background ordinary Minkowski space on which the respective theory deals with a massless spin two field with interactions provided by classical EH. The problem of power counting non-renormalizability is overcome in two steps. First we introduce the higher derivative terms $R^2,R^{\mu\nu} R_{\mu\nu}$ which make the model power counting renormalizable,
create however negative norm states, hence can only be considered as a Pauli-Villars regularization. Then there are two spin two fields in the model, their combined propagator yielding dynamic dimension $0$ to the combined field $h$.
In a second step we perform momentum space subtractions according to the Bogoliubov-Parasiuk-Hepp-Zimmermann-Lowenstein scheme, treating the $R$-term as an oversubtracted normal product with subtraction degrees $d=r=4$. This takes correctly into account the vanishing naive dimension of the combined field $h$.\\
Since this model is closed under renormalization we have at our disposal the full machinery of the BPHZL scheme, in particular the action principle, which admits the systematic construction and proof of the Slavnov-Taylor identity, i.e.\ 
formal (pseudo -)unitarity,
and parametric partial differential equations. Those are the Lowenstein-Zimmermann equation, which says that Green functions are independent of the auxiliary mass term $M$ which belongs to the scheme. Further there are the renormalization group and Callan-Symanzik equation. These control completeness of the parametrization and scaling, respectively.\\
The final step of establishing a quantized EH-theory cannot be taken since the regulators cannot be eliminated in a controlled way.  
The model has to stay as such, which suggests that the higher derivative terms in the action constitute an essential part of the theory, for which traces of the Einstein-Hilbert action have to be extracted.
However physical states for the EH theory can be constructed, according to the standard quartet mechanism of \cite{KuOj}: projecting out states with negative norm and then forming equivalence classes of states with vanishing norm.
The full $S$-matrix, which is derived from ST, is thus restricted to EH theory, but its unitarity is questionable.
Even if the latter would hold, the dependence on the parameters $c_1$ and $c_2$ presumably prevail nevertheless.\\
Next we mention a few items in which the present paper differs from previous attempts to solve the quantization problem. First of all we do not rely on an invariant regularization, i.e.\ the regularization employed in dimensional renormalization, which, it seems, has been used exclusively in the past. The BPHZL renormalization scheme requires that power counting is such that convergence results, e.g.\ for Green functions. This  we provide here. Then the study of anomalies is constructively possible.
We can thus safely use results obtained in the past in many papers by purely algebraic reasoning (cf.\ \cite{Baulieu:1983tg,Dragon:2012au}).
Those can now be completed with a power counting based, ``analytic'' treatment. This refers not only to anomaly discussions, but also to the so-called Batalin-Vilkovisky formalism (in quantum field theory).
The latter has been invoked for quantum gravity, specifically also for EH, in \cite{Brunetti:2013maa}. 
Although therein many innovative concepts have been introduced the construction suffers from the lack of renormalizability. 
In the presumably simplest context we present a solution for this, which is lacking a proof of unitarity though.
The hope then is that this example is fruitful in that wider range. For instance, when invoking the principle of generalized covariance (cf.\ \cite{Brunetti:2001dx}) one always relates two systems of manifold plus metric. One of them
could then just be ours with Minkowski space plus metric, and fluctuations around it.\\  
Another aspect concerns the field variable $h^{\mu\nu}$. In the literature most commonly used is the Goldberg variable $h^{\mu\nu}=\sqrt{-g}g^{\mu\nu}-\eta^{\mu\nu}$,
whereas we use $h^{\mu\nu}=g^{\mu\nu}-\eta^{\mu\nu}$. These variables are not equivalent (in the sense of point transformations), but differ by unphysical 
degrees of freedom. Our variable has the advantage that two-point functions (1PI and propagator) have fewer components in the spin expansion to be dealt with. \\
Let us also recall that our way of proceeding forced us to treat the fundamental field $h$ as a field of vanishing canonical dimension. It is then mandatory to discuss non-linear field redefinitions. They are quite analogous to those which one has to face in a power counting non-renormalizable formulation, but can here be handled in a completely controlled manner like in supersymmetric Yang-Mills theories when supersymmetry is linearly realized.

In the context of the CS-equation and in view of the RG-equation one comes in the vicinity of the concept of ``asymptotic safety'' \cite{Reuter:2012id},
where one deals directly with the infinite dimensional space of interactions with
arbitrarily high dimension which we (by purpose) avoided. It would be interesting to see where our proposal is to be detected there.
Similarly one could repeat the analysis 
of \cite{Fradkin:1981iu} under the present auspices. There one worked in Euclidean space and with the full, non-unitary model.

By its very nature our approach differs from the treatment as effective theory \cite{Donoghue:1995cz}, where one tries to find quantum effects of gravity without constructing a fundamental quantized model of it -- as one can formulate a model of mesons and hadrons without recurrence to QCD with its unsolved problem of confinement.

Extension of the present work to include matter seems to be most straightforward for scalar fields. 
Then one could contribute to the study of observables \cite{Frob:2017gyj} and spontaneous scale symmetry breaking \cite{Kubo:2020fdd}, having at one's disposal a power-counting renormalizable model. 
Adding vector fields of matter would also not require serious changes. Once fermions are introduced
one should employ the vierbein-formalism. In that context it should be particularly rewarding that one can now safely discuss chiral anomalies which are otherwise not easily handled. Also supergravity theories would deserve new interest.

Some new ideas or methods seem to be required, if one wants to go over to curved background. In particular normalization conditions and  asymptotic limits pose problems which in the present, flat background case are absent. 
A recent study on the formulation of perturbative gravity in presence of a cosmological constant \cite{Anselmi:2019ukt} tackles the challenge of developing new tools and uses a prescription to treat new degrees of freedom, which is described in \cite{Anselmi:2017ygm}.
Another candidate as far as methods are concerned is provided by the fairly recent work of one of the present authors (SP) \cite{Pottel:2017mnc}. 
There the BPHZ scheme has been extended to analytic (curved) spacetimes. I.e.\ propagators, power counting and the like are those of curved spacetime. 
Massive and massless models can be treated on an equal footing. 
For a graviton field details would have to be worked out. 
The problem of normalization conditions seems to be linked to asymptotic properties of the spacetime manifold which, regarding physics, is absolutely plausible.
This could be an interesting area of future research. 

\appendix
\section{Notation and Conventions\label{se:notation}}
\subsection{Geometry\label{se:geometry}}
In this work, we are employing the conventions below, which are the ``timelike conventions'' of Landau-Lifschitz (cf. \cite{Misner:1974qy}).
\[
\renewcommand{\arraystretch}{1.5}
\begin{array}{llcl}
{\rm flat \,\, metric}&\eta^{\mu\nu}&=&{\rm diag}\,\,\,(+1,-1,-1,-1)\\
{\rm Christoffel}&\Gamma^\sigma_{\mu\nu}&=&\frac{1}{2}
             g^{\sigma\rho}(\partial_\nu g_{\rho\mu}
                           +\partial_\mu g_{\rho\nu}-\partial_\rho g_{\mu\nu})\\
{\rm Riemann}    &\tensor{R}{^\lambda_{\nu\rho\sigma}} &=&\partial_\rho  \Gamma^\lambda_{\nu\sigma}
                           -\partial_\sigma\Gamma^\lambda_{\nu\rho}
                           +\Gamma^\lambda_{\tau\rho}  \Gamma^\tau_{\nu\sigma}
                           -\Gamma^\lambda_{\tau\sigma}\Gamma^\tau_{\nu\rho}\\
{\rm Ricci}&R_{\mu\nu}&=&\partial_\sigma\Gamma^\sigma_{\mu\nu} 
                        -\partial_\nu\Gamma^\sigma_{\mu\sigma}
                        +\Gamma^\sigma_{\mu\nu} \Gamma^\rho_{\sigma\rho}
                        -\Gamma^\rho_{\mu\sigma}\Gamma^\sigma_{\nu\rho}\\
{\rm curvature\,\,scalar}&R&=&g^{\mu\nu}R_{\mu\nu}                             
\end{array}
\]

\subsection{Projection operators\label{se:projectionoperators}} 
In order to cope with the spin properties of the field $h^{\mu\nu}$ it is 
useful to introduce projection operators. They are known at least since 
\cite{VanNieuwenhuizen:1973fi} and we shall use a notation due to \cite{Biazotti:2013yda}.
Based on the transverse and longitudinal projectors for vectors
\be\label{provec}
     \theta_{\mu\nu}\equiv \eta_{\mu\nu}-\frac{p_\mu p_\nu}{p^2} \qquad \qquad
\omega_{\mu \nu} \equiv \frac{p_\mu p_\nu}{p^2}
\ee
 the projectors are defined as 
\begin{eqnarray}\label{projs}
P^{(2)}_{{\rm TT}\mu\nu\rho\sigma} &\equiv&
 \frac{1}{2}(\theta_{\mu\rho}\theta_{\nu\sigma} +\theta_{\mu\sigma}\theta_{\nu\rho})
                                -\frac{1}{3}\theta_{\mu\nu}\theta_{\rho\sigma}\\
P^{(1)}_{{\rm SS}\mu\nu\rho\sigma} &\equiv&
	\frac{1}{2}(\theta_{\mu\rho}\omega_{\nu\sigma} 
	+\theta_{\mu\sigma}\omega_{\nu\rho} +\theta_{\nu\rho}\omega_{\mu\sigma}
	                   +\theta_{\nu\sigma}\omega_{\mu\rho})\\
P^{(0)}_{{\rm TT}\mu\nu\rho\sigma} &\equiv& 
		   \frac{1}{3}(\theta_{\mu\nu}\theta_{\rho\sigma})\\ 
P^{(0)}_{{\rm WW}\mu\nu\rho\sigma} &\equiv& 
		 \omega_{\mu\nu}\omega_{\rho\sigma}\\ 
P^{(0)}_{{\rm TW}\mu\nu\rho\sigma} &\equiv&
		 \frac{1}{\sqrt{3}}\theta_{\mu\nu}\omega_{\rho\sigma}\\ 
P^{(0)}_{{\rm WT}\mu\nu\rho\sigma} &\equiv&
		 \frac{1}{\sqrt{3}}\omega_{\mu\nu}\theta_{\rho\sigma} .
\end{eqnarray}
They satisfy the closure relation
\be\label{clsre}
(P^{(2)}_{TT} + P^{(1)}_{SS} + P^{(0)}_{TT} + P^{(0)}_{WW})_{\mu\nu\rho\sigma} =
   \frac{1}{2}(\eta_{\mu\rho}\eta_{\nu\sigma}+\eta_{\mu\sigma}\eta_{\nu\rho}).
\ee

\subsection{Tables\label{se:tables}}
We list dimensions $d$, $\phi\pi$-charge $Q_{\phi\pi}$ of (functions of) fields and parameters in the theory. For propagating and external fields we have
\[
\begin{array}{lccccccc}
            &h^{\mu\nu} &c  &\bar{c}&b   &K_{\mu\nu}&H_{\mu\nu}&L_\rho\\ \hline
 d          &0          &0  &2      &2   &3         & 3        &3     \\ 
 Q_{\phi\pi}&0          &+1 &-1     &0   &-1        &-1        &-2  
 \end{array}
 \]
Functions of (external-) fields have
 \[
 \begin{array}{lccccccc}
             &\hat{h}^{\mu\nu}&G^{\mu\nu}&\hat{H}_{\mu\nu} &M^{\mu\nu}&N^{\mu\nu}&\mathcal{M}^{(')}&\mathcal{N}^{(')}\\ \hline
  d          &0               &0         &3                &0         &1  & 4  &3\\
  Q_{\phi\pi}&0               &0         &-1               &0         &-1 & 0  &-1
 \end{array}
 \]
The parameters follow
\[
\begin{array}{lccccc}
            &\kappa&\alpha_0 &\chi  &z_{nk} &\chi_{nk}\\ \hline
d           & -1   &0        &0     &0      & 0  \\
Q_{\phi\pi} &  0   &0        &-1    &0      &-1
\end{array}
\]

\newpage

\subsection{Acronyms\label{se:acronyms}}
The following acronyms are used throughout the text:
\begin{table}[h]
  \begin{tabular}{ll}
    1PI & One-Particle Irreducible \\
    BPHZL & Bogoliubov-Parasiuk-Hepp-Zimmermann-Lowenstein \\
    BRST & Becchi-Rouet-Stora-Tyutin \\
    CS & Callan-Symanzik (Equation) \\
    EH & Einstein-Hilbert \\
    FP & Faddeev-Popov \\
    IR & Infrared \\
    LZ & Lowenstein-Zimmermann (Equation) \\
    RG & Renormalization Group (Equation) \\
    ST & Slavnov-Taylor (Identity) \\
    UV & Ultraviolet \\
    YM & Yang-Mills\\
    ZI & Zimmermann Identity
  \end{tabular}
\end{table}

\section{$\brsts_0$-Invariance\label{se:brst0inv}}
In Sect.\ \ref{se:push} we need the fact that 
$\Gamma^{\rm inv}_{hh}(m^2)$ is invariant under 
the abelian BRST transformation  
\be\label{bbrst}
\brsts_0h^{\mu\nu}=-\kappa(\partial^\mu c^\nu+\partial^\nu c^\mu).
\ee
We check this for 
\be\label{prjps}
\Gamma^{\rm inv}_{hh}=\int h\,(\sum_{rKL}\gamma^{(r)}_{KL}P^{(r)}_{\rm KL})\,h,
\ee
the projectors understood as expression in terms of differential operators which (as seen from Fourier transform) admits integration by parts.
For $r=2,0$ it is readily derived that the variation vanishes
due to the transversality of the projectors. Hence the respective $\gamma$'s are not restricted. For the other components we find
\begin{eqnarray}\label{prjps2}
 \brsts_0\int h(\gamma^{(1)}_{\rm SS}P^{(1)}_{\rm SS})h&=&-4\int\gamma^{(1)}_{\rm SS}\,
                       (\theta_{\nu\rho}\partial_\sigma c^\nu+
                        \theta_{\nu\sigma}\partial_\rho c^\nu) h^{\rho\sigma}\\
 \brsts_0\int h(\gamma^{(0)}_{\rm  WW}P^{(0)}_{\rm WW})h &=& -4\int\gamma^{(0)}_{\rm WW}\,
                       \partial_\lambda c^\lambda\omega_{\mu\nu}h^{\mu\nu}\\
 \brsts_0\int h(\gamma^{(0)}_{\rm TW}P^{(0)}_{\rm TW})h&=&-\int\gamma^{(0)}_{\rm TW}\, 
                     h^{\mu\nu}\theta_{\mu\nu}\frac{2}{\sqrt{3}}\partial_\lambda c^\lambda\\
 \brsts_0\int h(\gamma^{(0)}_{\rm WT}P^{(0)}_{\rm WT})h&=&-\int\gamma^{(0)}_{\rm WT}\,
                     \frac{2}{3}\partial_\lambda c^\lambda\theta_{\rho\sigma}h^{\rho\sigma} .
 \end{eqnarray}
Cancellation between different spin components can not take place, hence these 
$\gamma^{(r)}_{\rm KL}$ vanish. But this situation is precisely realized in the tree approximation.
Let us remark that for the Goldberg variable $h^{\mu\nu}= g^{\mu\nu}-\eta^{\mu\nu}$ and its respective $\brsts_0$-variation an analogous result can be derived.
However a relation between the $TT$-components $r=2,0$ will be only established by the $\brsts_1$-variation which is non-linear.

\section{Partial Fractions}\label{se:partfr}
In Landau gauge the free propagators have only two non-vanishing spin components
\be\label{frprsp}
\langle hh \rangle^{(2)}_{\rm TT}\,=\frac{-i}{p^2-m^2}\cdot\frac{1}{c_1p^2-c_3\kappa^{-2}}\qquad
\langle hh \rangle^{(0)}_{\rm TT}\,=\frac{i}{p^2-m^2}\cdot\frac{1}{(3c_2+c_1)p^2 +\frac{1}{2}c_3\kappa^{-2}}.
\ee
Their decomposition into partial fractions reads
\begin{eqnarray}\label{pfrprs}
\langle hh \rangle^{(2)}_{\rm TT}&=&\frac{1}{c_3\kappa^{-2}-c_1m^2}\cdot\frac{i}{p^2-m^2}
         \quad+\frac{-i}{c_3\kappa^{-2} (p^2-\frac{c_3}{c_1\kappa^2})}\nonumber\\
 {\rm pole}&:&\, p^2=m^2 \quad{\rm res}_{|{m^2=0}}=\frac{i}{c_3\kappa^{-2}}
 \quad {\rm pole:}\, p^2=\frac{c_3}{c_1\kappa^2} \quad {\rm res}=\frac{-i}{c_3\kappa^{-2}}\\                 
\langle hh \rangle^{(0)}_{\rm TT}&=&\frac{-1}{c_3\kappa^{-2}-2(3c_2+c_1)m^2}\cdot\frac{2i}{p^2-m^2}           \quad+\frac{1}{c_3\kappa^{-2}}\cdot\frac{2i}{p^2+\frac{c_3\kappa^{-2}}{2(3c_2+c_1)}}\nonumber\\
{\rm pole}&:&\,p^2=m^2 \quad{\rm res}_{|{m^2=0}}=\frac{-2i\kappa^2}{c_3}
\quad {\rm pole:}\, p^2=-\frac{c_3\kappa^{-2}}{2(3c_2+c_1)}\quad {\rm res}=\frac{2i\kappa^2}{c_3}
\end{eqnarray}
In the spin two parts the massless pole has positive residue, the massive pole instead has negative residue. Hence the first is physical, the second not. In the spin zero contribution the situation is reversed. When projecting to the massless contributions in the asymptotic limit this spin zero part belongs to the negative metric contribution and has to be canceled in the quartet mechanism.

\section*{Acknowledgements}
KS is deeply indebted to Elisabeth Kraus and Olivier Piguet for many years of joint work. Quite a few parts of the present paper are based on it. He is very grateful to Manfred Salmhofer for encouragement. 
SP gratefully acknowledges the hospitality of the Max Planck Institute for Mathematics in the Sciences (Leipzig), where parts of this work have been concluded.

\bibliographystyle{alpha}
\bibliography{operator_weyl}

\newcommand{\etalchar}[1]{$^{#1}$}
\begin{thebibliography}{KKL{\etalchar{+}}20}

\bibitem[ALS97]{Asorey:1996hz}
M.~Asorey, J.~L. Lopez, and I.~L. Shapiro.
\newblock {Some remarks on high derivative quantum gravity}.
\newblock {\em Int. J. Mod. Phys. A}, 12:5711--5734, 1997.

\bibitem[Ans17]{Anselmi:2017ygm}
Damiano Anselmi.
\newblock {On the quantum field theory of the gravitational interactions}.
\newblock {\em JHEP}, 06:086, 2017.

\bibitem[Ans19]{Anselmi:2019ukt}
Damiano Anselmi.
\newblock {Fakeons, unitarity, massive gravitons and the cosmological
  constant}.
\newblock {\em JHEP}, 12:027, 2019.

\bibitem[Bau85]{Baulieu:1983tg}
L.~Baulieu.
\newblock {Perturbative Gauge Theories}.
\newblock {\em Phys. Rept.}, 129:1, 1985.

\bibitem[BBH95a]{Barnich:1994kj}
G.~Barnich, F.~Brandt, and M.~Henneaux.
\newblock {General solution of the Wess-Zumino consistency condition for
  Einstein gravity}.
\newblock {\em Phys. Rev. D}, 51:1435--1439, 1995.

\bibitem[BBH95b]{Barnich:1995ap}
G.~Barnich, F.~Brandt, and M.~Henneaux.
\newblock {Local BRST cohomology in Einstein Yang-Mills theory}.
\newblock {\em Nucl. Phys. B}, 455:357--408, 1995.

\bibitem[BDdG14]{Biazotti:2013yda}
H.A. Biazotti, D.~Dalmazi, and G.B. de~Gracia.
\newblock {Dimensional reduction of the massless limit of the linearized 'New
  Massive Gravity`}.
\newblock {\em Eur. Phys. J. C}, 74(2):2747, 2014.

\bibitem[Bec85]{Becchi:1985bd}
C.~Becchi.
\newblock {Lectures on the Renormalization of Gauge Theories}.
\newblock In {\em {Les Houches Summer School on Theoretical Physics:
  Relativity, Groups and Topology}}, pages 787--821, 1 1985.

\bibitem[BFR16]{Brunetti:2013maa}
R.~Brunetti, K.~Fredenhagen, and K.~Rejzner.
\newblock {Quantum gravity from the point of view of locally covariant quantum
  field theory}.
\newblock {\em Commun. Math. Phys.}, 345(3):741--779, 2016.

\bibitem[BFV03]{Brunetti:2001dx}
R.~Brunetti, K.~Fredenhagen, and R.~Verch.
\newblock {The Generally covariant locality principle: A New paradigm for local
  quantum field theory}.
\newblock {\em Commun. Math. Phys.}, 237:31--68, 2003.

\bibitem[BS59]{Bogolyubov:1980nc}
N.N. Bogolyubov and D.V. Shirkov.
\newblock {\em {Introduction to the Theory of Quantized Fields}}, volume~3.
\newblock 1959.

\bibitem[CL76]{Clark:1976ym}
T.~E. Clark and J.~H. Lowenstein.
\newblock {Generalization of Zimmermann's Normal-Product Identity}.
\newblock {\em Nucl. Phys. B}, 113:109--134, 1976.

\bibitem[DB12]{Dragon:2012au}
N.~Dragon and F.~Brandt.
\newblock {\em {BRST Symmetry and Cohomology}}, pages 3--86.
\newblock 5 2012.

\bibitem[Don95]{Donoghue:1995cz}
J.~F. Donoghue.
\newblock {Introduction to the effective field theory description of gravity}.
\newblock In {\em {Advanced School on Effective Theories}}, 6 1995.

\bibitem[FL18]{Frob:2017gyj}
M.~B. Fr\"ob and W.~C.~C. Lima.
\newblock {Propagators for gauge-invariant observables in cosmology}.
\newblock {\em Class. Quant. Grav.}, 35(9):095010, 2018.

\bibitem[FT82]{Fradkin:1981iu}
E.S. Fradkin and Arkady~A. Tseytlin.
\newblock {Renormalizable asymptotically free quantum theory of gravity}.
\newblock {\em Nucl. Phys. B}, 201:469--491, 1982.

\bibitem[GS86]{Goroff:1985th}
Marc~H. Goroff and Augusto Sagnotti.
\newblock {The Ultraviolet Behavior of Einstein Gravity}.
\newblock {\em Nucl. Phys. B}, 266:709--736, 1986.

\bibitem[HO20]{Hehl:2020hhp}
Friedrich~W. Hehl and Yuri~N. Obukhov.
\newblock {Conservation of Energy-Momentum of Matter as the Basis for the Gauge
  Theory of Gravitation}.
\newblock {\em Fundam. Theor. Phys.}, 199:217--252, 2020.

\bibitem[IZ80]{Itzykson:1980rh}
C.~Itzykson and J.B. Zuber.
\newblock {\em {Quantum Field Theory}}.
\newblock International Series In Pure and Applied Physics. McGraw-Hill, New
  York, 1980.

\bibitem[KKL{\etalchar{+}}20]{Kubo:2020fdd}
J.~Kubo, J.~Kuntz, M.~Lindner, J.~Rezacek, P.~Saake, and A.~Trautner.
\newblock {Unified Emergence of Energy Scales and Cosmic Inflation}.
\newblock 12 2020.

\bibitem[KO78]{KuOj}
T.~Kugo and I.~Ojima.
\newblock {Subsidiary Conditions and Physical S Matrix Unitarity in Indefinite
  Metric Quantum Gravitational Theory}.
\newblock {\em Nucl. Phys. B}, 144:234--252, 1978.

\bibitem[KS92a]{EKKSII}
E.~Kraus and K.~Sibold.
\newblock {Conformal transformation properties of the energy momentum tensor in
  four-dimensions}.
\newblock {\em Nucl. Phys. B}, 372:113--144, 1992.

\bibitem[KS92b]{EKKSIII}
E.~Kraus and K.~Sibold.
\newblock {The General transformation law of the gravitational field via
  Noether's procedure}.
\newblock {\em Annals Phys.}, 219:349--363, 1992.

\bibitem[KS93]{EKKSI}
E.~Kraus and K.~Sibold.
\newblock {Local couplings, double insertions and the Weyl consistency
  condition}.
\newblock {\em Nucl. Phys. B}, 398:125--154, 1993.

\bibitem[Low71]{Lowenstein:1971jk}
J.H. Lowenstein.
\newblock {Differential vertex operations in Lagrangian field theory}.
\newblock {\em Commun. Math. Phys.}, 24:1--21, 1971.

\bibitem[Low75]{Lowenstein:1975ug}
J.~H. Lowenstein.
\newblock {BPHZ Renormalization}.
\newblock In {\em {International School of Mathematical Physics, 2nd course:
  Renormalization Theory}}, 12 1975.

\bibitem[Low76]{Lowenstein:1975ps}
J.H. Lowenstein.
\newblock {Convergence Theorems for Renormalized Feynman Integrals with
  Zero-Mass Propagators}.
\newblock {\em Commun. Math. Phys.}, 47:53--68, 1976.

\bibitem[LS76]{Lowenstein:1975ku}
J.~H. Lowenstein and E.~R. Speer.
\newblock {Distributional Limits of Renormalized Feynman Integrals with
  Zero-Mass Denominators}.
\newblock {\em Commun. Math. Phys.}, 47:43--51, 1976.

\bibitem[MTW73]{Misner:1974qy}
C.~W. Misner, K.S. Thorne, and J.A. Wheeler.
\newblock {\em {Gravitation}}.
\newblock W. H. Freeman, San Francisco, 1973.

\bibitem[Pot17]{Pottel:2017mnc}
S.~Pottel.
\newblock {Configuration Space BPHZ Renormalization on Analytic Spacetimes}.
\newblock 2017.

\bibitem[PS84]{Piguet:1984mv}
O.~Piguet and K.~Sibold.
\newblock {Gauge Independence in $N=1$ Supersymmetric \{Yang-Mills\} Theories}.
\newblock {\em Nucl. Phys. B}, 248:301, 1984.

\bibitem[PS85]{Piguet:1984js}
O.~Piguet and K.~Sibold.
\newblock {Gauge Independence in Ordinary \{Yang-Mills\} Theories}.
\newblock {\em Nucl. Phys. B}, 253:517--540, 1985.

\bibitem[PS86]{Piguet:1986ug}
O.~Piguet and K.~Sibold.
\newblock {\em {Renormalized Supersymmetry. The Perturbation Theory of N=1
  Supersymmetric Theories in Flat Space-Time}}, volume~12.
\newblock 1986.

\bibitem[RS12]{Reuter:2012id}
M.~Reuter and F.~Saueressig.
\newblock {Quantum Einstein Gravity}.
\newblock {\em New J. Phys.}, 14:055022, 2012.

\bibitem[Ste77]{Stelle}
K.S. Stelle.
\newblock {Renormalization of Higher Derivative Quantum Gravity}.
\newblock {\em Phys. Rev. D}, 16:953--969, 1977.

\bibitem[tHV74]{tHooft:1974toh}
G.~'t~Hooft and M.J.G. Veltman.
\newblock {One loop divergencies in the theory of gravitation}.
\newblock {\em Ann. Inst. H. Poincare Phys. Theor. A}, 20:69--94, 1974.

\bibitem[vdV92]{vandeVen:1991gw}
A.E.M. van~de Ven.
\newblock {Two loop quantum gravity}.
\newblock {\em Nucl. Phys. B}, 378:309--366, 1992.

\bibitem[VN73]{VanNieuwenhuizen:1973fi}
P.~Van~Nieuwenhuizen.
\newblock {On ghost-free tensor lagrangians and linearized gravitation}.
\newblock {\em Nucl. Phys. B}, 60:478--492, 1973.

\bibitem[Zim69]{Zimmermann:1969jj}
W.~Zimmermann.
\newblock {Convergence of Bogolyubov's method of renormalization in momentum
  space}.
\newblock {\em Commun. Math. Phys.}, 15:208--234, 1969.

\bibitem[Zim73a]{Zimmermann:1972te}
W.~Zimmermann.
\newblock {Composite operators in the perturbation theory of renormalizable
  interactions}.
\newblock {\em Annals Phys.}, 77:536--569, 1973.

\bibitem[Zim73b]{Zimmermann:1972tv}
W.~Zimmermann.
\newblock {Normal products and the short distance expansion in the perturbation
  theory of renormalizable interactions}.
\newblock {\em Annals Phys.}, 77:570--601, 1973.

\bibitem[Zim75]{Zimmermann:1975gk}
W.~Zimmermann.
\newblock {Remark on Equivalent Formulations for Bogolyubov's Method of
  Renormalization}.
\newblock In {\em {International School of Mathematical Physics, 2nd course:
  Renormalization Theory}}, pages 161--170, 1 1975.

\bibitem[Zim80]{Zimmermann:1979fd}
W.~Zimmermann.
\newblock {The Renormalization Group of the Model of $A^4$ Coupling in the
  Abstract Approach of Quantum Field Theory}.
\newblock {\em Commun. Math. Phys.}, 76:39, 1980.

\end{thebibliography}

\end{document}